\documentclass[fleqn,usenatbib]{mnras}

\usepackage{newtxtext,newtxmath}

\usepackage[T1]{fontenc}

\DeclareRobustCommand{\VAN}[3]{#2}
\let\VANthebibliography\thebibliography
\def\thebibliography{\DeclareRobustCommand{\VAN}[3]{##3}\VANthebibliography}

\usepackage{lineno}
\usepackage{graphicx}	
\usepackage{amsmath}	
\usepackage{xcolor}

\definecolor{purple}{RGB}{128, 0, 128}

\usepackage{soul}
\usepackage{mathtools}
\newcounter{tableeqn}[table]
\renewcommand{\thetableeqn}{\thetable.\arabic{tableeqn}}
\newcounter{tablesubeqn}[tableeqn]

\usepackage{amsmath} 
\usepackage{hyperref}
\usepackage{orcidlink}

\title[He triplet evolution of gas giants around K-dwarfs] 
{Evolution of helium triplet transits of close-in gas giants orbiting K-dwarfs} 

\author[A. P. Allan et al]{Andrew P. Allan\orcidlink{0000-0002-3900-5111},$^{1}$\thanks{E-mail: allan@strw.leidenuniv.nl} 
Aline A. Vidotto\orcidlink{0000-0001-5371-2675},$^{1}$
Carolina Villarreal D’Angelo\orcidlink{0000-0003-1701-7143},$^{2}$
Leonardo A. Dos Santos\orcidlink{0000-0002-2248-3838},$^{3}$
\newauthor
Florian A. Driessen\orcidlink{0000-0003-3005-7377}$^{1}$
\\
$^{1}$Leiden Observatory, Leiden University, P.O. Box 9513, 2300 RA Leiden,
The Netherlands\\ 
$^{2}$Instituto de Astronom\'ia Te\'orica y Experimental. Laprida 854, X5000BGR Córdoba, Argentina \\
$^{3}$Space Telescope Science Institute, 3700 San Martin Drive, Baltimore, MD 21218, USA
}

\date{Accepted 2023 October 30. Received 2023 October 18; in original form 2023 March 17}

\pubyear{2023}

\begin{document}
\label{firstpage}
\pagerange{\pageref{firstpage}--\pageref{lastpage}}
\maketitle


\begin{abstract}
Atmospheric escape in exoplanets has traditionally been observed using hydrogen Lyman-$\alpha$ and H-$\alpha$ transmission spectroscopy, but more recent detections have utilised the metastable helium triplet at 1083$~$nm. Since this feature is accessible from the ground, it offers new possibilities for studying atmospheric escape. Our goal is to understand how the observability of escaping helium evolves during the lifetime of a highly irradiated gas giant. We extend our previous work on 1-D self-consistent hydrodynamic escape from hydrogen-only atmospheres as a function of planetary evolution to the first evolution-focused study of escaping hydrogen-helium atmospheres. Additionally, using these novel models we perform helium triplet transmission spectroscopy. We adapt our previous hydrodynamic escape model to now account for both hydrogen and helium heating and cooling processes and simultaneously solve for the population of helium in the triplet state. To account for the planetary evolution, we utilise evolving predictions of planetary radii for a close-in 0.3$~M_{\rm Jup}$ gas giant and its received stellar flux in X-ray, hard and soft EUV, and mid-UV wavelength bins assuming a K dwarf stellar host. We find that the helium triplet signature diminishes with evolution. Our models suggest that young ($\lesssim 150$~Myr), close-in gas giants ($\sim 1$ to $2~R_{\rm Jup}$) should produce helium 1083$~$nm transit absorptions of $\sim 4\%$ or $\sim 7\%$, for a slow or fast-rotating K dwarf, respectively, assuming a 2\% helium abundance.
 
\end{abstract}




\section{Introduction}

Highly irradiated exoplanets undergo extreme atmospheric escape. This occurs mainly through the process of hydrodynamic escape, in which photoionisations of hydrogen (and, to a lesser extent, helium) atoms by X-ray and extreme-ultraviolet (hereafter XUV) photons heat the atmosphere causing its expansion and ultimately a bulk outflow of the material. This escape process can be modelled by treating the atmosphere as a collisional fluid \citep{yelle_2004, Murray-Clay2009}. In addition to the stellar XUV heating, heating from the planet interior can also contribute to the atmospheric escape mechanism \citep{Ginzburg_2018_core_powered_mdot, 2020MNRAS.499...77K, 2021MNRAS.504.2034K, Kubyshkina_Fossati2022}.

Intense atmospheric escape can affect a planet's overall evolution and structure. The level of atmospheric escape in highly irradiated gas giant planets declines substantially with planetary evolution \citep[e.g.][]{Owen_EVOL_ATM_ESCAP_REVIEW_2019, Allan_Vidotto_2019, 2021A&A...654L...5P}. A reduction in the stellar XUV flux received with age and a growing planetary gravitational force are responsible for this. Failure to account for how atmospheric escape varies with evolution can lead to significant inaccuracies in the predicted atmospheric structure and escape. Strong atmospheric escape over the lifetime of a planet is one of the explanations offered for apparent trends in the detected population of exoplanets, namely the `hot-Neptune desert' \citep{Mazeh_2016, Owen_Dong_2018} and the `radius valley' \citep{Fulton_2017}. Planetary evolution not only impacts the physical process of escape but additionally its corresponding observable signatures. In \citet{Allan_Vidotto_2019}, we found that the hydrogen Lyman-$\alpha$ and H-$\alpha$ signatures of atmospheric escape weaken significantly over the evolution of highly irradiated gas giants with primordial hydrogen atmospheres. 

Since our previous evolution study, there has been an explosion of detections of atmospheric escape using a new spectroscopic signature, the metastable helium triplet at 1083$~$nm \citep[for theoretical studies, see][]{Seager_Sasselov_2000, Oklopcic_2018_10839_window}. This signature's ability to be observed using ground-based (in addition to space-based) facilities is greatly responsible for its popularity, and it presents a great advantage over the more traditional hydrogen Lyman-$\alpha$ signature, which is observable only from space. Additionally, the Lyman-$\alpha$ line suffers from strong interstellar medium absorption and contamination by geocoronal emission, rendering the line core unusable in atmospheric escape analysis.

Since the first detections of escaping helium \citep{Nortmann_2018, Spake_2018_NATURE_He_in_atm, Allart_2018_warm_Nep_Hat_p_11b_He_obs}, there have been over a dozen detections \citep[e.g.,][]{Allart_2019_WASP_107b, Alonso_Floriano_Snellen_2019_HD209458b_He_obs, 2020_Kirk_tail_107b, Spake_2021_tail_wasp107b, 2020_Guilluy_GAPS_He_189, Fu_2022_Hatp11b_JWST}, as well as many constraint-setting non-detections, \citep[e.g.,][]{Zhang_2021_55_cnc_e_non_detection, Krishnamurthy_2021_non_detections_trappist, Fosatti_2022_non_detect_He_Wasp_80b}. For a recent compilation of detections and non-detections, we refer the reader to \citet{Dos_Santos_2022_iauga_obs_of_pl_winds_outflows}.

Of the helium 1083$~$nm detections, K-type host stars appear to be favoured. This has been explained by their relatively low mid-UV flux, which depopulates the helium triplet state, and a high EUV flux, which populates the state through photoionisations followed by recombinations \citep{Oklopcic_2019_dep_st_rad}.
Due to the favourability of K-type hosts, both models and observations of their spectra are important for interpreting escaping helium detections \citep{ France_2016, Johnstone_2021,  Poppenhaeger_2022, Richey_Yowell_2022}. An important consideration is that there is no `typical' XUV flux that fits all K dwarfs \citep[see, e.g.,][]{2016ApJ...824..102L, 2016ApJ...824..101Y, 2022AN....34320008S}. In cool dwarf stars, the high-energy radiation originates in the chromosphere and corona, and they are ultimately related to the magnetic activity of the star \citep[e.g.][]{Vidotto_2014, 2022ApJ...927..179T,2023arXiv230210376N}. Because activity decays substantially with age, the XUV flux of K dwarfs is significantly affected with evolution \citep{Johnstone_2021, 2021A&A...650A.108P}.

In the context of predicting helium transit signatures, many of the existing models adopt a one-dimensional (1-D) Parker wind solution \citep{Parker}, assuming either an isothermal atmosphere or a constant speed of sound \citep{Oklopcic_2018_10839_window, Lampon_2021, pwinds}, and thus neglecting the self-consistent calculation of the energetics of the flow. However, such models often result in a degeneracy between the temperature and the mass-loss rate. \citet{Linssen2022} reduce this degeneracy, by linking isothermal Parker wind solutions with the detailed photoionisation code \textsc{Cloudy} \citep{Ferland1998, Ferland2017}, in order to rule out regions of the explored temperature -mass loss parameter space, without requiring the added complexity of self-consistently modelling the escape, by solving the coupled hydrodynamic and radiative processes consistently. %
The latter is instead the approach of the 1-D models of \textsc{TPCI} \citep{Salz2015TPCI} and \textsc{ATES} \citep{Caldiroli2021}, 
which is more similar to our approach.
 
 In this paper, we compute atmospheric hydrodynamic escape solutions using a single, non-isothermal hydrodynamic model, which self-consistently incorporates energetic balance through different heating and cooling mechanisms. As will be detailed, we have modified the hydrogen-only model of \citet{Allan_Vidotto_2019} to achieve this. Our model still assumes spherical symmetry (i.e, it is 1-D). Because three-dimensional (3-D) models  naturally account for asymmetries arising from a variety of processes, such as interactions with the stellar wind, magnetism and day-to-night side variations \citep{2014MNRAS.438.1654V, 2018MNRAS.479.3115V, 2019ApJ...873...89M, 2020MNRAS.498L..53C, Carolan_2021}, they are more realistic than 1-D models. A wealth of information on the helium signature of escape has already been gained from 3-D models \citep{Shaikhislamov_2021_GJ3470b, Khodachenko_2021_107b_3d, MacLeod_Oklopcic_2021, Wang_Dai2021_WASP107b, Wang_Dai2021_WASP-69b, Rumenskikh_2022_3D_HD189733b, Yan_2022_Wasp_52b}. However, their long computational time leaves 3-D models poorly suited for evolution studies, like ours, that require numerous models at differing ages.

Here, we model the long-term evolution of atmospheric escape of planets orbiting K dwarfs, taking into account both the evolution of the host star and the planet. The main goal is to study the evolution of the helium 1083$~$nm signature. To the best of our knowledge, this study is the first with this goal. Our model computes self-consistently the ionisation of hydrogen, as well as both the single and double ionisation of helium and its population in the $1^1S$, $2^1S$ and $2^3S$ states. In section \ref{sec:model_hydro}, we describe this model. Section \ref{sec:evol_dep_params} outlines our planetary evolution set-up, while section \ref{sec:escape_results} showcases how the physical process of atmospheric escape varies with evolution. In section \ref{sec:observabilty_he_transits}, we present our ray-tracing technique for simulating helium 1083$~$nm transmission spectroscopic transits and the resulting transit predictions with evolution. Finally, section \ref{sec:conclusions} summarises the main findings of our study. 

\section{Atmospheric escape model through photoionisation}
\label{sec:model_hydro}

One of the main drawbacks of describing atmospheric escape using a Parker wind model is that the solution of these types of models are highly sensitive to the assumed isothermal temperature \citep{Parker}. Additionally, the escape rates also depend strongly on the assumed base densities. While Parker-type models are extremely useful, allowing for quick predictions for atmospheric escape \citep[e.g.][]{pwinds}, they  
omit important physics such as heating and cooling processes, which produce non-isothermal temperature structures, which can then affect the derived signatures of spectroscopic transits. 

To model the evolution of atmospheric escape, we update the photoionisation-driven escape model described in \citet[][see also \citealt{Murray-Clay2009}]{Allan_Vidotto_2019} to now include helium in addition to hydrogen. Original and current model versions are both non-isothermal, 1-D, spherically symmetric and treat the escaping atmosphere as a fluid. They solve the equations of fluid dynamics in a co-rotating frame using a shooting method approach  \citep[e.g.][]{2006ApJ...639..416V}. Additionally, the new helium-incorporated version presented in this work

\begin{itemize}
  
\item tracks the state of helium, accounting for transitions between the helium $1^1S, ~ 2^1S, ~ 2^3S, $ singly and doubly ionised states (displayed in Figure \ref{fig:sketch}).

\item Considers heating and cooling due to hydrogen and helium.

\item Reads separate X-ray, mid-UV, hard and soft EUV fluxes. %

\item Considers photons emitted in direct recombinations and radiative decays to the helium $(1^1S)$ state.

\end{itemize}

Previously, we considered only photoionisations arising from a given monochromatic flux of EUV photons, implying that the input flux of EUV photons was concentrated at a representative photon energy of 20~eV. Here, we drop the monochromatic approximation and instead utilise four separate energy flux bins, each with their own representative energies ($e_{\lambda}$). Going beyond a monochromatic flux is necessary for our present work due to the differing minimum photoionisation energies of hydrogen and the considered helium states. Furthermore, the individual flux bins each have their own unique dependencies on age. Table \ref{tab:photoion_constants} lists the hydrogen and helium photoionisations we consider and their respective cross-sections ($\sigma_{\lambda}$) and excess kinetic energy factors ($\epsilon_{\lambda}$). The wavelength ranges covered by these chosen flux bins are X-ray (0.517-12.4 nm), hard EUV (hEUV, 10-36$~$nm), soft-EUV (hereafter sEUV, 36-92 nm) and mid-UV (91.2-320 nm). We set their representative energies $e_{\lambda}$ 
as 248, 40, 20 and 7~eV, respectively. Some of our bin selections are consistent with values used in other models in the literature such as \citet{Murray-Clay2009, Daria_2018_grid_1_40_earths, Wang_Dai2021_WASP-69b, Wang_Dai2021_WASP107b}. Table \ref{tab:photoion_constants} also lists parameters for three additional photoionising sources, photons with energies 24.6, 21.2 and 10.3$~$eV arising from helium state transitions in the planetary atmosphere, rather than originating directly from the assumed stellar source. Photoionisations in the model are discussed in more detail in section \ref{sec:photo_and_heat}.

\begin{table}
\caption{Considered photoionisation processes and their respective excess kinetic energy factors $\epsilon_{\lambda}$, cross-section $\sigma_\lambda$ and their relevant reference. References correspond to: \citep{spitzer_78}$^\text{a}$, \citep{2006agna.book.....O}$^\text{b}$, \citep{Brown1971}$^\text{c}$, 
\citep{Norcross1971}$^\text{d}$,
\citep{Verner1996}$^\text{e}$,  \citep{Jacobs_1974_NASA_cross_sec_exc_He}$^\text{f}$.  }  
\label{tab:photoion_constants}
  \centering
  \begin{tabular}{|c|c|c|c|c|c|}
  		\hline 
  		ID & absorber & photon energy &  $ \epsilon_{\lambda} $  & $ \sigma_{\lambda} $ (cm$^2$)  & ref. \\
  		\hline 	\hline
\refstepcounter{tableeqn} (\thetableeqn)\label{photoion_1} & 
H$^0$ & sEUV: 20$~$eV & 0.32 & 2.21 $\times$ 10$^{-18}$  & a,b  \\    

\refstepcounter{tableeqn} (\thetableeqn)\label{photoion_2} & 
H$^0$ & hEUV: 40$~$eV & 0.66 & 8.29 $\times$ 10$^{-20}$  & a,b \\  

\refstepcounter{tableeqn} (\thetableeqn)\label{photoion_3} & 
H$^0$ & X-ray: 248$~$eV & 0.95 & 1.10 $\times$ 10$^{-21}$  & a,b \\ 

\refstepcounter{tableeqn} (\thetableeqn)\label{photoion_4} & 
H$^0$  & 24.6$~$eV & 0.45 & 1.24 $\times 10^{-18}$   &  a,b   \\

\refstepcounter{tableeqn} (\thetableeqn)\label{photoion_5} & 
H$^0$  & 21.2$~$eV & 0.36 & 1.87 $\times 10^{-18}$    & a,b  \\

\hline

\refstepcounter{tableeqn} (\thetableeqn)\label{photoion_6} & 
He($1^1S)$ & hEUV: 40$~$eV & 0.36 & 2.15 $\times$ 10$^{-18}$  & c \\ 
\refstepcounter{tableeqn} (\thetableeqn)\label{photoion_7} & 
He($1^1S)$ & X-ray: 248$~$eV & 0.90 & 3.00 $\times$ 10$^{-20}$   & c \\

\refstepcounter{tableeqn} (\thetableeqn)\label{photoion_8} & 
He($1^1S)$  & 24.6$~$eV & 0 & 2.42 $\times 10^{-19} $   &  c   \\

\hline 

\refstepcounter{tableeqn} (\thetableeqn)\label{photoion_9} & 
He($2^3S)$ & mid-UV: 7$~$eV & 0.31 & 3.68 $\times$ 10$^{-18}$   & d \\ 

\refstepcounter{tableeqn} (\thetableeqn)\label{photoion_10} & 
He($2^3S)$ & sEUV: 20$~$eV & 0.76 & 5.48 $\times 10^{-19}$  & d \\  

\refstepcounter{tableeqn} (\thetableeqn)\label{photoion_11} & 
He($2^3S)$ & hEUV: 40$~$eV & 0.88 & 9.00 $\times 10^{-19}$  & d \\

\refstepcounter{tableeqn} (\thetableeqn)\label{photoion_12} & 
He($2^3S)$ & 24.6$~$eV & 0.80 &  4.26 $\times$ 10$^{-19}$    &  d   \\

\refstepcounter{tableeqn} (\thetableeqn)\label{photoion_13} & 
He($2^3S)$  & 21.2eV & 0.77 &  4.97 $\times$ 10$^{-19}$    & d   \\

\refstepcounter{tableeqn} (\thetableeqn)\label{photoion_14} & 
He($2^3S)$  & 10.3eV & 0.54 &  1.95 $\times$ 10$^{-18}$    &  d   \\

\hline

\refstepcounter{tableeqn} (\thetableeqn)\label{photoion_15} & 
He($2^1S)$  & mid-UV: 7$~$eV & 0.43 &  3.63 $\times$ 10$^{-18}$    & d   \\

\refstepcounter{tableeqn} (\thetableeqn)\label{photoion_16} & 
He($2^1S)$  & sEUV: 20$~$eV & 0.80 &  3.35 $\times$ 10$^{-19}$    &  d   \\

\refstepcounter{tableeqn} (\thetableeqn)\label{photoion_17} & 
He($2^1S)$  & hEUV: 40$~$eV & 0.90 &  1.00 $\times$ 10$^{-18}$    &  d   \\

\refstepcounter{tableeqn} (\thetableeqn)\label{photoion_18} & 
He($2^1S)$  & 24.6$~$eV & 0.84 & 2.42 $\times$ 10$^{-19}$    &  d   \\

\refstepcounter{tableeqn} (\thetableeqn)\label{photoion_19} & 
He($2^1S)$  & 21.2eV & 0.81 &  3.03 $\times$ 10$^{-19}$    &  d   \\

\refstepcounter{tableeqn} (\thetableeqn)\label{photoion_20} & 
He($2^1S)$  & 10.3eV & 0.61 & 1.71 $\times$ 10$^{-18}$   &  d   \\

\hline 

\refstepcounter{tableeqn} (\thetableeqn)\label{photoion_21} & 
He$^+$  & X-ray: 248$~$eV & 0.78 &  2.09 $\times$ 10$^{-20}$   &  e   \\

\hline

\end{tabular}
\end{table}

\subsection{Fluid dynamic equations}
\label{sec:fluid_dyn}
We model the escaping planetary atmosphere by treating it as a non-isothermal outflow, utilising the equations of fluid dynamics. These equations enforce the conservation of momentum, energy, and mass. In steady state, the momentum equation in spherical symmetry can be written as 
\begin{equation}\label{eq:con_mom}
{\varv}\frac{d ({\varv})}{d r } = -\frac{1}{\rho}\frac{d P}{d r}   -    \frac{G M_{ \rm pl}}{r^2}+\frac{3GM_*r}{a^3}  ,
\end{equation}
where $r$ is the radial coordinate from the centre of the planet, $\rho$ and $\varv$ represent the atmospheric mass density and velocity, respectively, and $P$ is the thermal pressure. The masses of star and planet are $M_*$ and $M_{ \rm pl}$, and the orbital distance is $a$. The first term on the right hand side of Equation \ref{eq:con_mom} is  the thermal pressure gradient while the second term represents attraction due to gravity. This equation is analogous to the momentum equation for a stellar wind \citep{Parker} with an additional term on the right side due to tidal effects. The tidal term is the sum of the centrifugal force and differential stellar gravity along the line between the planet and star \citep{muno}.

Conservation of energy requires that
\begin{equation}\label{eq:cons_energy}
    \rho {\varv} \frac{d}{d r} \left(\frac{k_BT}{(\gamma - 1)m} \right) = \frac{k_BT}{m}{\varv} \frac{d\rho}{d r }+Q-C  ,
\end{equation}
where $k_B$ is the Boltzmann constant, $T$ is the gas temperature, $\gamma = {5}/{3}$ and $m$ is the mean  particle mass. The term on the left indicates the change in the internal energy of the fluid. The first term on the right represents cooling due to gas expansion (adiabatic cooling), while the second ($Q$) and third ($C$) are heating and cooling terms, which in this work accounts for a number of physical process involving hydrogen and helium species (see sections \ref{sec:photo_and_heat} and \ref{sec:cooling}). This equation is, again, typically used in stellar wind models, although with different heating and cooling terms \citep{2006ApJ...639..416V}.

The conservation of mass requires that 
\begin{equation}\label{eq:cons_mass_1st}
\frac{d (r^2 \rho {\varv}) }{d r } = 0  . 
\end{equation}
Our simulated planetary outflow originates from the sub-stellar point of the planet, which is the point closest to the star. We then apply our calculated solution over 4$\pi$ steradians rendering it an upper limit to atmospheric escape \citep{Murray-Clay2009, 2015ApJ...815L..12J}. Therefore, from Equation \ref{eq:cons_mass_1st}, we have that the mass-loss rate of the escaping atmosphere is $\dot{m} = 4\pi r^2 \rho {\varv} $.

We also solve the equation of ionisation balance of hydrogen 
 \begin{equation}\label{eq: ion_bal}
\varv \frac{d (n_{\text{H}^+}) }{d r}
= 
\Phi\left[\text{H}^0\right] + \Psi\left[\text{H}^0\right]  - \alpha_{B}\left[\text{H}^0\right] . 
\end{equation}
\\
\\
 This equation states that the advection of hydrogen ions (first term) is balanced by the combined rate of photo-ionisation ($\Phi\left[\text{H}^0\right]$, see Equation \ref{eq:photoion_rate} and Table \ref{tab:photoion_constants}) and collisional-ionisation ($\Psi\left[\text{H}^0\right]$, Table \ref{tab:pop_depop}) of neutral hydrogen with the radiative recombinations ($\alpha_B\left[\text{H}^0\right]$, also see Table \ref{tab:pop_depop}).

Altogether, the fluid dynamic equations (Equations \ref{eq:con_mom}-\ref{eq:cons_mass_1st}), the hydrogen ionisation balance equation (Equation \ref{eq: ion_bal}) and the four additional helium population equations (see next section, Equations \ref{eq:He_sin} - \ref{eq:He_ion}) form a system of coupled differential equations. There is only one physical solution for this system: a transonic wind beginning at the base of the atmosphere with a subsonic velocity, passing through the sonic point and thereby reaching supersonic velocities \citep{Parker}. Contrary to isothermal wind models, we do not know a priori the location of the sonic point nor the initial wind velocity, therefore we follow a numerical approach in solving the coupled differential equations utilising a `shooting method’ combined with a fourth order Runge-Kutta solver and an iteration scheme. Our initial solution assumes a hydrostatic atmosphere for the density, and the shooting method finds the solution that passes through the sonic point. This new solution provides a different density profile than initially assumed, so using this newly found density structure, we again search for the solution that passes through the sonic point through the shooting method. In this way, the density, temperature, velocity and optical depth profiles are updated with each set of iterations of the shooting method, iteratively approaching to a converged solution. 
We consider this solution to have been reached once the predicted mass-loss rate and terminal velocity of two subsequent model results agree to within 1\%. All atmospheric profiles are calculated with a resolution of $10^{-6}R_{\rm pl}$ below the sonic point, and $2 \times 10^{-5}R_{\rm pl}$ above.

The necessary free parameters of our model include the temperature $T_0=1\,000~$K and density $\rho_0=4 \times 10^{-14}$ g cm$^{-3}$ at the base of the atmosphere, chosen to be at 1$~R_{\rm pl}$. \citet{Allan_Vidotto_2019} and \citet{Murray-Clay2009} found that large variations in these values had a negligible effect on the resulting simulated escape. This is similar in our models as will be further discussed in Appendix A. Our model must also be given initial guesses for the abundances by number at the base of the atmosphere for ionised hydrogen, singly- and doubly- ionised helium as well as its $1^1S$, $2^1S$ and $2^3S$ states. These initial guesses were $10^{-5}$, $10^{-5}$, $10^{-7}$, $\sim 1$, $\sim 0$ and $\sim 0$, meaning that initially both hydrogen and helium are predominantly neutral, with helium being nearly entirely in its singlet state. To ensure the validity of the hydrodynamic escape we model, once the solution is found, we confirm that the atmosphere remains within the collisional regime with a Knudsen number (ratio of mean free path to atmospheric scale height) $< 1$. In the non-collisional regime, particle models such as those used in \citet{Bourrier2013, Bourrier+2016_GJ436, Allart_2018_warm_Nep_Hat_p_11b_He_obs, Spake_2018_NATURE_He_in_atm, Allart_2019_WASP_107b} are better suited to model atmospheric escape. 
Our models are computed all the way to $10~R_{\rm pl}$, but we note that in reality, the extent of the atmosphere could be larger or smaller, as its boundary should be set by the interactions with the wind of the host star \citep[e.g.][]{2020MNRAS.494.2417V}, neglected in our study.

\subsection{Calculating helium state populations}
\label{sec:he_state_pop}

\begin{figure*}
 \includegraphics[width=2.1\columnwidth]{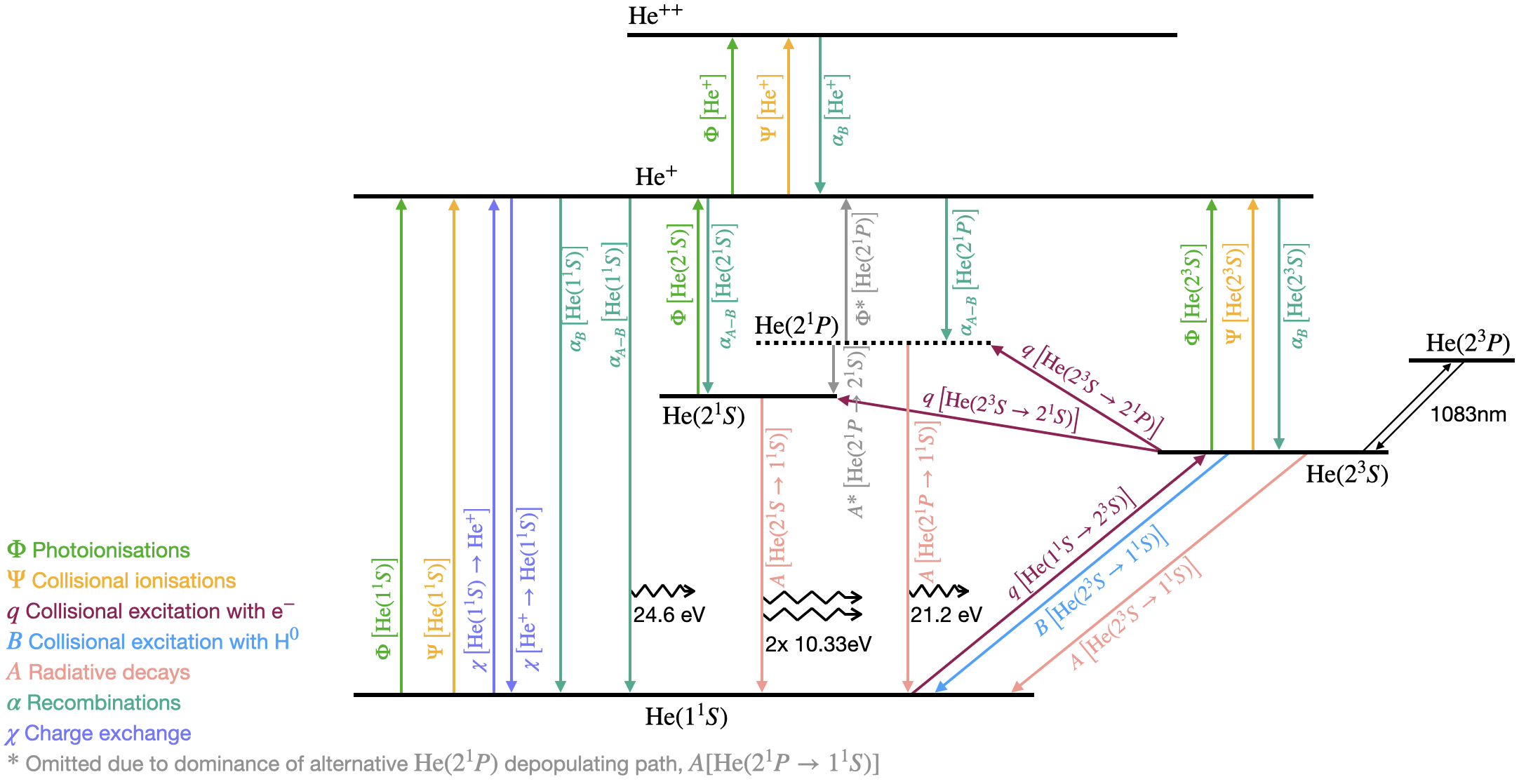}
    \caption{Schematic displaying the various helium processes considered in our modelling. See Table \ref{tab:pop_depop} for the rate equations and their relevant references. } \label{fig:sketch}
\end{figure*}
In addition to Equations \ref{eq:con_mom} to \ref{eq: ion_bal}, we also solve four continuity equations for the populations of helium, in the helium singlet ground state, He($1^1S$)
\begin{equation}\label{eq:He_sin}
\begin{aligned}
\varv \frac{ d (f_{1^1S}) }{d r } =  &   \alpha_B\left[\text{He}(1^1S)\right] + \alpha_{A-B}\left[\text{He}(1^1S)\right]  \\
& +B\left[\text{He}(2^3S \rightarrow 1^1S)\right] \\
& +A\left[\text{He}(2^3S, ~ 2^1S, ~2^1P \rightarrow 1^1S)\right] \\
& + \chi\left[\text{He}^+ \rightarrow \text{He}(1^1S)\right] 
 -  \chi\left[\text{He}(1^1S) \rightarrow \text{He}^{+}\right] \\
& - q\left[\text{He}(1^1S \rightarrow 2^3S)\right] 
 -\Phi\left[\text{He}(1^1S)\right]
 - \Psi\left[\text{He}(1^1S)\right]  , 
\end{aligned}
\end{equation} 
the helium triplet state, He($2^3S$) 
\begin{equation}\label{eq:He_trip}
\begin{aligned}
\varv \frac{ d (f_{2^3S}) }{d r } = &   \alpha_B\left[\text{He}(2^3S)\right]   
+ q\left[\text{He}(1^1S \rightarrow 2^3S)\right] \\ 
& - q\left[\text{He}(2^3S \rightarrow 2^1S, 2^1P)\right] \\
& -A\left[\text{He}(2^3S \rightarrow 1^1S)\right] -B\left[\text{He}(2^3S \rightarrow 1^1S)\right] \\
  & -\Phi\left[\text{He}(2^3S)\right] 
 - \Psi\left[\text{He}(2^3S)\right]   , 
\end{aligned}
\end{equation}
the helium $2^1S$ state, He($2^1S$)
\begin{equation}\label{eq:He_21s} 
\begin{aligned}
\varv \frac{ d (f_{2^1S}) }{d r } = & \alpha_{A-B}\left[\text{He}(2^1S)\right] +
 q\left[\text{He}(2^3S \rightarrow 2^1S)\right] \\
 &  - A\left[\text{He}(2^1S \rightarrow 1^1S)\right] 
 - \Phi\left[\text{He}(2^1S)\right],
 \end{aligned}
\end{equation} 
and the singly ionised helium state, He$^+$
    \begin{equation}\label{eq:He_ion}
    \begin{aligned}
\varv \frac{ d (f_{\text{He}^+}) }{d r } =  & \Phi\left[\text{He}(1^1S, ~ 2^1S, ~ 2^3S )\right] +  \Psi\left[\text{He}(1^1S, ~ 2^3S) \right]  \\
& +  \alpha_B\left[\text{He}^{+}\right]     
- \alpha_B\left[\text{He}(1^1S)\right] - \alpha_{A-B}\left[\text{He}(1^1S)\right] \\
& - \alpha_{A-B}\left[\text{He}(2^1S)\right] -  \alpha_{A-B}\left[\text{He}(2^1P)\right]   \\
& - \alpha_B\left[\text{He}(2^3S)\right] -\Phi\left[\text{He}^{+}\right]-\Psi\left[\text{He}^{+}\right] \\
& +  \chi\left[\text{He}(1^1S) \rightarrow \text{He}^{+}\right]-\chi\left[\text{He}^+ \rightarrow \text{He}(1^1S)\right] .
\end{aligned}
\end{equation}

These state that the fractions of helium in its $1^1S$, $2^3S$, $2^1S$, and singly ionised states $f_{1^1S}$, $f_{2^3S}$, $f_{2^1S}$ and $f_{\text{He}^+}$ respectively, are determined by the balance of the processes directly populating and depopulating these states. These processes (de-)populating the helium states are defined as follows 

\begin{itemize}

\item  $\Phi$ - photoionisations (out of the state in parentheses) by capable stellar XUV and mid-UV photons as well as by photons released in the planetary atmosphere through helium transitions, 
 
\item  $\psi$ - collisional ionisations (out of the state in parentheses), 

\item  $\alpha$ - recombinations (into the state in parentheses), 
 
\item  $q$ - collisional excitation with a free electron, 
 
\item  $B$ - collisional excitation with H$^0$, 
 
\item  $A$ - radiative decays, 
 
\item  $\chi$ - charge exchange. 

\end{itemize}
 
Table \ref{tab:pop_depop} lists all of their corresponding rate equations for these helium states in addition to that for hydrogen. Figure \ref{fig:sketch} visualises the considered paths for the helium states. The helium number densities in each state are obtained by multiplying their respective fraction by the total number density of helium $n_{\rm He}$. Our model considers free electrons produced from both hydrogen and helium ionisations while maintaining a neutral net charge for the global modleled atmosphere. We also calculate the fraction of doubly-ionised helium, assuming its number density to be $n_{\rm He^{++}} =n_{\rm He} - n_{1^1S} - n_{2^3S} - n_{2^1S} - n_{\rm He^{+}}$. However, including He$^{++}$ is found to have a negligible effect on the modelled hydrodynamics and observability of atmospheric escape.

Being a non-metastable and hence short-lived state \citep{Wiese_Fuhr_2009_NIST_lines}, our model does not solve for the fraction of helium in the $2^1P$ state. Rather, we follow \citet{Oklopcic_2018_10839_window} in allowing electron collisional excitation from $2^3S$ to $2^1P$, $q\left[\text{He}(2^3S \rightarrow 2^1P)\right]$ to populate the $1^1S$ state on account of the rapid decay, $A\left[\text{He}(2^1P \rightarrow 1^1S)\right]$. \citet{Lampon_2020} and \citet{pwinds} also follow this assumption, which most importantly accounts for the depopulation of the $2^3S$ state. Additionally, our model considers recombinations into $2^1P$, $\alpha_{A-B}\left[\text{He}(2^1P)\right]$, which is also set to populate $1^1S$ in our model's helium equations. It should be noted that the rapid radiative decay from the $2^1P$ to $1^1S$ state $A\left[\text{He}(2^1S \rightarrow 1^1S)\right]$ releases a 21.2~eV photon \citep{Eikema_1996, Sun_HU_2020_helium_spec}, capable of photoionising hydrogen, helium $2^3S$ and $2^1S$ in our model\footnote{The released 21.2~eV is also capable of photoionising He($2^1P$) however our model does not account for this particular photoionisation as we do not solve for the He($2^1P$) fraction. Given that the rapid radiative decay $A\left[\text{He}(2^1P \rightarrow 1^1S)\right]$ is an alternative depopulating path for He($2^1P$), we consider omitting these photoionisations to be reasonable. Furthermore, the cross section for the photoionisation of He($2^1P$) is between one and two orders of magnitudes less that that for $2^1S$ and $2^3S$ as shown in Figure \ref{fig:hd85512}.}. As will be discussed in section \ref{sec:photo_and_heat}, our model considers such photoionisations.

We use Case-B coefficients in our calculations of hydrogen and the helium  $2^3S, ~ 2^1S, ~ 2^1P$ states. Direct recombinations (i.e. Case-A $-$ Case-B) are not considered for these mentioned states, given their low energies and the high probability for the photon released in the process to photoionise a species identical to that formed in the recombination, resulting in no net variation in the ionisation fraction. Direct recombinations to He($1^1S$) state however are different in that they release a higher energy 24.6$~$eV photon \citep{Benjamin_1999, 2006agna.book.....O}, capable of photoionising both the more abundant hydrogen in addition to the various helium states. Accordingly, we calculate both direct and non-direct recombinations for He($1^1S$), in order to obtain the rate of the 24.6~eV photoionisations, as will be discussed in section \ref{sec:photo_and_heat}.

\begin{table*}
\caption{The populating and depopulating processes for the hydrogen and helium states considered in our model. Densities are given in cm$^{-3}$ and temperature in K. He(2$^1P$)$^*$ is marked with an asterisk as a reminder that we do not solve for the fraction of helium in the $2^1P$ state, as it is short-lived (see section \ref{sec:he_state_pop}). $\alpha_{B^*}\left[\text{He}(1^1S)\right]$ is marked with an asterisk to draw attention to the removal of helium $2^1P$ and $2^1S$ recombinations from this term. Errors present in a previous version of this table (arXiv:2311.01313v1) are corrected here as described in Appendix \ref{sec:correction}. Letters correspond to the following references:  \citep{Brown1971}$^\text{a}$,  \citep{Norcross1971}$^\text{b}$,  \citep{Verner1996}$^\text{c}$, \citep{spitzer_78}$^\text{d}$, \citep{2006agna.book.....O}$^\text{e}$,  \citep{Benjamin_1999}$^\text{f}$,  \citep{Wiese_Fuhr_2009_NIST_lines}$^\text{g}$,  \citep{Eikema_1996}$^\text{h}$, 
\citep{Bergeson_1998_He21S}$^\text{i}$, 
\citep{Hui_Gnedin_1997}$^\text{j}$, \citep{Caldiroli_2021_ATES}$^\text{k}$,  \citep{Storey_Hummer1995}$^\text{l}$, \citep{Bray2000}$^\text{m}$,
\citep{Lampon_2020}$^\text{n}$, 
\citep{Roberge_Dalgarno1982}$^\text{o}$,  
\citep{Oklopcic_2018_10839_window}$^\text{p}$, 
\citep{Cen_1992}$^\text{q}$, 
\citep{Koskinen_2013}$^\text{r}$,  
\citep{Drake1971}$^\text{s}$.   
}  
 \label{tab:pop_depop}
  \centering
  \begin{tabular}{|c|c|c|c|}
  		\hline		    
  		 populates & depopulates  & rates (s$^{-1}$) & reference \\
  	 \hline \hline


 & & { \bf photoionisation by stellar XUV and mid-UV photons}   & \\

He$^+$ & He(1$^1S$) & 
Equation \ref{eq:photoion_rate} and Table \ref{tab:photoion_constants} for rate, caused by $\lambda_{\rm hEUV/X-ray}$ &  
a \\ 

He$^+$ & He(2$^3S$) & 
Equation \ref{eq:photoion_rate} and Table \ref{tab:photoion_constants} for rate, caused by $\lambda_{\rm s/hEUV/mid-UV}$ &  
b \\  

He$^+$ & He(2$^1S)$ & 
Equation \ref{eq:photoion_rate} and Table \ref{tab:photoion_constants} for rate, caused by  $\lambda_{\rm s/hEUV/mid-UV}$   &  
b \\  

 He$^{++}$ & He$^+$ & 
Equation \ref{eq:photoion_rate} and Table \ref{tab:photoion_constants} for rate, caused by $\lambda_{\rm X-ray}$  &  
c \\  

H$^+$ & H$^0$ & 
Equation \ref{eq:photoion_rate} and Table \ref{tab:photoion_constants} for rate, caused by $\lambda_{\rm s/hEUV/X-ray}$ &  
d,e \\ 

\hline

 & &  { \bf photoionisation by planetary-atmosphere-produced photons}   & \\ 

H$^+$~/~He$^+$ & H$^0$~/~He$(1^1S, 2^1S, 2^3S)$  & 
  $  \zeta_{\rm \text{sp}, 24.6~eV} ~ \alpha_{A-B}\left[\text{He}(1^1S)\right]  $ &  
e, f \\

H$^+$~/~He$^+$ & H$^0$~/~He$(2^1S, 2^3S)$  & 
$ \zeta_{\rm sp, 21.2eV} ~ A\left[\text{He}(2^1P \rightarrow 1^1S)\right]  $ &  
g, h \\ 

He$^+$ &  He$(2^1S, 2^3S)$  & 
$ 2 ~ \zeta_{\rm sp, 10.3eV} ~ A\left[\text{He}(2^1S \rightarrow 1^1S)\right]  $ &  
e, i \\

\hline

 &  & { \bf recombination }   & \\
He(2$^3S$) & He$^+$ &  
$ \alpha_B\left[\text{He}(2^3S)\right]= 2.10 \times 10^{-13}  \left(T/10^4\right)^{-0.778} n_{\text{e}} f_{\rm He^+}  $  & 
f \\  

He(2$^1S$) &  He$^+$  &    
$ \alpha_{A-B}\left[\text{He}(2^1S)\right]= 5.55 \times 10^{-15} \left(T/10^4\right)^{-0.451}  n_{\text{e}} f_{\rm He^+}  $  & 
f \\ %

He(2$^1P$)$^*$ &  He$^+$  &    
$ \alpha_{A-B}\left[\text{He}(2^1P)\right]= 1.26 \times 10^{-14} \left(T/10^4\right)^{-0.695}  n_{\text{e}} f_{\rm He^+} $  & 
f \\ %

He(1$^1S$) &  He$^+$  &    
$ \alpha_{B^*}\left[\text{He}(1^1S)\right]= 6.23 \times 10^{-14} \left(T/10^4\right)^{-0.827}  n_{\text{e}} f_{\rm He^+}$ $- \alpha_{A-B}\left[\text{He}(2^1P)\right] - \alpha_{A-B}\left[\text{He}(2^1S)\right]  $  & 
f \\ %

He(1$^1S$) & He$^+$ &    
$\alpha_{A-B}\left[\text{He}(1^1S)\right]=1.54 \times 10^{-13}  \left(T/10^4\right)^{-0.486} n_{\text{e}} f_{\rm He^+} $  (emits 24.6$~$eV photon)  & 
f \\ 

He$^+$ & He$^{++}$ &    
$ \alpha_B\left[\text{He}^+\right]=5.506\times10^{-14} ~ (1263030/T)^{1.5} ~ \left(1+(460960/T)^{0.407}\right)^{-2.242} n_{\text{e}}f_{\text{He}^{++}}$
 & 
j, k %
  \\

H$^0$ & H$^+$ &    
$ \alpha_B\left[\text{H}^0\right]=2.7\times10^{-13}\left(T/10^4\right)^{-0.9}$ $n_{\text{e}}f_{\text{H}^+}$ & 
l, e   \\  %

\hline 
  & & { \bf collisional (de-)excitation }  & \\

He(2$^1S$) & He(2$^3S$) &  
$ q\left[\text{He}(2^3S \rightarrow 2^1S)\right] = 2.6 \times  10^{-8} n_{\text{e}} f_{2^3S} $ &  
m, p \\

He(2$^1P$)$^*$ & He(2$^3S$) & 
$ q\left[\text{He}(2^3S \rightarrow 2^1P)\right] = 4.0 \times  10^{-9} n_{\text{e}} f_{2^3S} $ &  
m, p \\

He(1$^1S$) & He(2$^3S$) &  
$ B\left[\text{He}(2^3S \rightarrow 1^1S)\right] = 5.0 \times 10^{-10} n_{\text{H}^0} f_{2^3S} $  &  
n, p \\ %

He(2$^3S$) & He(1$^1S$) &  
$ q\left[\text{He}(1^1S \rightarrow 2^3S)\right] = 4.5 \times 10^{-20}  n_{\text{e}} f_{1^1S} $   &  
m, p \\

\hline
 & &  { \bf collisional ionisation}    & \\ 
H$^+$  & H$^0$ &  
 $ \Psi\left[\text{H}^0\right] = \left. \left( 1.27 \times 10^{-21} ~ T^{0.5} ~ \exp{\left[ -157809.1/ ~ T\right]}  n_{\text{e}} f_{\rm H^0} \right)  \middle/  \left( e_{\text{ion,~H}^0}\left[\text{erg}\right]\right) \right. $  &  
q \\ %

He$^+$  &  He(1$^1$S) &  
 $ \Psi\left[\text{He}(1^1S)\right] =  \left. \left( 9.38 \times 10^{-22}~ T^{0.5} ~ \exp{\left[ -285335.4/ ~ T\right]}  n_{\text{e}} f_{1^1S}  \right) \middle/  \left( e_{\text{ion,~He}(1^1S)}\left[\text{erg}\right]\right) \right. $  &  
q  \\ %

He$^+$ & He$(2^3S)$ &  
$ \Psi\left[\text{He}(2^3S)\right] = \left. \left(  6.41 \times 10^{-21}~ T^{0.5} ~ \exp{\left[ -55338/ ~ T\right]}  n_{\text{e}} f_{2^3S}  \right) \middle/  \left( e_{\text{ion,~He}(2^3S)}\left[\text{erg}\right]\right) \right. $  &  
q  \\ %

He$^{++}$ & He$^{+}$ &  
 $ \Psi\left[\text{He}^+\right] = \left. \left(  4.95 \times 10^{-22}~ T^{0.5} ~ \exp{\left[ -631515/ ~ T\right]}  n_{\text{e}} f_{\text{He}^{+}} \right) \middle/   \left( e_{\text{ion,~He}^+}\left[\text{erg}\right]\right) \right.  $  & 
q  \\ %

\hline
 & & { \bf charge exchange }  & \\

He(1$^1S$), H$^{+}$ & He$^+$, H$^0$ &  
$ \chi\left[\text{He}^+ \rightarrow \text{He}(1^1S)\right] = 1.25 \times 10^{-15} (300/T)^{-0.25} n_{\text{H}^0} f_{\rm He^+}   $  &  
r, n \\ 

He$^+$, H$^0$ & He(1$^1S$), H$^+$ & 
$ \chi\left[\text{He}(1^1S) \rightarrow \text{He}^+\right] = 1.75 \times 10^{-11} (300/T)^{0.75} \exp[-128000/T] n_{\text{H}^+} f_{1^1S} $  & 
r, n \\ 

\hline 
 & & { \bf radiative decay}  & \\ 

He(1$^1S$) & He(2$^3S$) &  
$ A\left[\text{He}(2^3S \rightarrow 1^1S)\right]  = 1.272  \times 10^{-4} f_{2^3S} $    &  
s \\ %

He(1$^1S$) & He(2$^1S$) &  
 $ A\left[\text{He}(2^1S \rightarrow 1^1S)\right]= 51.3 ~ f_{2^1S} $  \text{(emits two 10.3~eV photons)}  &  
e \\  

He(1$^1S$) & He(2$^1P$)$^*$ &  
$ A\left[\text{He}(2^1P \rightarrow 1^1S)\right] =$ \st{$1.7989  \times 10^{9} f_{2^1P} $}  &  
g \\  %

 &  &  
 $ \approx  q\left[\text{He}(2^3S \rightarrow 2^1P)\right] + \alpha_{A-B}\left[\text{He}(2^1P)\right] $ \text{(emits 21.2~eV photon)}  &  
assumed here \\

\hline

\end{tabular}

\end{table*}

In short, our model considers hydrogen and helium photoionisation, collisional ionisation, recombination, collisional (de-) excitation, charge exchange and radiative decays. Our model solves the four helium populations equations (Equations \ref{eq:He_sin} to \ref{eq:He_ion}) simultaneously with Equations \ref{eq:con_mom} to \ref{eq: ion_bal}.

To model the escape of a primordial hydrogen and helium atmosphere, the helium to hydrogen number abundance must be known. Models have typically assumed this to be constant throughout their atmosphere, which we also assume. 
Some typically derived values are: He$/$H $\sim 0.02/0.98$ for HD209458b \citep{Lampon_2020, Khodachenko_2021_B_field_hd209458b};  He$/$H $\sim 0.016/0.984$ \citep{Shaikhislamov_2021_GJ3470b} and He$/$H $\sim 0.015/0.985$ \citep{Lampon_2021} for GJ3470b; He$/$H$\sim0.008/0.992$ \citep{Lampon_2021} and He$/$H$\sim0.005/0.995$ \citep{Rumenskikh_2022_3D_HD189733b} for HD189733b. All of the mentioned helium abundances fall below the solar value of approximately 0.1. Models of WASP-107b were the first to obtain a He$/$H number abundance close to the solar value, with a predicted helium abundance of 0.075-0.15 \citep[][]{Khodachenko_2021_107b_3d}. In our models, we use helium number abundances of 0.02 and 0.1 to investigate how this affects the predicted transits.

\subsection{Photoionisation and heating processes with hydrogen and helium}
\label{sec:photo_and_heat}

Photoionisations are not only important for determining the population of atmospheric hydrogen and helium states. If a given photon has an energy in excess of the photoionisation energy, this excess is transferred to an electron in the form of kinetic energy. This collisional process acts to heat up the atmosphere.

The amount of photoionisations and heating that is deposited in the atmosphere depends on the individual optical depths $\tau_{\lambda}$ for each of the considered photoionisations
\begin{equation}
\tau_{\lambda} = \int^{1R_{\rm pl}}_{\rm top} ~ n ~ \sigma_{\lambda} dr \, ,
\end{equation}
where $n$ is the number density  of the absorber (neutral hydrogen, helium singlet or triplet) and $\sigma_{\lambda}$ is the absorber- and wavelength-dependent photoionisation cross section (Table \ref{tab:photoion_constants}). The integration is performed for distances $r$ starting at the top of the atmosphere ($10 R_{\rm pl}$, in our models) until the bottom of the atmosphere at $1 R_{\rm pl}$. The rate of one of the considered photoionisations is then
\begin{equation}
\label{eq:photoion_rate}
\Phi_{\lambda}=  F_{\lambda}   \; \zeta_{\rm sp, \lambda} \;  e^{-\tau_{\lambda}} \;  \sigma_{\lambda}   \; \frac{1}{e_{\lambda}} ,
\end{equation}
where $F_{\lambda}$ is the energy flux at $e_{\lambda}$ and $\zeta_{\rm sp, \lambda}$ is a weighting factor we introduce to prevent the same photon from being able to photoionise both hydrogen and helium, where `sp' refers to the specific species being photoionised. The fraction of photons which photoionise hydrogen rather than helium from a given flux bin with $e_{\lambda}$ is (see Equation 2.21 of \citealt{2006agna.book.....O})
\begin{equation}
\label{eq:avail_photons}
\zeta_{\rm H^0, \lambda}  =  \frac{n_{\text{H}^0} \sigma_{\rm H^0, \lambda}}{n_{\text{H}^0} \sigma_{\rm H^0, \lambda}  + n_{\rm He}  \sigma_{\rm He, \lambda}}.
\end{equation}
We use similar weighting factor expressions for each considered photoionisation process in our model. For example, Table \ref{tab:photoion_constants} shows that the X-ray flux bin can photoionise H$^0$, He($1^1S$) and He$^+$. Hence, the weighting factor for an X-ray photon to photoionise He($1^1S$), $\zeta_{\rm He(1^1S),X-ray}$ in our model is given by
\begin{equation*}
   \frac{n_{\text{He}(1^1S)} \sigma_{\rm He(1^1S),X-ray}}{ 
   n_{\text{He}(1^1S)} \sigma_{\rm He(1^1S),X-ray}
   + n_{\text{H}^0} \sigma_{\rm H^0,X-ray}  
   + n_{\text{He}^+} \sigma_{\rm He^+,X-ray}  
   }  .
\end{equation*}
This important consideration to prevent from double-counting photons is often not included in hydrogen-helium atmospheric escape models. The rate of the heating associated with a photoionisation is calculated as 
\begin{equation}
Q_{\lambda}   = \epsilon_{\lambda}  e_{\lambda}  \Phi_{\lambda} \; n =  \epsilon_{\lambda} F_{\lambda}   \; \zeta_{\rm sp, \lambda} \; e^{-\tau_{\lambda} }  \; \sigma_{\lambda}   \; n  \,    \, .
\end{equation}
Combined, the terms $F_{\lambda} \zeta_{\rm sp, \lambda} e^{-\tau_{\lambda} }$ account for the remaining supply of photoionising-capable photons at a given distance into the atmosphere. $\epsilon_{\lambda} = 1 - (e_{\text{ion}} / e_{\lambda})$ accounts for how much energy the photon with $e_{\lambda}$ has in excess of the photoionisation energy cost ($e_{\text{ion}}=$13.6, 24.6, 4.0 4.8, 54.4~eV for H$^0$, helium $1^1S$, $2^1S$, $2^3S$ and He$^+$, respectively). Values of $\epsilon_{\lambda}$ are listed in Table \ref{tab:photoion_constants}. Finally, the combined $ \sigma_{\lambda}  \; n  $ terms account for the availability of potential absorbers. The total photoionisation heating $Q$, included in Equation \ref{eq:cons_energy}, is then the sum of the heating due to each of the considered species.

In addition to the considered stellar flux, our model also accounts for 24.6~eV and 21.2~eV photons released in recombinations and radiative decay from He($2^1P$) to He($1^1S$). It also accounts for the release of two 10.3~eV \citep{Bergeson_1998_He21S} photons released in the radiative decay from helium's $2^1S$ to $1^1S$ state. We assume the photoionisation rates by these photons to be given by the rate of the process from which they are released ($\times$ 2 for the two-photon process), as shown in Table \ref{tab:pop_depop}. In doing so, we assume that all of these photons are re-absorbed by the planetary atmosphere and do not escape. Given that these photons are predominantly released in the optically thick inner-most region of the atmosphere, we consider this assumption to be reasonable. To determine the fractions of which of the viable atmospheric species are absorbed by these photons, we again utilise the weighting factor given by Equation \ref{eq:avail_photons}. For example, the modelled rate of the released 24.6~eV photons which photoionise hydrogen is  $  \zeta_{\rm H^0, 24.6~eV} ~ \alpha_{A-B}\left[\text{He}(1^1S)\right] $. As will be shown, this contributes non-negligibly to the overall photoionisation of hydrogen and heating in the inner-most region of the modelled planetary atmosphere.

\citet{Cecchi_Pestellini2006} showed that X-rays can play a large role in heating the upper layers of planetary atmospheres, particularly those which orbit closely to young (high X-ray flux) stars.  Owing to their large photon energies, primary X-ray photoionisations are followed by subsequent photoionisations caused by released energetic secondary photo- and Auger-electrons \citep{Gudel_summerschool2015Xray}. Heating by X-rays is dominated by such secondary photo-electron generation from the K-shells of metals, with oxygen and carbon being the most important \citep{Owen_Jackson_2012}. Few works \citep{Cecchi_Pestellini2006, Shematovich2014, Danielle_Locci_2022} have modelled the secondary electron photoionisation cascade. As our model considers only hydrogen and helium and not heavier elements important for such processes, secondary photoionisations are omitted from our model (as is the case for the models of \citealt{Oklopcic_2018_10839_window, Lampon_2020, pwinds}). Hence, it should be noted that our models may underestimate the number of photo-inisations by X-rays, particularly when the host star is a young, initially fast rotator. A recent work by \citet{Gillet_2023_photoelectrons} studies the effects of secondary ionisation by photoelectrons and stresses that such ionisations should be taken into account in 1-D hydrodynamic modelling, reporting a reduction in the predicted mass-loss rate of 43\% upon its inclusion for a 0.69$M_{\text{Jup}}$ planet orbiting the K1V host HD97658 at 0.074~au.

\subsection{Cooling processes with hydrogen and helium}
\label{sec:cooling}

In addition to heating, our energy equation (Equation \ref{eq:cons_energy})  also includes various hydrogen and helium cooling contributions. These are collisional ionisation, collisional excitation, recombination and Bremsstrahlung. The volumetric cooling rates were obtained from \citet{Black_1981} and \citet{Cen_1992} and are listed in Table \ref{tab:cool}.
To obtain the stated cooling contribution processes (\ref{cool_4}) and (\ref{cool_10}), we rearranged the respective rate equations given in Table 3 of \citet{Black_1981} so as to use our own helium triplet number density rather than an approximation given by Equation 11 of \citet{Black_1981}. Rearranging and substituting this equation into their rate equations gives the cooling rate equations listed in Table \ref{tab:cool}. The sum of all the volumetric cooling rates listed in Table \ref{tab:cool} results in $C$, which is included in Equation \ref{eq:cons_energy}.

\begin{table*}
\caption{Cooling processes included in our model. The sum of the cooling rates ($C$) are included in Equation \ref{eq:cons_energy}. Densities are given in cm$^{-3}$ and temperatures in K. References: \citep{Black_1981}$^\text{a}$,
\citep{Cen_1992}$^\text{b}$,  
\citep{Murray-Clay2009}$^\text{c}$,
\citep{Allan_Vidotto_2019}$^\text{d}$.}  
 \label{tab:cool}
\centering
\begin{tabular}{|c|c|c|}
\hline 
ID  & volumetric cooling rates [erg cm$^{-3}$ s$^{-1}$] & reference \\

\hline \hline 
& Collisional ionisation cooling &  \\

\refstepcounter{tableeqn} (\thetableeqn) \label{cool_1} &  
$1.27 \times 10^{-21} ~ T^{0.5} ~ \exp{\left[ -157809.1/ ~ T\right]} ~ n_{\rm e} ~ n_{\text{H}^0}$ & a,b \\

\refstepcounter{tableeqn} (\thetableeqn)\label{cool_2}  &  
$9.38 \times 10^{-22} ~ T^{0.5} ~ \exp[ -285335.4/ ~ T] ~ n_{\rm e} ~ n_{ \text{He}(1^1S)}$ & a,b \\

\refstepcounter{tableeqn} (\thetableeqn)\label{cool_3}     &  
$ 4.95 \times 10^{-22} ~ T^{0.5} ~ \exp[ -631515/ ~ T] ~ n_{\rm e} ~ n_{\rm He^+}$ & a,b \\

\refstepcounter{tableeqn} (\thetableeqn) \label{cool_4}    &  
$ 6.41 \times 10^{-21} ~ T^{0.5} ~  \exp[ -55338/ ~ T]    ~  n_{\rm e}    ~ n_{\rm He(2^3S)} $ & a,b, see Section \ref{sec:cooling} \\

\hline 
& Recombination cooling &  \\

\refstepcounter{tableeqn} (\thetableeqn) \label{cool_5}  &  
$\left.  \left( 8.70 \times 10^{-27} ~ T^{0.5} ~ \left(\frac{T}{10^3}\right)^{-0.2}\right)  \middle/ \left( 1+\left(\frac{T}{10^6}\right)^{0.7}\right)  ~ n_{\rm e} ~ n_{\text{H}^+} \right. $ & b \\  [0.3cm] %

\refstepcounter{tableeqn} (\thetableeqn) \label{cool_6}   &   

$1.55 \times 10^{-26} ~ T^{0.3647}   ~ n_{\rm e} ~ n_{\rm He^+}$  & a,b \\

\refstepcounter{tableeqn} (\thetableeqn) \label{cool_7}   &   
$\left.  \left( 3.48 \times 10^{-26} ~ T^{0.5}  ~ \left(\frac{T}{10^3}\right)^{-0.2} \right)  \middle/ \left( 1+\left(\frac{T}{10^6}\right)^{0.7} \right)  ~ n_{\rm e} ~ n_{\text{He}^{++}} \right. $   & b \\

\hline 
& Dielectronic recombination cooling & \\

\refstepcounter{tableeqn} (\thetableeqn) \label{cool_8}  & 
$1.24 \times 10^{-13} ~ T^{-1.5} ~ \exp[ -470000/ ~ T] ~ (1+0.3 ~\exp[ -94000/ ~ T] )    ~ n_{\rm e} ~ n_{\rm He^+}$ & a,b \\

\hline 
& Collisional excitation cooling &  \\

\refstepcounter{tableeqn} (\thetableeqn) \label{cool_9}  & 
$7.5 \times  10^{-19}  ~ \exp[ - 118348  / ~ T] ~ n_{\rm e} ~ n_{\text{H}^0}$ & a,b,c,d \\

\refstepcounter{tableeqn} (\thetableeqn) \label{cool_10}       &  
$5.54 \times 10^{-17} ~ T^{-0.397} ~ \exp[ -473638/ ~ T] ~ n_{\rm e} ~ n_{\rm He^+} $ & a,b \\

\refstepcounter{tableeqn} (\thetableeqn) \label{cool_11}       &  
$1.16 \times 10^{-20} ~ T^{0.5} ~ \exp[ -13179/ ~ T] ~ n_{\rm e} ~ n_{\rm He(2^3S)} $  & a,b, see Section \ref{sec:cooling} \\

\hline 
& Free-free emission (Bremsstrahlung) &  \\

\refstepcounter{tableeqn} (\thetableeqn) \label{cool_12}       &
 $1.42 \times 10^{-27} ~ T^{0.5} ~ (3/2) ~ \left[n_{\text{H}+} + n_{\rm He^+} + 4 n_{\rm He^++} \right] n_{\rm e}  $  & a,b \\
\hline

\end{tabular}

\end{table*}

\subsection{Test case: solving helium populations in post-processing}
\label{sec:test_post-proc}

For comparative purposes, we also present a representative model in which the helium equations are solved in a post-processing step (as is the case for the models of \citealt{Oklopcic_2018_10839_window,Lampon_2020,pwinds}). If solved in a post-processing step, helium heating and cooling contributions can not be included in the energy equation (Equation \ref{eq:cons_energy}) as to calculate these contributions requires knowledge of the helium populations. Hence, whether the helium population equations are solved simultaneously with the fluid dynamic equations or in a post-processing step can impact the predicted atmospheric escape, the results of which will be discussed in sections \ref{sec:eff_He_energetics} and \ref{sec:eff_post_p_obs}.

Our post-processing test model

\begin{itemize}

  \item assumes an initially fast rotating stellar host. 

  \item Assumes a constant helium number abundance of 10\%. 

  \item Allows heating by photoionisation of hydrogen only.

  \item Allows photoionisation by stellar XUV and mid-UV photons only, ignoring photons released in the planetary atmosphere as a result of helium transitions.
  
  \item Neglects photoionisation weighting factors (see Equation \ref{eq:avail_photons}), meaning that a single photon could possibly photoionise hydrogen, as well as any of the viable considered helium states in the model.

  \item Includes only three cooling processes: recombination, collisional ionisation and excitation of hydrogen.
  
  \item Considers only electrons arising from hydrogen photoionisations. 
  
\end{itemize}

While these terms are chosen to be similar to the post-processing 1-D models of \citet{Oklopcic_2018_10839_window,Lampon_2020,pwinds}, the specific modelling differs from each of these works in numerous ways. For example, they assume a Parker wind in their modelling of the atmospheric escape and hence do not calculate the heating due to photoionisations. However, \textsc{p-winds} \citep{pwinds} allows the user to input a more complex and self-consistent atmospheric structure, rather than assume a Parker wind, as done for HD~189733~b in \citet{2023AJ....166...89D}. Additionally, there are some minor variations in a number of the assumed rates between states (see Table \ref{tab:pop_depop}).

\section{Model inputs: Evolution of high-energy stellar flux and planetary radii}
\label{sec:evol_dep_params}

\citet{Allan_Vidotto_2019} previously showed that the evolution of atmospheric escape for a close-in planet depends on two important factors: 
\begin{enumerate}
\item as the host star evolves, its activity declines due to spin down \citep{Skumanich_1972, Kawaler_1988, Vidotto_2014}, resulting in declining fluxes in the XUV \citep{Jackson_2012_X_ray_age_relation, Johnstone2015a, Johnstone_2021} and
\item as the planet evolves, cooling causes it to contract with time \citep{Fortney2010}.
\end{enumerate}

The level of atmospheric escape and consequently the observational signatures of escaping hydrogen in the Lyman-$\alpha$ and H-$\alpha$ lines were found to vary strongly with the evolution of the modelled planets, with younger planets exhibiting greater escape and deeper absorptions in both lines. This is the result of a favourable combination of higher irradiation fluxes and weaker gravities at young ages. In a continuation of this work, we now study how the helium 1083$~$nm signature evolves over the lifetime of a planet.

\begin{figure}
 	 \includegraphics[width=1.0\columnwidth]{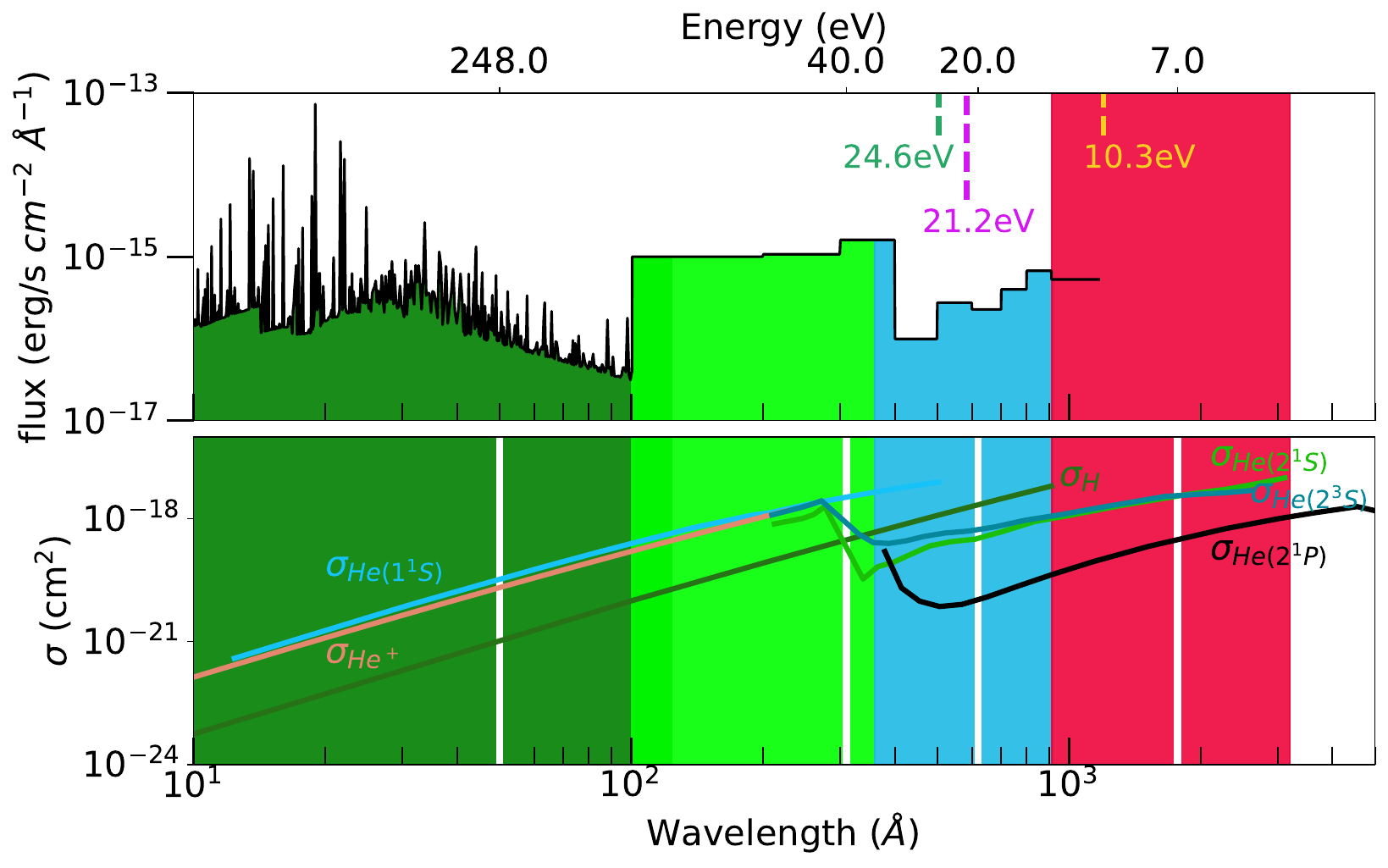} 

\caption{[Upper-panel] MUSCLES spectrum of HD~85512 used in to normalise the high-energy evolution tracks of \citet{Johnstone_2021}, shown in Figure \ref{fig:evol_inputs}. The energy of the three considered photons produced due to helium transitions in the planetary atmosphere are marked by the dashed lines. [Lower-panel] Photo-ionisation cross sections for hydrogen and considered helium states. Note that while He($2^1P$) photoionisations are not considered in the model, its cross-section profile is shown here for comparison to that of the $2^1S$ and $2^3S$ helium states. The four shaded regions indicate the wavelength channels of X-ray, hEUV, sEUV and mid-UV going from left to right. The representative energies of each of these flux bins is marked by the vertical solid white lines in the lower-panel.} \label{fig:hd85512}
\end{figure}

\begin{figure}
\begin{center}
\includegraphics[width=0.48\textwidth]{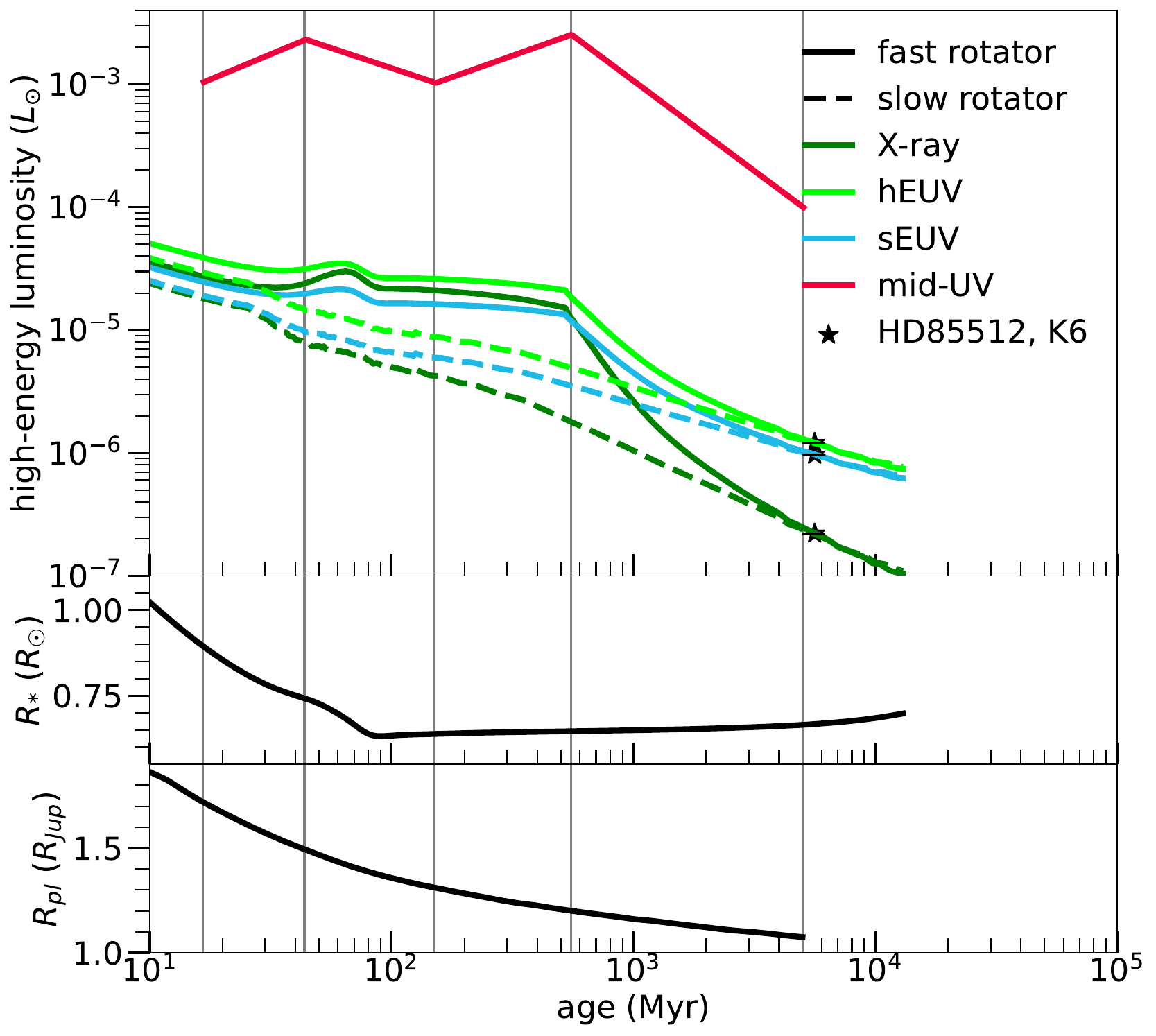}
 \caption{[Upper-panel] Evolution of high-energy stellar luminosity, where hEUV, sEUV and X-ray wavelength bins were obtained by normalising the predictions of \citet{Johnstone_2021} for a 0.7-$M_{\sun}$ by the K dwarf HD85512 (star symbols, from Figure \ref{fig:hd85512}). The mid-UV luminosity tracks combines the near-UV and far-UV fluxes from \citet{Richey_Yowell_2022}. [Central-panel] Stellar radius evolution for the same star, also obtained from the model of \citet{Johnstone_2021}. [Lower-panel] Planetary radius with respect to age for a 0.3-$M_{\rm jup}$ planet orbiting a solar-like star at 0.045 au \citep{Fortney2010}. The vertical grey lines indicate the sampled ages for our various classes of models.  } \label{fig:evol_inputs}
\end{center}
\end{figure}

In order to incorporate the decline in XUV flux in this current work, we look to the physical rotational evolution model of \citet{Johnstone_2021} that is constrained by observed rotation distributions in young stellar clusters. Their study offers publicly available evolutionary tracks of stellar flux for a wide variety of stellar mass, age and initial rotation. We use their tracks of X-ray, hEUV, sEUV flux and stellar radius evolution for a 0.7-$M_{\sun}$ star, chosen to be representative of a K-dwarf star as detections of 1083$~$nm seem to favour such a host. We obtain tracks for both their `slow' and `fast' initial stellar rotation definitions corresponding to the 5th and 95th percentiles of their observed 150~Myr rotation distribution. In \citet{Allan_Vidotto_2019}, we normalised the XUV flux tracks of \citet{Johnstone2015a} so that the solar XUV flux was reproduced at the solar age. Similarly, we now normalise the \citet{Johnstone_2021} XUV flux tracks so that they reproduce the XUV flux of spectral type K6 star HD~85512, utilising a spectrum obtained from the Measurements of the Ultraviolet Spectral Characteristics of Low-mass Exoplanetary Systems (MUSCLES) Treasury Survey \citep{France_2016}, shown in Figure \ref{fig:hd85512}. MUSCLES combines Hubble and ground-based observations and where necessary (such as in the mostly inaccessible EUV wavelength range) stellar spectral models. While other K dwarf spectra were obtained by MUSCLES, we choose HD85512 to perform this normalisation based on its relatively late age. By this age, the slow and fast rotators have converged to the same rotation rate, meaning that normalising the flux tracks at this age retains the original track ratios between slow and fast rotator. We assume an age of $\sim 5600$~Myr for HD~85512 \citep{2011A&A...534A..58P} and a distance of 11.28 pc \citep{GAIA_2016_MISSION, Gaia_2022_DR3} to calculate its surface flux in each of the considered bins (see star symbols in Figure \ref{fig:evol_inputs}).

Our resulting evolution X-ray, hEUV and sEUV input fluxes \citep{Johnstone_2021} are shown in the upper-panel of Figure \ref{fig:evol_inputs}. Solid and dashed tracks correspond to slow and fast initial rotators, respectively.  For the near-UV and far-UV fluxes,  we use the results from \citet{Richey_Yowell_2022}, but in our modelling we merge their near- and far-UV fluxes into a single `mid-UV flux'. We include the mid-UV flux due to the ability of these photons to photoionise helium out of the triplet state. 

The central-panel of Figure \ref{fig:evol_inputs} shows the evolution of the stellar radius, also obtained from the model of \citet{Johnstone_2021}. Our evolving planetary radii inputs are shown in the lower-panel of Figure \ref{fig:evol_inputs}. These are the same radii used in \citet{Allan_Vidotto_2019}, corresponding to a 0.3~$M_{\rm jup}$ gas giant planet orbiting a solar-like star at 0.045~au \citep{Fortney2010}. Note that these models neglect planetary-radius inflation. The close proximity to its host and the planetary mass puts this planet at the upper edge of the hot-Neptune desert \citep{Mazeh_2016}. It should be noted that there is an inconsistency between the assumed stellar parameters in our flux (K dwarf) and radius evolution input (which assumes a G dwarf). 
In the future, it would be interesting to couple our hydrodynamic escape model to a planetary thermal evolution model as is done in \citet{2020MNRAS.499...77K}. This would allow us to also consider the decrease in planetary mass, which could be significant in case of strong atmospheric escape over time. This mass reduction would act to reduce the planetary gravity with evolution. As discussed in \citet{Allan_Vidotto_2019}, omitting this shrinking mass with evolution in our modelling leads to our models slightly underestimating the true atmospheric escape progressively with planetary evolution, as the planetary gravity is progressively overestimated. This mass loss underestimation is minor due to the stronger dependence of the gravitational force with planetary radius compared to planetary mass, and planetary radius varying more than the mass with evolution.

\section{The evolution of atmospheric escape} \label{sec:escape_results}

Table \ref{tab:models} describes the four sets of evolution-sampled models we  consider. This shows the initial stellar rotation, the He$/$H number abundance and whether or not helium heating and cooling processes are included. Hydrogen heating and cooling processes are included in all models. Helium heating and cooling, as well as the helium populations are self-consistently included in the top 3 models, while the model F10\%PostProc does not include any helium heating or cooling terms (in the hydrodynamics model, the presence of the helium particles only affects the mean molecular weight of the gas) -- in this case, the helium population is computed in a post-processing step. All models share the following common parameters: stellar mass $M_{*}=0.7M_{\sun}$, orbital distance $a=0.045~$au and planetary mass $M_{\rm pl}=0.3~M_{\rm Jup}$, while the received stellar flux and the planetary and stellar radii evolve as given by Figure \ref{fig:evol_inputs}. 

\begin{table*}
\caption{Description of the evolution model sets presented in this paper. Rotation refers to the initial stellar rotation (fast rotators have higher high-energy fluxes during their youth). The third column indicates the helium number abundance, with the remainder of the atmosphere being hydrogen. The listed heating and cooling processes correspond to the row IDs of Tables \ref{tab:photoion_constants} and \ref{tab:cool}. Model F10\%PostProc does not incorporate helium energetics in the hydrodynamic escape (although the mass of the helium particles affect the mean molecular weight) 
and instead computes the helium state fractions post-processingly, as described throughout section \ref{sec:model_hydro}. All models share the following common parameters; $M_{*}=0.7M_{\sun}$, $a=0.045~$au, $M_{\rm pl}=0.3~M_{\rm Jup}$. The received stellar flux and radius and the planetary radius vary with evolution following Figure \ref{fig:evol_inputs}.}  
\label{tab:models}
\centering
\begin{tabular}{cccccccccccc}
  		\hline 
      		model  &  stellar  & He  & heating  & cooling & helium    \\
  		 title &   rotation & (\%) &  eqs & eqs & population    \\
  		\hline \hline
S2\% & slow & 2 & \ref{photoion_1} -- \ref{photoion_21} &  \ref{cool_1} -- \ref{cool_12}   & self-consistent   \\  
F2\% & fast & 2 & \ref{photoion_1} -- \ref{photoion_21} &  \ref{cool_1} -- \ref{cool_12}  & self-consistent  \\	
F10\% & fast & 10 & \ref{photoion_1} -- \ref{photoion_21} &  \ref{cool_1} -- \ref{cool_12}  & self-consistent    \\
F10\%PostProc & fast & 10 & \ref{photoion_1} -- \ref{photoion_3} &    \ref{cool_1}, \ref{cool_5}, \ref{cool_9}   & post-processing    \\
\hline   
\end{tabular} 
\end{table*}

\begin{figure}
 \includegraphics[width=0.98\columnwidth]{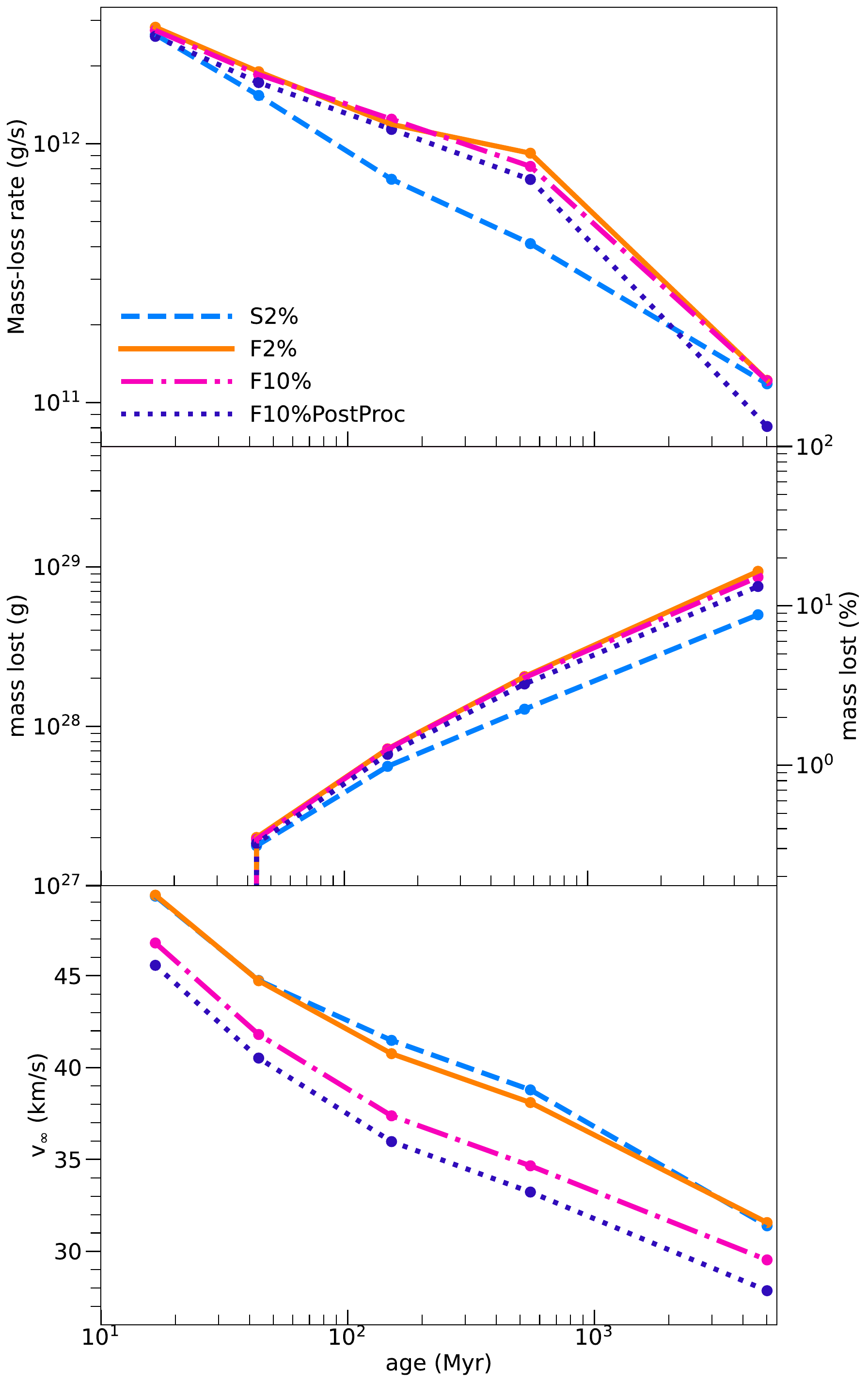}
 \caption{ Mass-loss rate, cumulative mass lost and the terminal velocity of the escaping atmosphere as a function of planetary age are shown in the upper, central and lower-panels, respectively. The evolution model sets are described in Table \ref{tab:models}. } \label{fig:mdot_vterm}
\end{figure}

Figure \ref{fig:mdot_vterm} displays how the predicted mass-loss rate (upper-panel), cumulative mass lost (central-panel) and terminal velocity of the escaping atmosphere (lower-panel) vary as a function of age for each of our model sets. We calculate the cumulative mass lost by integrating the mass-loss rate profile with respect to planetary age. As found in \citet{Allan_Vidotto_2019}, the diminishing XUV flux required to heat the planetary atmosphere combined with the growing gravitational force due to the shrinking planetary radii lead to the weakening of atmospheric escape as the planet evolves across all models. Variations with evolution are more extreme for the mass-loss rate compared to the terminal velocity. As discussed below, this is the result of the majority of heating occurring in the sub-sonic region of the atmosphere.

\begin{figure*}
\includegraphics[width=1.9\columnwidth]{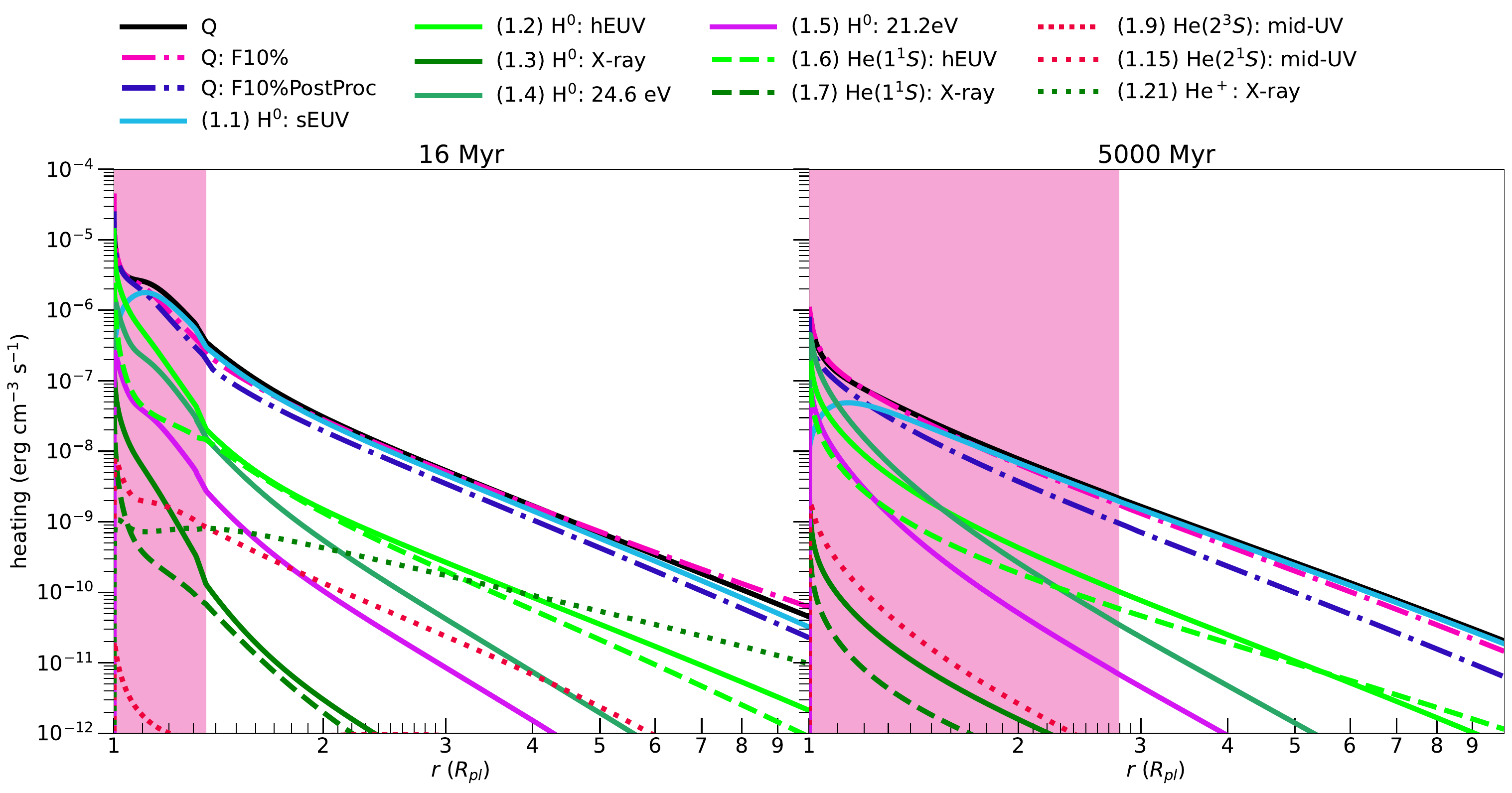}
\includegraphics[width=1.9\columnwidth]{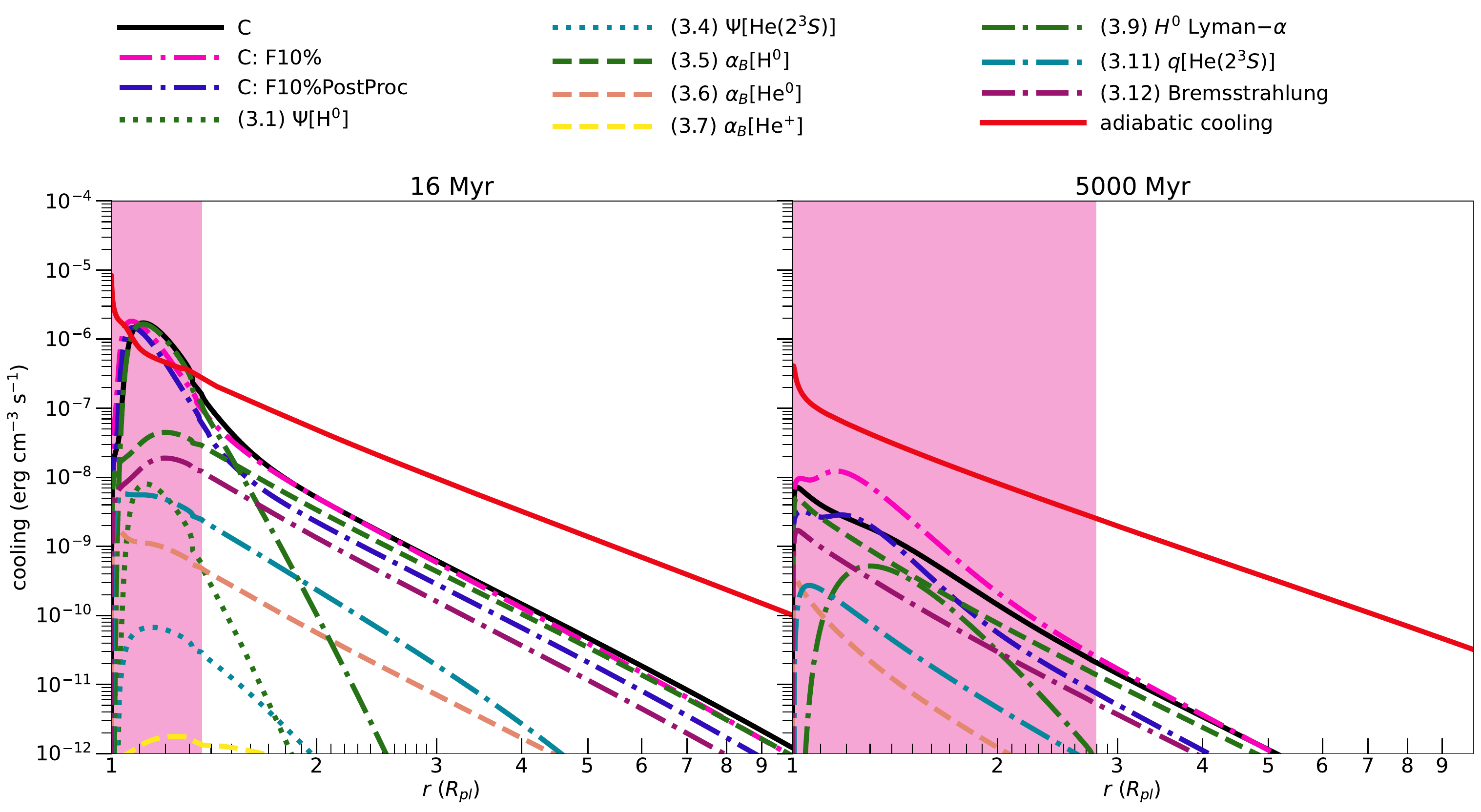}
\caption{Volumetric heating rates (upper-panel) and volumetric cooling rate (lower-panel) profiles at young (left) and old (right) planetary ages. Shaded regions on the left of each panel highlight the sub-sonic region of the atmosphere. In both legend sets, the numbers in parentheses relate each process to their corresponding process number in Table \ref{tab:photoion_constants} for heating and Table \ref{tab:cool} for cooling. In all displayed panels there are processes that fall below the chosen $y$-axis range and hence are not included in the legend. The tables on the other hand list all of the included processes. The displayed individual heating and cooling contributions correspond to our F2\% model (see Table \ref{tab:models}). $Q$ and $C$ correspond to this model's net heating source and cooling sink contributions (see Equation \ref{eq:cons_energy}). For comparison, we show also the $Q$ and $C$ of models F10\% and F10\%PostProc. } \label{fig:heating_opt_depth}
\end{figure*}

To investigate the evolution of the energetics of our models, the upper-panels of Figure \ref{fig:heating_opt_depth} show how the considered photoionisation heating processes differ at young (left) compared to old (right) planetary ages as a function of distance. The lower-panels show the dominant cooling processes. The displayed individual heating and cooling contributions correspond to our F2\% model (see Table \ref{tab:models}). $Q$ and $C$ correspond to the total heating source and cooling sink contributions (see Equation \ref{eq:cons_energy}) of this model. In each panel, the outflow velocity of the F2\% model remains sub-sonic within the shaded regions, reaching supersonic velocities beyond. For comparison, we show also the total heating and cooling contributions of models F10\% and F10\%PostProc. It should be noted that the F10\%PostProc heating profile could not simply be reproduced by summing the displayed hydrogen-only heating contributions of the F10\% model. Firstly, this is because we account for the limited availability of photons as explained previously with Equation \ref{eq:avail_photons}. Secondly, the heating affects the density distribution of neutrals which in turn affects the heating contributions.

We see in Figure \ref{fig:heating_opt_depth} that the same heating process remains dominant above $\sim 1.1 R_{\text{pl}}$ throughout the full planetary evolution: heating due to the photoionisation of neutral hydrogen by sEUV photons (ID: \ref{photoion_1} of Table \ref{tab:photoion_constants}). Below $\sim 1.1 R_{\text{pl}}$, heating from hydrogen photoionisations by hEUV photons (ID: \ref{photoion_2}) dominate over sEUV photons for the young planet, whereas at the oldest age, hydrogen photoionisation by the 24.6$~$eV photon (ID: \ref{photoion_4}), itself ejected in a direct recombination to He($1^1S$), becomes the dominant heater below $\sim 1.1 R_{\text{pl}}$. We found that the contribution of photoionisation of singlet state helium to the heating process, although not dominant, is not negligible for abundances of 2\%, and is more important for higher abundances. Sections \ref{sec:eff_He_frac} and \ref{sec:eff_He_energetics} will discuss the resulting effects of altering the assumed He$/$H number abundance and omitting helium energetics. Heating arising from photoionisations out of the helium triplet state is negligible.

\begin{figure}
 	\includegraphics[width=0.98\columnwidth]{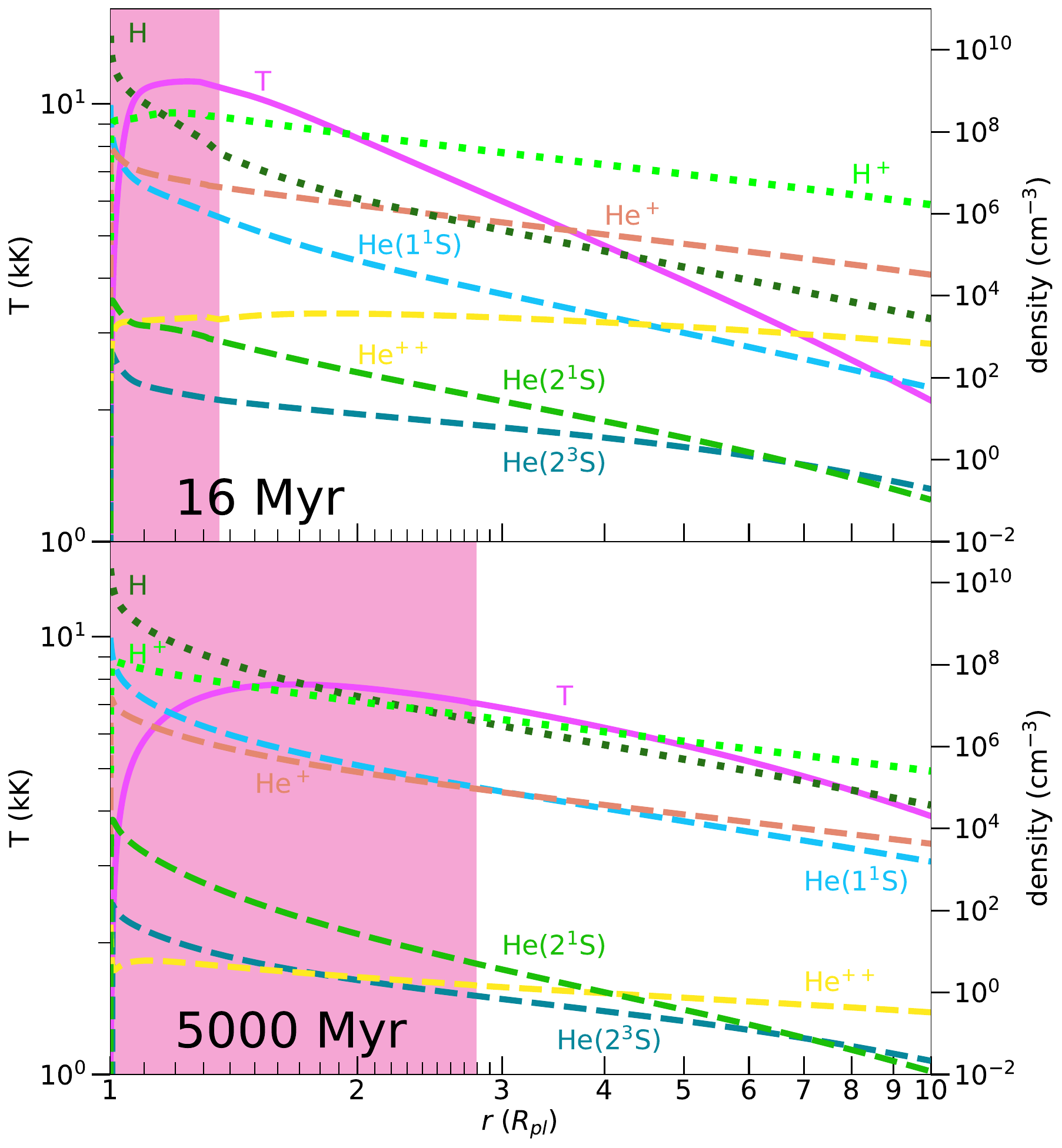}
    \caption{Temperature (left axis, solid line-style) and various number density profiles (right axis, dashed and dotted for helium and hydrogen, respectively). The upper-panel corresponds to an age of 16$~$Myr while the lower-panel shows 5\,000$~$Myr. The planetary parameters are that of our F2\% model (see Table \ref{tab:models}). Similar figures for our F10\% and F10\%PostProc are given in the Appendix. } \label{fig:t_density_struct}
\end{figure}

Clearly, the majority of heating affects the sub-sonic region of the atmosphere. The heating contribution peaks within the sub-sonic region for the displayed young and old models. Accordingly, the temperature profile follows a similar behaviour, reaching peaks of approximately 11\,250~K and 7\,750~K respectively as will later be shown in Figure \ref{fig:t_density_struct}. Due to the heating being mostly deposited in the sub-sonic region, heating variations with planetary evolution have a greater effect on the mass-loss rate, while having a lesser effect on the velocity of the outflowing atmosphere. This is indeed verified by Figure \ref{fig:mdot_vterm}, with a decrease of over one order of magnitude in mass-loss rate while the terminal velocity decreases by 35\% over the full evolution of our F2\% model.

The lower-panels of Figure \ref{fig:heating_opt_depth} show the dominant cooling processes. Similar to the heating processes, the net cooling peaks in the sub-sonic region of the atmosphere. Adiabatic cooling due to the expansion of the escaping atmosphere is more important than the cooling sink terms (composing $C$ in Equation \ref{eq:cons_energy}) for the old 5\,000-Myr model. Although not shown, this is also true for models F10\% and F10\%PostProc at the same age. At the young age of 16$~$Myr however, we see that collisional excitation of neutral hydrogen (dash-dotted green, ID: \ref{cool_9} in Table \ref{tab:cool}) more commonly referred to as Lyman-$\alpha$ cooling dominates at distances between 1.1 to $1.3~R_{\rm pl}$. Outside of this distance cooling due to adiabatic expansion again dominates. In their modelling with \textsc{ATES}, \citet{Caldiroli_2021_ATES} show that adiabatic expansion dominates the atmospheric cooling in the case of their low-irradiation model of HD~97658~b, similar to our 5\,000~Myr model, while for their high-irradiation planet WASP-43~b, radiative cooling, predominantly Lyman-$\alpha$ cooling, dominates below $1.5~R_{\text{pl}}$, similar to our $16$~Myr planet. Our model's predicted heating and cooling behaviour also agrees reasonably well with that of \citet{Zhang_2022_TOI_560} obtained with \textsc{TPCI} \citep{Salz2015TPCI} for the young mini-Neptune TOI~560.01. Assuming solar metalicity, they find adiabatic cooling to be more important than radiative cooling at nearly all radii, with radiative cooling overtaking in the $\sim 2-3$~$R_{\text{pl}}$ region of the atmosphere. Their predicted heating is also dominated by neutral hydrogen photoionisations with smaller but non-negligible contributions from helium in agreement with our findings. Below $1.3~R_{\text{pl}}$, line heating and photoionsation of metals dominate their modelled heating.

\begin{figure*}
 	\includegraphics[width=0.9\textwidth]{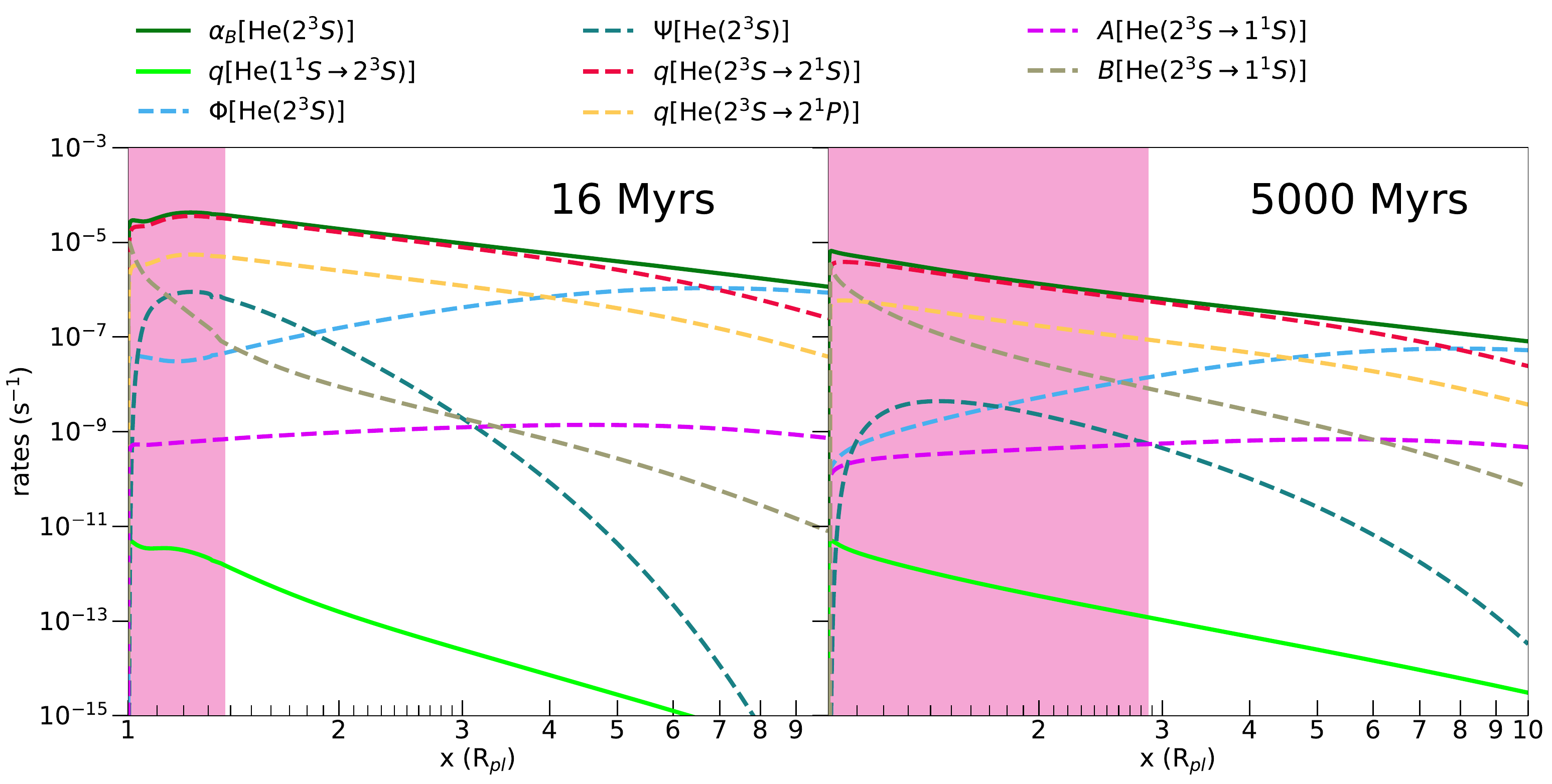}
    \caption{Rates of processes directly populating (solid) and depopulating (dashed) the helium triplet state at young (left) and old (right-panel) ages as a function of distance. Table \ref{tab:pop_depop} lists each considered transition. Here, only rates within the displayed $y-$axis range are listed in the Figure legend. The assumed planetary parameters are that of our F2\% model (Table \ref{tab:models}).} \label{fig:df_dtrip}
\end{figure*}

Figure \ref{fig:df_dtrip} shows the rates of the processes that directly populate (solid lines) or depopulate (dashed lines) the helium triplet state. Left and right panels correspond to ages 16 and 5\,000$~$Myr, respectively. Again, only the dominant processes are displayed in the figure, the exhaustive list of considered processes is reserved for Table \ref{tab:pop_depop}. While the relative rates between processes vary slightly with age, it is clear that the same triplet (de-)populating processes dominate throughout the planetary evolution. This also holds true at the intermediate ages not shown. At all distances, the dominant triplet-populating process is the recombination of ionised helium into the triplet state, $\alpha_B$[He$(2^3S)$].

Mostly balancing this process, is the de-excitation of triplet-state helium into the $2^1S$ state by means of a collision with a free electron ($q\left[\text{He}(2^3S \rightarrow 2^1S)\right]$). The general behaviour of the mentioned processes agree with the 1-D model of \citet{Oklopcic_2018_10839_window} and the 3-D models of \citet{Khodachenko_2021_107b_3d} and \citet{Rumenskikh_2022_3D_HD189733b} for WASP-107b and HD189733b, respectively. The depopulation of the triplet state due to both the photoionisation from mid-UV and sEUV photons is not a dominant process except in the very outer, more tenuous atmosphere, consistent with the findings of \citet{Oklopcic_2019_dep_st_rad} who shows that for stars cooler than spectral type G, collisional de-excitation from triplet to singlet state dominate. This is due to cooler K-dwarf stars having a smaller ratio of mid-UV$~:~$EUV flux, meaning a smaller ratio of triplet depopulating$~:~$escape driving flux, compared to warmer G type stars.

The triplet number density profile resulting from the balance of the mentioned populating and depopulating processes is shown in Figure \ref{fig:t_density_struct}, at young (upper-panel) and old (lower-panel) ages. The number densities of the other modelled helium states as well as neutral and ionised hydrogen (dotted profiles) are also included. The atmospheric temperature is shown by the solid profile. The temperature reached by the younger planet's atmosphere is larger due to the higher level of XUV flux received: the atmosphere of the 16-Myr old planet reaches a maximum temperature of 11\,250~K, while the atmosphere of the 5-Gyr-old planet reaches a maximum temperature of 7\,750$~$K. The level of hydrogen and helium ionisation is also larger at younger ages on account of the relatively larger incident flux. The number density of helium in the triplet state is greater for the younger planet. This is due to a higher rate of helium recombinations due to a larger availability of ionised helium at younger ages (Figure \ref{fig:df_dtrip}). 

\subsection{Effect of helium abundance on hydrodynamics}
\label{sec:eff_He_frac}

To investigate how the abundance of helium affects the hydrodynamic escape models, we run two set of models where we adopt helium number abundances of 0.02 (F2\%) and 0.1 (F10\%). 

In Figure \ref{fig:heating_opt_depth}, individual heating contributions were shown only for the F2\% model, with only the total $Q$ profile of F10\% on display. A deeper analysis into the heating for the F10\% model reveals however that heating due to helium unsurprisingly plays a more significant role for a greater helium abundance (see Appendix \ref{sec:append_He_10} and Figure \ref{fig:app1}). 

The nature of the atmospheric escape with evolution for models F2\% and F10\% is shown by their mass-loss rate and terminal velocities given in Figure \ref{fig:mdot_vterm} by the solid (F2\%) and dash-dotted (F10\%) tracks. Clearly, they both lead to very similar mass-loss rates, with both models resulting in similar cumulative loss of 15\%  (F10\%) and 16\% (F2\%) of the planet's initial mass over its entire evolution. The larger helium abundance produces also a slight decrease of 5-9\% in terminal velocity  maintained throughout the evolution as seen in the bottom panel of Figure \ref{fig:mdot_vterm}. 
Slower terminal velocities are reached due to the heavier mean molecular weight combined with similar total heating and cooling profiles (see Figure \ref{fig:heating_opt_depth}). In conclusion, we find that increasing the assumed helium abundance leads to slightly slower outflows, which undergo slightly less atmospheric mass losses. 

While the He$/$H range we consider here appears only to affect the hydrodynamics of escape trivially, there are two important effects resulting from the choice of the abundance. Firstly, the observability of helium transit is strongly dependent on the choice of helium abundance (Section \ref{sec:eff_He_frac_obs}). Secondly, if we omit the calculation of helium energetics, the choice of helium abundance can significantly affect the atmospheric escape. This will now be discussed.

\subsection{Effect of helium energetics on hydrodynamics}
\label{sec:eff_He_energetics}
The inclusion of helium affects the atmospheric dynamics in two ways. Firstly, it raises the mean molecular weight above that of a hydrogen-pure atmosphere. Intuitively, this effect acts to weaken atmospheric escape by raising the gravitational force experienced by the atmospheric material. This escape-reducing effect is straightforward to include in the modelling of atmospheric escape. Secondly, and more complex to model, is the effect of additional heating processes due to the inclusion of helium. In our modelling, we self-consistently incorporate the helium energetics, i.e. heating due to photoionisation from the $1^1S$, $2^1S$, $2^3S$ and singly ionised states, as well as cooling due to collisional excitation, and ionisation, recombination, and Bremsstrahlung in the fluid dynamic equations (Equations \ref{eq:con_mom}-\ref{eq:cons_mass_1st}). To investigate the effects that self-consistently including helium energetics has, we compute the model set F10\%PostProc which omits such effects for comparison, as described previously in section \ref{sec:test_post-proc}.

The top panels of Figure \ref{fig:heating_opt_depth} show the total $Q$ heating profiles of the model sets F10\% (dash-dotted fuchsia) and F10\%PostProc (dash-dotted purple). Comparing these, we see that close to the planet, the heating rates are comparable, but further away, the heating profile of the F10\%PostProc model is considerably less. This leads to similar peak atmospheric temperatures, with models F10\% and F10\%PostProc reaching 11.2(8.8)$~$kK and 11.3(8.3)$~$kK at ages 16 (5\,000)$~$Myr, respectively, but a faster temperature decay with distance when omitting helium energetics (see Appendix \ref{sec:append_He_energetic}, Figure \ref{fig:app2}).
In order to better understand how the contribution of heating is affected by the specific model set-up, we  calculate the net heating contributions for each model by integrating their volumetric heating rates over volume. Doing so reveals a $\sim$21\% heating reduction for the F10\%PostProc model relative to the F10\% model at the youngest age, growing to 45\% for the oldest model age of 5$~$Gyr.

As reduced atmospheric heating drives less escape, our F10\%PostProc model (purple dotted tracks in Figure \ref{fig:mdot_vterm}) undergoes less mass loss compared to the F10\% model (fuchsia dash-dotted), with respective losses of 13 and 15\% of their initial mass over their evolution. The terminal velocity is also reduced by the omission of helium energetics, with reductions between 2.5 to 5.5\% throughout the planetary evolution. The main barrier to the inclusion of helium energetics is that the helium populations must be solved simultaneously with the fluid dynamics equations rather than in a post-processing step. This is because knowledge of the helium populations are required for calculating the helium heating and cooling terms, which ought to be included in the energy conservation equation. 
It is important to note however that even our model without helium heating is more sophisticated than Parker-type wind models, which assume ad-hoc constant temperatures throughout the atmosphere.

\section{The observability of helium transits} 
\label{sec:observabilty_he_transits}
\subsection{The ray-tracing model for transmission spectroscopy} \label{sec:trans_model}

Here, we model transmission spectroscopy of the helium triplet at 1083$~$nm to investigate how the observability of this signature varies with long-term evolution. Our ray-tracing technique for modelling transmission spectroscopy is a helium-adapted version of a previous model described in \citet[][Lyman-$\alpha$ and H-$\alpha$]{Allan_Vidotto_2019} and \citet[][O\,I$~130.22$ nm]{Vidotto_2018}. Hence, this section gives only a brief summary aimed at documenting the necessary modifications. 
For a more thorough description of the general model we refer the reader to \citet{Allan_Vidotto_2019}.


As our ray tracing model is 3-D, we symmetrically fill a 3-D grid centred on the planet with our 1-D hydrodynamic calculations of atmospheric temperature, velocity and density. One of the grid axes is aligned along the observer–star line, so that the grid seen in the plane of the sky is a square of 201 $\times$ 201 cells. We also account for the line-of-sight velocity related to the orbital motion of the planet (assuming circular orbit) -- this correction is more important for phases farther from mid-transit, as in mid-transit, the line-of-sight velocity of the planet's orbital motion goes to zero.

The helium triplet is comprised of three individual lines, with line-centre wavelengths in air of 1082.909, 1083.025, 1083.034$~$nm. Hence, we model three individual wavelength-dependent transits for each of the triplet lines following the process described in detail in \citet{Allan_Vidotto_2019} for Lyman-$\alpha$ and H-$\alpha$ lines. We use the NIST database\footnote{\url{https://physics.nist.gov/asd}} \citep{NIST_ASD} to obtain the following line properties: the Einstein coefficient is $A_{ki} = 1.0216 \times 10^7$ s$^{-1}$, the mentioned line centre wavelengths and their corresponding oscillator strengths $f_{ik}$, 0.059902, 0.17974, 0.29958. With these atomic parameters, we calculate a wavelength-dependent optical depth along a single ray in the direction connecting the observer to the star–planet system. In doing so, we use a Voigt line profile, a convolution of a Gaussian and Lorentzian line profile accounting for both Doppler and natural broadening
\begin{equation}
\phi_{\lambda_0} =\frac{\lambda_0}{\sqrt{\pi}\varv_{\rm th}}  \frac{\chi}{\pi}\int_{-{\infty}}^{{\infty}}\frac{e^{-w^2}}{X^2 +(\Delta {\varv}/ \varv_{\rm th}-w)^2 }dw \, ,
\end{equation}
where $\lambda_0$ is the wavelength at line centre for the specific helium triplet line, $X={A_{ki} \lambda_0}/({4 \pi \varv_{\rm th}})$ is the damping parameter, $\varv_{\rm th} = (2 k_{\rm B} T / m_{\rm He})^{1/2}$ is the thermal velocity with $m_{\rm He}$ being the mass of atomic helium. The velocity offset from the line centre is $\Delta {\varv} = {\varv}{\rm channel}-{\varv}_{\rm LOS}$,  where ${\varv}_{\rm LOS}$ is the line of sight flow velocity of the escaping wind and ${\varv}{\rm channel}$ represents the velocity `channel' (related to wavelength via Doppler shifts) of the measurement. We slice our velocity calculations in 71 channels from -100 km/s to +100 km/s.  In calculating this line profile, we make use of IDL’s inbuilt \texttt{voigt} function. From the wavelength-dependent optical depth, we calculate the wavelength-dependent absorption. By integrating this over all considered rays within the stellar disk, we find the total absorption. In this way, we calculate the transit depth contributions due to each of the three lines that make up the helium triplet. Finally, we sum these contributions to obtain the total helium triplet transit depth.

We assume a transit along the centre of the stellar disk (i.e., no impact parameter). We also neglect centre-to-limb variations in the stellar disc. With $M_{*}=0.7M_{\sun}$, $a=0.045~$au and the age-specific $R_{*}$ shown in Figure \ref{fig:evol_inputs}, we obtain a transit duration between 2 to 2.4 hours for our models (see Equation 17 of \citealt{Allan_Vidotto_2019}).

\subsection{The evolution of helium 1083~nm transmission spectroscopy}
\label{sec:trans_spec_res}

\begin{figure*}
\includegraphics[width=0.9\textwidth]{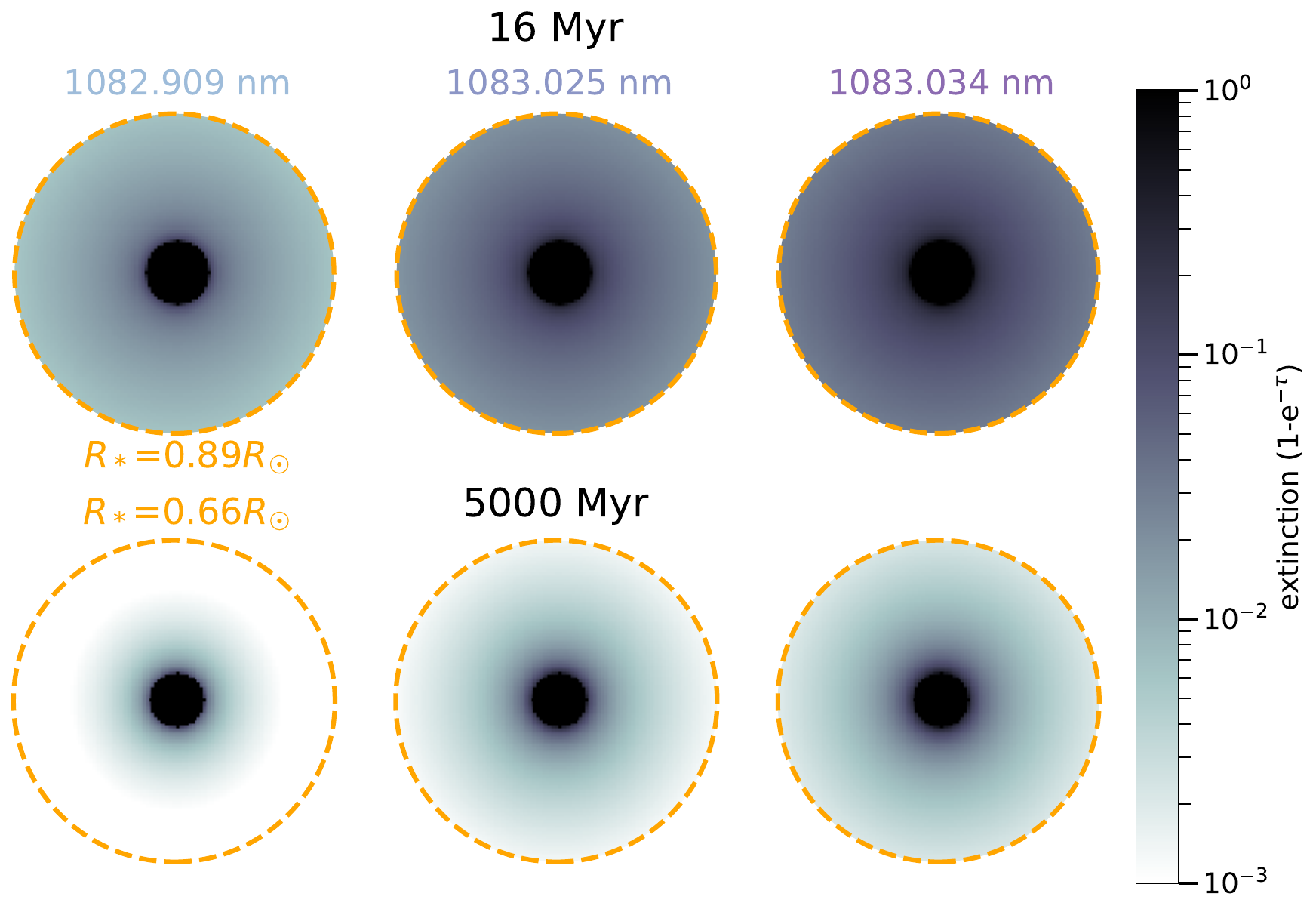
}
 \caption{Plane-of-the-sky extinction contributions at 16 (upper-row) and 5\,000~Myr (lower-row), shown at mid transit. Each column shows the contribution of one of the three lines of the helium triplet. The dashed orange circle marks the stellar disk with respective radii of 0.89 and 0.66 $~R_{\sun}$ at ages 16 and 5\,000~Myr. The planetary parameters are those of our F2\% model (see Table \ref{tab:models}).  } \label{fig:transit}
\end{figure*}

\begin{figure}
\includegraphics[width=0.48\textwidth]{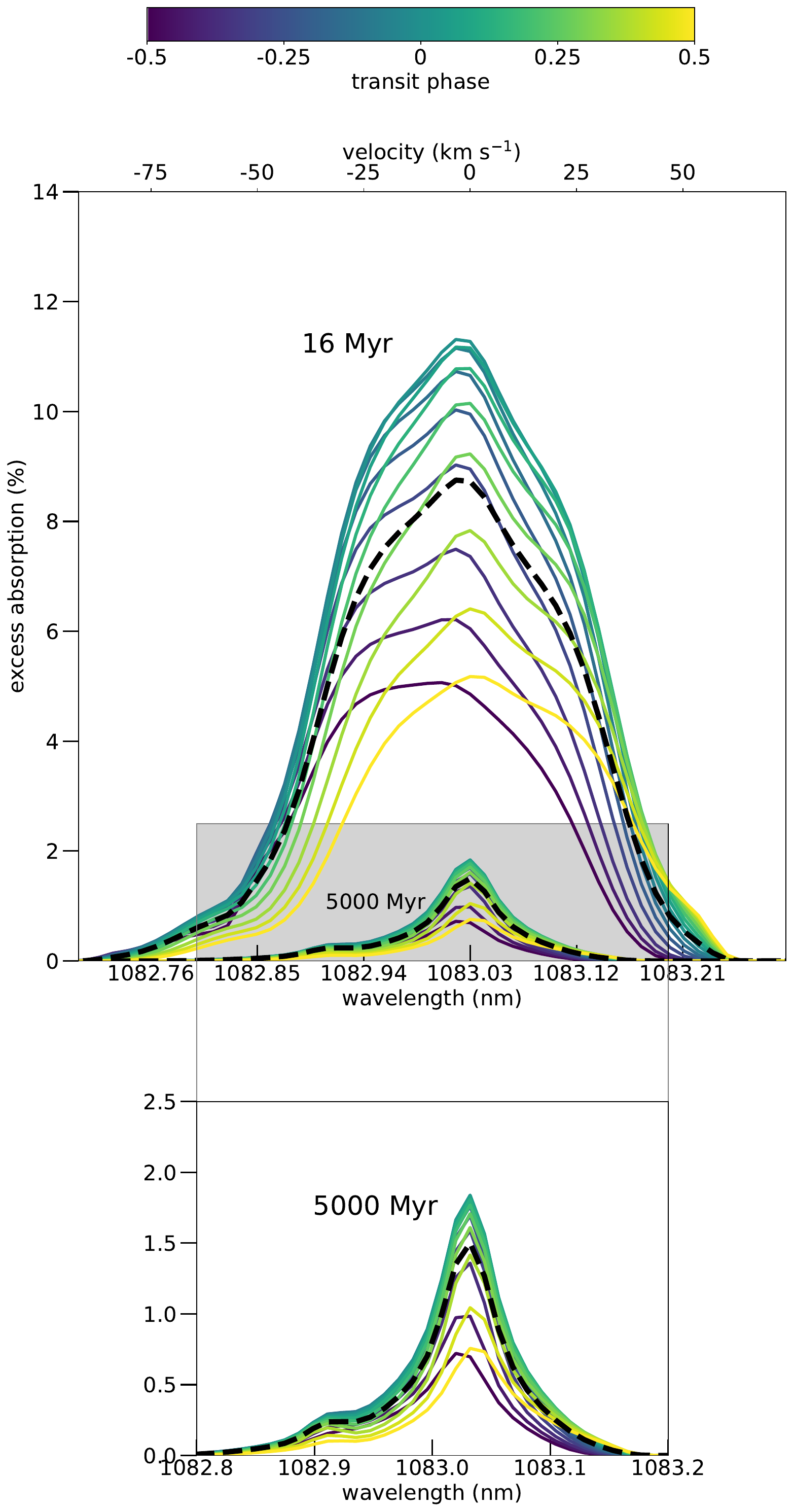}
 \caption{ Helium 1083$~$nm transmission spectra as a function of time/transit phase as indicated by the colour bar. Transit phases of -0.5 and +0.5 correspond to the time of first and fourth contacts, respectively. The mean average of all spectra between these two phases is shown in dashed black. The upper-panel shows the spectra at a planetary age of 16~Myr while the lower-panel offers a closer look at the spectra at age 5\,000~Myr. The assumed planetary parameters are that of our model-set entitled F2\% (see Table \ref{tab:models}).}
\label{transit_phase}
\end{figure}

Figure \ref{fig:transit} shows the plane-of-the-sky extinction computed for each of the individual lines of the helium triplet (shown here at mid-transit only). In each panel the stellar disk is outlined by the orange dashed circle. By integrating the extinction, we compute the excess absorption shown in Figure \ref{transit_phase}, where colour corresponds to various transit phases. At younger ages (upper-panel), the excess transit is larger and the transit depth is greatest at mid-transit as this is when the most dense atmospheric material occults the stellar disk. There is an asymmetry between pre- and post-mid-transit spectra as a result of the Doppler velocities of the atmospheric material due to the orbital motion. This asymmetry is greatest at times furthest from mid-transit as the majority of obscuring material is red-shifted (transit phase=+0.5, fourth contact `T4') or blue-shifted (transit phase=-0.5, first contact `T1'). The degree of asymmetry would be even greater had we considered the interaction with a stellar wind \citep{Carolan_2021}, as the outflowing atmosphere would be asymmetric, likely featuring a comet tail-like structure.

In practice, transit observations are integrated over a certain amount of time to obtain a sufficiently high signal-to-noise ratio. As the helium 1083$~$nm signature is then averaged out over time, the absorption from such an observation is weaker than the theoretical instantaneous 1083$~$nm signature at mid-transit. We account for this time-sampling effect by averaging our absorption profiles between T1 and T4, following \citet{pwinds}. 
The black dashed spectra in Figure \ref{transit_phase} show these calculated mean averages.

\begin{figure}
\includegraphics[width=0.98\columnwidth]{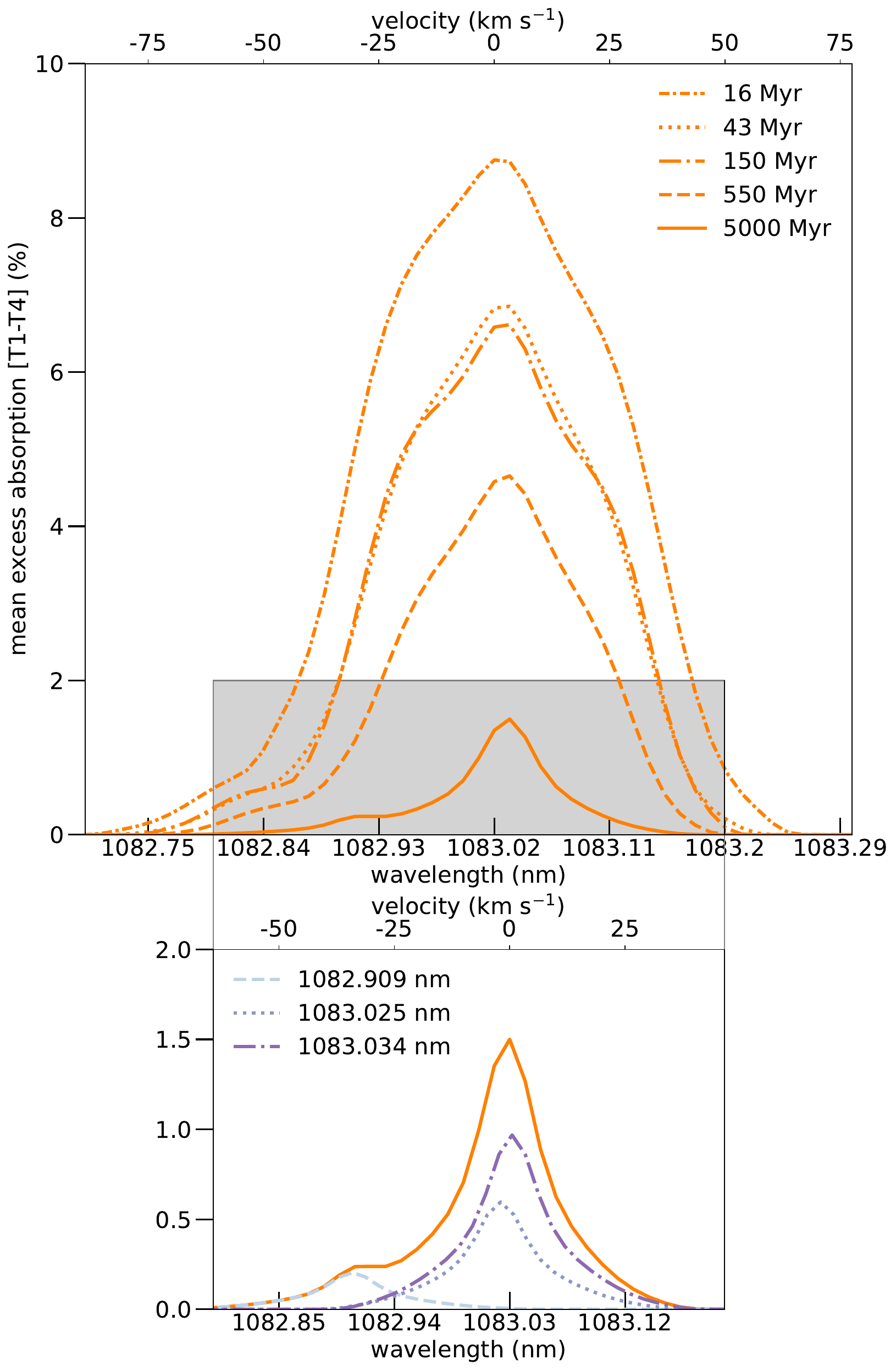}
 \caption{ [Upper-panel] The helium 1083$~$nm transmission spectra averaged over phases between first and fourth contacts. The line-style relates each spectrum to its planetary age. [Lower-panel] A zoom-in of the grey shaded region in the upper-panel. The 5\,000~Myr transmission spectrum is shown by the solid line as well as its individual line contributions as shown by the legend. The planetary parameters are those of our F2\% model (see Table \ref{tab:models}).    } \label{fig_1083}
\end{figure}

\begin{figure}
\includegraphics[width=0.45\textwidth]{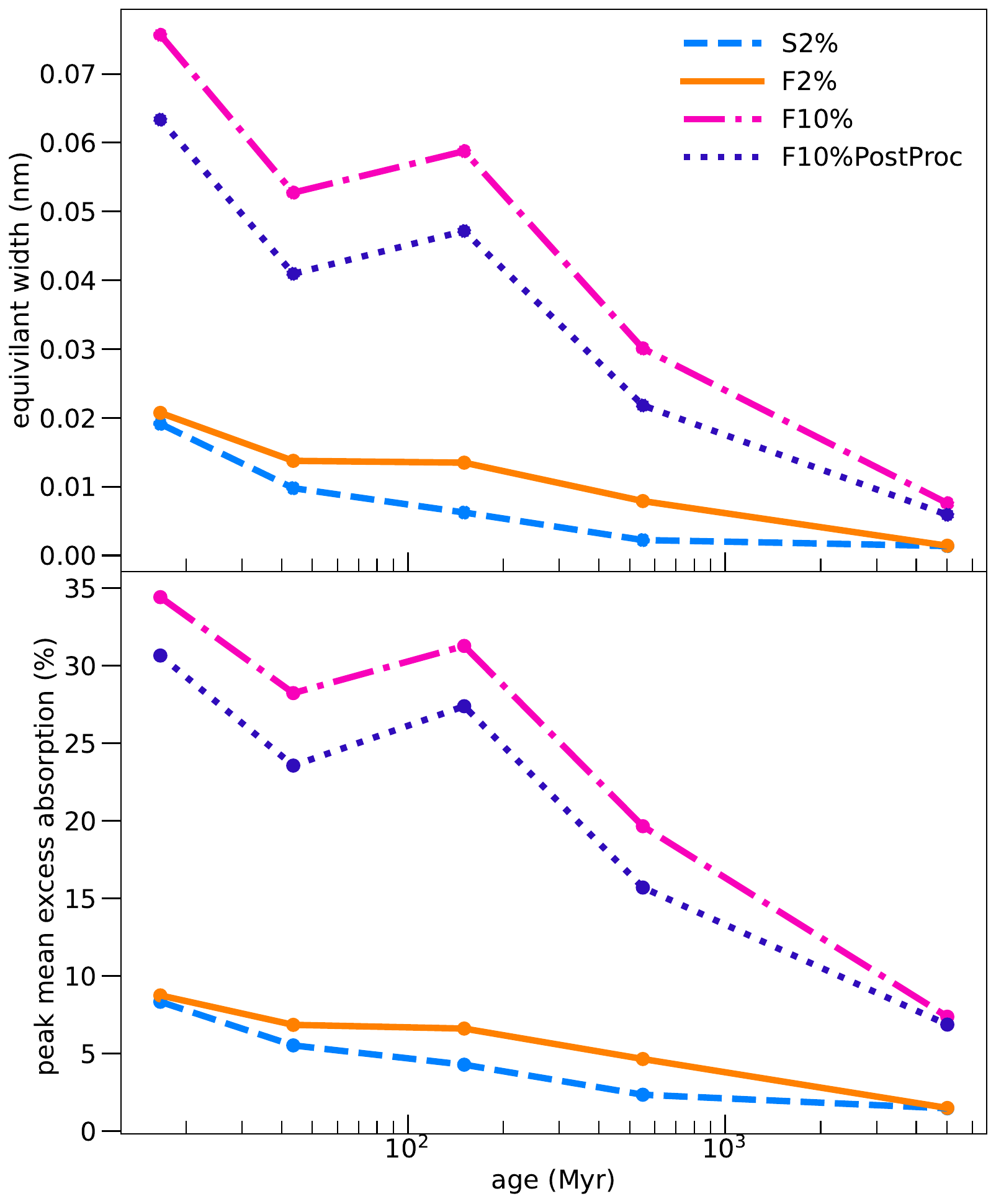}
 \caption{T1-T4 phase-averaged helium 1083$~$nm equivalent widths integrated over 1082.6-1083.35$~$nm (upper-panel) and peak excess absorptions (lower-panel) as a function of planetary evolution, for our model sets indicated in the legend (see Table \ref{tab:models}).  } \label{fig_EW_evol}
\end{figure}

The upper-panel of Figure \ref{fig_1083} shows the (T1-T4 phase-averaged) helium 1083$~$nm transmission spectra along planetary evolution. The lower-panel offers a closer look at the oldest age spectrum and its constituent contributions from each of the individual helium triplet lines. Figure \ref{fig_EW_evol} displays the evolution of the T1-T4 phase-averaged equivalent width (upper-panel) and peak excess absorption (lower-panel) of these helium 1083$~$nm signatures for all four of our model sets. 
In our models, thermal broadening is negligible and the broad profiles and large equivalent widths, especially more noticeable at younger ages, are due to Doppler broadening. This is because there is more absorbing material at larger distances (and thus with larger velocities) in the 16~Myr old model than in older systems. 
Additionally, we see that the helium 1083$~$nm signature weakens with planetary evolution. 
The main cause of the larger 1083$~$nm absorption at younger ages is the larger amount of material in the triplet state obscuring the stellar disk during planetary transit (see, e.g., the density profiles in Figure \ref{fig:t_density_struct} and the extinction in Figure \ref{fig:transit}). Interestingly, we see in Figure \ref{fig_1083} very little variation in the 1083$~$nm absorption profile between ages 43 and 150$~$Myr, despite the reduction in atmospheric escape. This is because the stellar radius also drops from 0.74 to 0.64$~R_\odot$, hence the ratio of obscuring He($2^3S$) to the stellar disk area remains similar at both ages.

\subsection{Effect of helium abundance on 1083~nm observability}
\label{sec:eff_He_frac_obs}
 When comparing models with different abundances F2\% and F10\% in Figure \ref{fig_EW_evol}, we see a large change in the equivalent widths and peak excess absorptions. For example, model-set F2\% (F10\%) produces an equivalent width and peak excess absorptions of 0.013~nm (0.058~nm) and 6.6\% (31.4\%) at an age of 150~Myr, while at the later age of 5\,000~Myr, the corresponding values are 0.0014~nm (0.007~nm) and 1.5\% (7.4\%). Although these models exhibited similar atmospheric escape hydrodynamics (section \ref{sec:eff_He_frac}), our synthetic transit observations are very sensitive to the adopted helium abundance.

The F10\% and F10\%PostProc model sets show increased 1083$~$nm absorption going from 43 to 150$~$Myr while F2\% maintains roughly the same absorption between these two ages, as discussed in the previous subsection. The F2\% model set exhibits a similar evolution of atmospheric escape to the F10\% model, however unsurprisingly the latter has overall more He($2^3S$) extinction and accordingly greater 1083$~$nm absorption. The larger ratio of obscuring He($2^3S$) material relative to the stellar disk for F10\% compared to the F2\% model set emphasises the observational effect of the more drastic stellar disk area variation coinciding with a lesser variation in atmospheric escape between 43 and 150$~$Myr, leading to the bump in Figure \ref{fig_EW_evol} at 150$~$Myr for F10\% and F10\%PostProc models.

\subsection{Effect of helium energetics on 1083~nm observability}
\label{sec:eff_post_p_obs}

Figure \ref{fig_EW_evol} shows similar transit properties for models F10\% and F10\%PostProc, with F10\%PostProc having smaller and narrower absorptions. Interestingly, the inclusion of helium energetics impacts the resulting 1083$~$nm signature, however, as for the modelled hydrodynamics (section \ref{sec:eff_He_energetics}), the resulting variation is small over the modelled parameter space. 

When excluding helium energetics, the missed heating from helium photoionisation and hydrogen photoionisation by helium produced photons is partially balanced by a larger availability of (stellar) photons capable of photoionising hydrogen, contributing to atmospheric heating (the F10\%PostProc sets all photoionisation weighting factors to 1). 100\% of the received stellar photons (bar those in the mid-UV due to their low energy) are then available to photoionise hydrogen. For a particular photoionisation, there is greater excess kinetic energy to heat the atmosphere than if the same photon was allowed to ionise He($1^1S$). At older ages, the variation between the self-consistent and post-processing model is smaller on account of the reduced level of photoionisations.

\section{Discussion and Conclusions}
\label{sec:conclusions}
\subsection{Summary of main findings}
In this paper, we investigated the evolution of atmospheric escape for highly irradiated 0.3$M_{\rm Jup}$ exoplanets with primordial hydrogen/helium atmospheres orbiting a K-dwarf star at 0.045~au, with the goal of predicting the evolution of their helium 1083$~$nm transits. Our model self-consistently solves the fluid dynamic equations in addition to the coupled equations for hydrogen ionisation balance and the helium $1^1S$, $2^1S$, $2^3S$ and He$^+$ states. For the transmission spectroscopy modelling of the helium triplet signature at 1083$~$nm, we use a ray tracing technique previously applied to Lyman-$\alpha$ and H-$\alpha$ transmission spectroscopy modelling \citep{Allan_Vidotto_2019}. We utilise evolving predictions of stellar flux and radius \citep{Johnstone_2021} and planetary radius \citep{Fortney2010} as input for our evolution modeling. We also explored how the He$/$H abundances and self-consistent inclusion and omission of helium energetics affect both the atmospheric escape and its 1083$~$nm absorption signature.

Our main conclusions regarding the hydrodynamics of escaping atmosphere are listed below:
\begin{itemize}
\item The heating processes which dominate are: photoionisation of neutral hydrogen by sEUV (36-92$~$nm), hEUV (10-36$~$nm) photons as well as by the 24.6$~$eV photon produced in the direct recombination to the helium $1^1S$ state. Although not dominant, heating arrising from the photoionisation of He(1$^1$S) by hEUV photons is not negligible for abundances of 2\% (Figure \ref{fig:heating_opt_depth}), and is more important for higher abundances (Figure \ref{fig:app1}).

\item Neglecting helium heating contributions leads to a reduction of 21 and 45\% in the net heating contribution at young and old ages, respectively, assuming a helium abundance of 0.1. See Figure \ref{fig:heating_opt_depth}.

\item The majority of the heating affects the sub-sonic region of the atmosphere meaning that heating reductions with planetary evolution affects the mass-loss rate more so than the terminal velocity of the escaping atmosphere. 

\item Increasing the helium abundance from 2\% to 10\% results in a slightly slower outflow, with a smaller mass-loss rate (Figure \ref{fig:mdot_vterm}). This does not strongly affect the hydrodynamics, but is important in the observability of the triplet (as we will discuss below).

\item Omitting helium energetics in the hydrodynamics models leads to a lower mass loss and terminal velocity (Figure \ref{fig:mdot_vterm}).

\end{itemize}

Regarding the population of the helium triplet and its observability, we found the following main conclusions: 
\begin{itemize}
\item The dominant populating process for the helium triplet state is the recombination of ionised helium into the triplet state (Figure \ref{fig:df_dtrip}).
\item The dominant depopulating process for the helium triplet state helium is de-excitation into the helium singlet state through a collision with a free electron (Figure \ref{fig:df_dtrip}). 
\item Although increasing the helium abundance from 2\% to 10\% does not affect the hydrodynamics, it substantially changes the observability of the transiting atmosphere. At an age of 5~Gyr, the (T1-T4 phase-averaged) peak excess absorption increases from 1.5\% to 7.4\% (see Figure \ref{fig_EW_evol}).

\end{itemize}

When accounting for the evolution of the system, we showed that the diminishing XUV flux required to heat the planetary atmosphere combined with the growing gravitational force due to the shrinking planetary radii leads to the weakening of atmospheric escape as the planet evolves. As a consequence of a slower, less dense and less extended wind, the helium 1083$~$nm signature weakens overall with evolution (Figures \ref{fig:transit}, \ref{fig_1083}, \ref{fig_EW_evol}).

We found that the strongest helium 1083$~$nm absorption occurs at the youngest ages. Our models predict that a very young ($\lesssim$150$~$Myr) gas giant closely orbiting a K-dwarf star could produce a helium 1083$~$nm absorption of $\sim$4\% to $\sim$7\% assuming a helium abundance of 2\% (if orbiting a slow or fast rotating star). This weakening with evolution, however, is not necessarily monotonic. For example, the transit depths and equivalent width of the planet at age 150~Myr are larger than those at 40~Myr for models with He$/$H$=0.1$. This is due to a large drop in stellar radius corresponding with small decline of atmospheric escape.

\subsection{Do we observe larger helium absorption at younger ages?}

Our model suggests stronger helium 1083$~$nm absorption at younger ages. Is this higher absorption seen in the observations?

So far, there has been a limited number of helium 1083$~$nm transit observations of young systems. One difficulty is the greater stellar variability in helium 1083$~$nm at younger ages. Another difficulty is the small number of exoplanets known to transit a young ($<500~$Myr) star, and in particular a K-type star best suited for 1083$~$nm detections. 
This currently small sample consists of super-Earths and mini-Neptunes, whose escaping atmospheres are intrinsically harder to detect than for larger planets. 

In recent years,  helium transit observations of a few young systems resulted in 
either non-detections or tentative detections possibly of stellar rather than planetary origin, namely of AU Mic b  \citep{Hirano_2020_au_mic}, of the V1298 Tau planets \citep{Gaidos_2022V1298_Tau, 2021_Vissapragada_search_He_in_V1298_tau_system}, of TOI 1807b and 2076b \citep{Gaidos2023}, and of the 400-Myr old mini-Neptune HD 63433c \citep{2022_HD63433_upper_limit_non_detect}. Of these, only TOI 1807b and 2076b are thought to have K dwarf hosts, but these planets are rocky planets whose primary H-rich atmosphere might have already been lost. 
\citet{2022_Zhang_4_mini_Nep, Zhang_2022_TOI_560} recently reported successful helium 1083$~$nm detections for four young mini-Neptunes orbiting K dwarfs, with mean excess absorptions of $\sim$1\% in each.

It should be noted that the larger absorption predicted at young ages in our evolution models is not at odds with the current observations, as the current sample of observed young exoplanets orbiting K-dwarfs for which helium 1083~nm transmission spectroscopy has been performed does not yet include a gas giant planet. Hence, finding a system in which a larger radius $1-2 R_{\rm Jup}$ close-in ($<0.1~$au) exoplanet transits a young ($<150~$Myr), ideally K dwarf star would be a better test to our evolution models. In a forthcoming work, we apply our modelling techniques for atmospheric escape and helium 1083$~$nm transmission spectroscopy to smaller, mini-Netpunes, more consistent with the current helium 1083$~$nm detections of young exoplanets.

\section*{Acknowledgements}
APA and AAV acknowledge funding from the European Research Council (ERC) under the European Union's Horizon 2020 research and innovation programme (grant agreement No 817540, ASTROFLOW). We thank several colleagues for discussions over this project: S. Carolan, D. Kubyshkina, J.S. Pineda, I. F. Shaikhislamov, A. Oklopčić, R. Allart and M. L\'opez Puertas.
We gratefully thank the anonymous referee for their thorough analysis, insightful comments and suggestions that have greatly enhanced this work.
This work has made use of data from the European Space Agency (ESA) mission
{\it Gaia} (\url{https://www.cosmos.esa.int/gaia}), processed by the {\it Gaia}
Data Processing and Analysis Consortium (DPAC,
\url{https://www.cosmos.esa.int/web/gaia/dpac/consortium}). Funding for the DPAC
has been provided by national institutions, in particular the institutions
participating in the {\it Gaia} Multilateral Agreement.

\section*{Data Availability}
The data described in this article will be shared on reasonable request to the corresponding author.
%
%
%
%



\bibliographystyle{mnras}
\bibliography{example} 

\begin{thebibliography}{}
\makeatletter
\relax
\def\mn@urlcharsother{\let\do\@makeother \do\$\do\&\do\#\do\^\do\_\do\%\do\~}
\def\mn@doi{\begingroup\mn@urlcharsother \@ifnextchar [ {\mn@doi@}
  {\mn@doi@[]}}
\def\mn@doi@[#1]#2{\def\@tempa{#1}\ifx\@tempa\@empty \href
  {http://dx.doi.org/#2} {doi:#2}\else \href {http://dx.doi.org/#2} {#1}\fi
  \endgroup}
\def\mn@eprint#1#2{\mn@eprint@#1:#2::\@nil}
\def\mn@eprint@arXiv#1{\href {http://arxiv.org/abs/#1} {{\tt arXiv:#1}}}
\def\mn@eprint@dblp#1{\href {http://dblp.uni-trier.de/rec/bibtex/#1.xml}
  {dblp:#1}}
\def\mn@eprint@#1:#2:#3:#4\@nil{\def\@tempa {#1}\def\@tempb {#2}\def\@tempc
  {#3}\ifx \@tempc \@empty \let \@tempc \@tempb \let \@tempb \@tempa \fi \ifx
  \@tempb \@empty \def\@tempb {arXiv}\fi \@ifundefined
  {mn@eprint@\@tempb}{\@tempb:\@tempc}{\expandafter \expandafter \csname
  mn@eprint@\@tempb\endcsname \expandafter{\@tempc}}}

\bibitem[\protect\citeauthoryear{{Allan} \& {Vidotto}}{{Allan} \&
  {Vidotto}}{2019}]{Allan_Vidotto_2019}
{Allan} A.,  {Vidotto} A.~A.,  2019, \mn@doi [\mnras] {10.1093/mnras/stz2842},
  \href {https://ui.adsabs.harvard.edu/abs/2019MNRAS.490.3760A} {490, 3760}

\bibitem[\protect\citeauthoryear{{Allart} et~al.,}{{Allart}
  et~al.}{2018}]{Allart_2018_warm_Nep_Hat_p_11b_He_obs}
{Allart} R.,  et~al., 2018, \mn@doi [Science] {10.1126/science.aat5879}, \href
  {https://ui.adsabs.harvard.edu/abs/2018Sci...362.1384A} {362, 1384}

\bibitem[\protect\citeauthoryear{{Allart} et~al.,}{{Allart}
  et~al.}{2019}]{Allart_2019_WASP_107b}
{Allart} R.,  et~al., 2019, \mn@doi [\aap] {10.1051/0004-6361/201834917}, \href
  {https://ui.adsabs.harvard.edu/abs/2019A&A...623A..58A} {623, A58}

\bibitem[\protect\citeauthoryear{{Alonso-Floriano} et~al.,}{{Alonso-Floriano}
  et~al.}{2019}]{Alonso_Floriano_Snellen_2019_HD209458b_He_obs}
{Alonso-Floriano} F.~J.,  et~al., 2019, \mn@doi [\aap]
  {10.1051/0004-6361/201935979}, \href
  {https://ui.adsabs.harvard.edu/abs/2019A&A...629A.110A} {629, A110}

\bibitem[\protect\citeauthoryear{{Benjamin}, {Skillman}  \& {Smits}}{{Benjamin}
  et~al.}{1999}]{Benjamin_1999}
{Benjamin} R.~A.,  {Skillman} E.~D.,   {Smits} D.~P.,  1999, \mn@doi [\apj]
  {10.1086/306923}, \href
  {https://ui.adsabs.harvard.edu/abs/1999ApJ...514..307B} {514, 307}

\bibitem[\protect\citeauthoryear{Bergeson et~al.,}{Bergeson
  et~al.}{1998}]{Bergeson_1998_He21S}
Bergeson S.~D.,  et~al., 1998, \mn@doi [Phys. Rev. Lett.]
  {10.1103/PhysRevLett.80.3475}, 80, 3475

\bibitem[\protect\citeauthoryear{{Black}}{{Black}}{1981}]{Black_1981}
{Black} J.~H.,  1981, \mn@doi [\mnras] {10.1093/mnras/197.3.553}, \href
  {https://ui.adsabs.harvard.edu/abs/1981MNRAS.197..553B} {197, 553}

\bibitem[\protect\citeauthoryear{{Bourrier} \& {Lecavelier des
  Etangs}}{{Bourrier} \& {Lecavelier des Etangs}}{2013}]{Bourrier2013}
{Bourrier} V.,  {Lecavelier des Etangs} A.,  2013, \mn@doi [\aap]
  {10.1051/0004-6361/201321551}, \href
  {https://ui.adsabs.harvard.edu/abs/2013A&A...557A.124B} {557, A124}

\bibitem[\protect\citeauthoryear{{Bourrier}, {Lecavelier des Etangs},
  {Ehrenreich}, {Tanaka}  \& {Vidotto}}{{Bourrier}
  et~al.}{2016}]{Bourrier+2016_GJ436}
{Bourrier} V.,  {Lecavelier des Etangs} A.,  {Ehrenreich} D.,  {Tanaka} Y.~A.,
   {Vidotto} A.~A.,  2016, \mn@doi [\aap] {10.1051/0004-6361/201628362}, \href
  {https://ui.adsabs.harvard.edu/abs/2016A&A...591A.121B} {591, A121}

\bibitem[\protect\citeauthoryear{{Bray}, {Burgess}, {Fursa}  \& {Tully}}{{Bray}
  et~al.}{2000}]{Bray2000}
{Bray} I.,  {Burgess} A.,  {Fursa} D.~V.,   {Tully} J.~A.,  2000, \mn@doi
  [\aaps] {10.1051/aas:2000277}, \href
  {https://ui.adsabs.harvard.edu/abs/2000A&AS..146..481B} {146, 481}

\bibitem[\protect\citeauthoryear{{Brown}}{{Brown}}{1971}]{Brown1971}
{Brown} R.~L.,  1971, \mn@doi [\apj] {10.1086/150851}, \href
  {https://ui.adsabs.harvard.edu/abs/1971ApJ...164..387B} {164, 387}

\bibitem[\protect\citeauthoryear{{Caldiroli}, {Haardt}, {Gallo}, {Spinelli},
  {Malsky}  \& {Rauscher}}{{Caldiroli} et~al.}{2021a}]{Caldiroli2021}
{Caldiroli} A.,  {Haardt} F.,  {Gallo} E.,  {Spinelli} R.,  {Malsky} I.,
  {Rauscher} E.,  2021a, \mn@doi [\aap] {10.1051/0004-6361/202141497}, \href
  {https://ui.adsabs.harvard.edu/abs/2021A&A...655A..30C} {655, A30}

\bibitem[\protect\citeauthoryear{{Caldiroli}, {Haardt}, {Gallo}, {Spinelli},
  {Malsky}  \& {Rauscher}}{{Caldiroli} et~al.}{2021b}]{Caldiroli_2021_ATES}
{Caldiroli} A.,  {Haardt} F.,  {Gallo} E.,  {Spinelli} R.,  {Malsky} I.,
  {Rauscher} E.,  2021b, \mn@doi [\aap] {10.1051/0004-6361/202141497}, \href
  {https://ui.adsabs.harvard.edu/abs/2021A&A...655A..30C} {655, A30}

\bibitem[\protect\citeauthoryear{{Carolan}, {Vidotto}, {Plavchan}, {Villarreal
  D'Angelo}  \& {Hazra}}{{Carolan} et~al.}{2020}]{2020MNRAS.498L..53C}
{Carolan} S.,  {Vidotto} A.~A.,  {Plavchan} P.,  {Villarreal D'Angelo} C.,
  {Hazra} G.,  2020, \mn@doi [\mnras] {10.1093/mnrasl/slaa127}, \href
  {https://ui.adsabs.harvard.edu/abs/2020MNRAS.498L..53C} {498, L53}

\bibitem[\protect\citeauthoryear{{Carolan}, {Vidotto}, {Villarreal D'Angelo}
  \& {Hazra}}{{Carolan} et~al.}{2021}]{Carolan_2021}
{Carolan} S.,  {Vidotto} A.~A.,  {Villarreal D'Angelo} C.,   {Hazra} G.,  2021,
  \mn@doi [\mnras] {10.1093/mnras/staa3431}, \href
  {https://ui.adsabs.harvard.edu/abs/2021MNRAS.500.3382C} {500, 3382}

\bibitem[\protect\citeauthoryear{{Cecchi-Pestellini}, {Ciaravella}  \&
  {Micela}}{{Cecchi-Pestellini} et~al.}{2006}]{Cecchi_Pestellini2006}
{Cecchi-Pestellini} C.,  {Ciaravella} A.,   {Micela} G.,  2006, \mn@doi [\aap]
  {10.1051/0004-6361:20066093}, \href
  {https://ui.adsabs.harvard.edu/abs/2006A&A...458L..13C} {458, L13}

\bibitem[\protect\citeauthoryear{{Cen}}{{Cen}}{1992}]{Cen_1992}
{Cen} R.,  1992, \mn@doi [\apjs] {10.1086/191630}, \href
  {https://ui.adsabs.harvard.edu/abs/1992ApJS...78..341C} {78, 341}

\bibitem[\protect\citeauthoryear{{Dos Santos}}{{Dos
  Santos}}{2022}]{Dos_Santos_2022_iauga_obs_of_pl_winds_outflows}
{Dos Santos} L.~A.,  2022, \mn@doi [arXiv e-prints]
  {10.48550/arXiv.2211.16243}, \href
  {https://ui.adsabs.harvard.edu/abs/2022arXiv221116243D} {p. arXiv:2211.16243}

\bibitem[\protect\citeauthoryear{{Dos Santos} et~al.,}{{Dos Santos}
  et~al.}{2022}]{pwinds}
{Dos Santos} L.~A.,  et~al., 2022, \mn@doi [\aap]
  {10.1051/0004-6361/202142038}, \href
  {https://ui.adsabs.harvard.edu/abs/2022A&A...659A..62D} {659, A62}

\bibitem[\protect\citeauthoryear{{Dos Santos} et~al.,}{{Dos Santos}
  et~al.}{2023}]{2023AJ....166...89D}
{Dos Santos} L.~A.,  et~al., 2023, \mn@doi [\aj] {10.3847/1538-3881/ace445},
  \href {https://ui.adsabs.harvard.edu/abs/2023AJ....166...89D} {166, 89}

\bibitem[\protect\citeauthoryear{{Drake}}{{Drake}}{1971}]{Drake1971}
{Drake} G.~W.,  1971, \mn@doi [\pra] {10.1103/PhysRevA.3.908}, \href
  {https://ui.adsabs.harvard.edu/abs/1971PhRvA...3..908D} {3, 908}

\bibitem[\protect\citeauthoryear{{Eikema}, {Ubachs}, {Vassen}  \&
  {Hogervorst}}{{Eikema} et~al.}{1996}]{Eikema_1996}
{Eikema} K.~S.~E.,  {Ubachs} W.,  {Vassen} W.,   {Hogervorst} W.,  1996,
  \mn@doi [\prl] {10.1103/PhysRevLett.76.1216}, \href
  {https://ui.adsabs.harvard.edu/abs/1996PhRvL..76.1216E} {76, 1216}

\bibitem[\protect\citeauthoryear{{Ferland}, {Korista}, {Verner}, {Ferguson},
  {Kingdon}  \& {Verner}}{{Ferland} et~al.}{1998}]{Ferland1998}
{Ferland} G.~J.,  {Korista} K.~T.,  {Verner} D.~A.,  {Ferguson} J.~W.,
  {Kingdon} J.~B.,   {Verner} E.~M.,  1998, \mn@doi [\pasp] {10.1086/316190},
  \href {https://ui.adsabs.harvard.edu/abs/1998PASP..110..761F} {110, 761}

\bibitem[\protect\citeauthoryear{{Ferland} et~al.,}{{Ferland}
  et~al.}{2017}]{Ferland2017}
{Ferland} G.~J.,  et~al., 2017, \mn@doi [\rmxaa] {10.48550/arXiv.1705.10877},
  \href {https://ui.adsabs.harvard.edu/abs/2017RMxAA..53..385F} {53, 385}

\bibitem[\protect\citeauthoryear{Fortney \& Nettelmann}{Fortney \&
  Nettelmann}{2010}]{Fortney2010}
Fortney J.~J.,  Nettelmann N.,  2010, \mn@doi [Space Science Reviews]
  {10.1007/s11214-009-9582-x}, 152, 423

\bibitem[\protect\citeauthoryear{{Fossati} et~al.,}{{Fossati}
  et~al.}{2022}]{Fosatti_2022_non_detect_He_Wasp_80b}
{Fossati} L.,  et~al., 2022, \mn@doi [\aap] {10.1051/0004-6361/202142336},
  \href {https://ui.adsabs.harvard.edu/abs/2022A&A...658A.136F} {658, A136}

\bibitem[\protect\citeauthoryear{{France} et~al.,}{{France}
  et~al.}{2016}]{France_2016}
{France} K.,  et~al., 2016, \mn@doi [\apj] {10.3847/0004-637X/820/2/89}, \href
  {https://ui.adsabs.harvard.edu/abs/2016ApJ...820...89F} {820, 89}

\bibitem[\protect\citeauthoryear{{Fu} et~al.,}{{Fu}
  et~al.}{2022}]{Fu_2022_Hatp11b_JWST}
{Fu} G.,  et~al., 2022, arXiv e-prints, \href
  {https://ui.adsabs.harvard.edu/abs/2022arXiv221113761F} {p. arXiv:2211.13761}

\bibitem[\protect\citeauthoryear{{Fulton} et~al.,}{{Fulton}
  et~al.}{2017}]{Fulton_2017}
{Fulton} B.~J.,  et~al., 2017, \mn@doi [\aj] {10.3847/1538-3881/aa80eb}, \href
  {https://ui.adsabs.harvard.edu/abs/2017AJ....154..109F} {154, 109}

\bibitem[\protect\citeauthoryear{{Gaia Collaboration} et~al.,}{{Gaia
  Collaboration} et~al.}{2016}]{GAIA_2016_MISSION}
{Gaia Collaboration} et~al., 2016, \mn@doi [\aap]
  {10.1051/0004-6361/201629272}, \href
  {https://ui.adsabs.harvard.edu/abs/2016A&A...595A...1G} {595, A1}

\bibitem[\protect\citeauthoryear{{Gaia Collaboration} et~al.,}{{Gaia
  Collaboration} et~al.}{2022}]{Gaia_2022_DR3}
{Gaia Collaboration} et~al., 2022, arXiv e-prints, \href
  {https://ui.adsabs.harvard.edu/abs/2022arXiv220800211G} {p. arXiv:2208.00211}

\bibitem[\protect\citeauthoryear{{Gaidos} et~al.,}{{Gaidos}
  et~al.}{2022}]{Gaidos_2022V1298_Tau}
{Gaidos} E.,  et~al., 2022, \mn@doi [\mnras] {10.1093/mnras/stab3107}, \href
  {https://ui.adsabs.harvard.edu/abs/2022MNRAS.509.2969G} {509, 2969}

\bibitem[\protect\citeauthoryear{{Gaidos} et~al.,}{{Gaidos}
  et~al.}{2023}]{Gaidos2023}
{Gaidos} E.,  et~al., 2023, \mn@doi [\mnras] {10.1093/mnras/stac3301}, \href
  {https://ui.adsabs.harvard.edu/abs/2023MNRAS.518.3777G} {518, 3777}

\bibitem[\protect\citeauthoryear{{Garc{\'{\i}}a Mu{\~n}oz}}{{Garc{\'{\i}}a
  Mu{\~n}oz}}{2007}]{muno}
{Garc{\'{\i}}a Mu{\~n}oz} A.,  2007, \mn@doi [planss]
  {10.1016/j.pss.2007.03.007}, \href
  {http://adsabs.harvard.edu/abs/2007P%26SS...55.1426G} {55, 1426}

\bibitem[\protect\citeauthoryear{{Gillet}, {Garcia Munoz}  \&
  {Strugarek}}{{Gillet} et~al.}{2023}]{Gillet_2023_photoelectrons}
{Gillet} A.,  {Garcia Munoz} A.,   {Strugarek} A.,  2023, \mn@doi [arXiv
  e-prints] {10.48550/arXiv.2309.08390}, \href
  {https://ui.adsabs.harvard.edu/abs/2023arXiv230908390G} {p. arXiv:2309.08390}

\bibitem[\protect\citeauthoryear{{Ginzburg}, {Schlichting}  \&
  {Sari}}{{Ginzburg} et~al.}{2018}]{Ginzburg_2018_core_powered_mdot}
{Ginzburg} S.,  {Schlichting} H.~E.,   {Sari} R.,  2018, \mn@doi [\mnras]
  {10.1093/mnras/sty290}, \href
  {https://ui.adsabs.harvard.edu/abs/2018MNRAS.476..759G} {476, 759}

\bibitem[\protect\citeauthoryear{{G{\"u}del}}{{G{\"u}del}}{2015}]{Gudel_summerschool2015Xray}
{G{\"u}del} M.,  2015, in European Physical Journal Web of Conferences. p.
  00015, \mn@doi{10.1051/epjconf/201510200015}

\bibitem[\protect\citeauthoryear{{Guilluy} et~al.,}{{Guilluy}
  et~al.}{2020}]{2020_Guilluy_GAPS_He_189}
{Guilluy} G.,  et~al., 2020, \mn@doi [\aap] {10.1051/0004-6361/202037644},
  \href {https://ui.adsabs.harvard.edu/abs/2020A&A...639A..49G} {639, A49}

\bibitem[\protect\citeauthoryear{{Hirano} et~al.,}{{Hirano}
  et~al.}{2020}]{Hirano_2020_au_mic}
{Hirano} T.,  et~al., 2020, \mn@doi [\apjl] {10.3847/2041-8213/aba6eb}, \href
  {https://ui.adsabs.harvard.edu/abs/2020ApJ...899L..13H} {899, L13}

\bibitem[\protect\citeauthoryear{{Hui} \& {Gnedin}}{{Hui} \&
  {Gnedin}}{1997}]{Hui_Gnedin_1997}
{Hui} L.,  {Gnedin} N.~Y.,  1997, \mn@doi [\mnras] {10.1093/mnras/292.1.27},
  \href {https://ui.adsabs.harvard.edu/abs/1997MNRAS.292...27H} {292, 27}

\bibitem[\protect\citeauthoryear{{Jackson}, {Davis}  \& {Wheatley}}{{Jackson}
  et~al.}{2012}]{Jackson_2012_X_ray_age_relation}
{Jackson} A.~P.,  {Davis} T.~A.,   {Wheatley} P.~J.,  2012, \mn@doi [\mnras]
  {10.1111/j.1365-2966.2012.20657.x}, \href
  {https://ui.adsabs.harvard.edu/abs/2012MNRAS.422.2024J} {422, 2024}

\bibitem[\protect\citeauthoryear{Jacobs}{Jacobs}{1974}]{Jacobs_1974_NASA_cross_sec_exc_He}
Jacobs V.~L.,  1974, \mn@doi [Phys. Rev. A] {10.1103/PhysRevA.9.1938}, 9, 1938

\bibitem[\protect\citeauthoryear{Johnstone et~al.,}{Johnstone
  et~al.}{2015a}]{Johnstone2015a}
Johnstone C.~P.,  et~al., 2015a, \mn@doi [Astrophysical Journal Letters]
  {10.1088/2041-8205/815/1/L12}, 815, 1

\bibitem[\protect\citeauthoryear{{Johnstone} et~al.,}{{Johnstone}
  et~al.}{2015b}]{2015ApJ...815L..12J}
{Johnstone} C.~P.,  et~al., 2015b, \mn@doi [\apjl]
  {10.1088/2041-8205/815/1/L12}, \href
  {https://ui.adsabs.harvard.edu/abs/2015ApJ...815L..12J} {815, L12}

\bibitem[\protect\citeauthoryear{{Johnstone}, {Bartel}  \&
  {G{\"u}del}}{{Johnstone} et~al.}{2021}]{Johnstone_2021}
{Johnstone} C.~P.,  {Bartel} M.,   {G{\"u}del} M.,  2021, \mn@doi [\aap]
  {10.1051/0004-6361/202038407}, \href
  {https://ui.adsabs.harvard.edu/abs/2021A&A...649A..96J} {649, A96}

\bibitem[\protect\citeauthoryear{{Kawaler}}{{Kawaler}}{1988}]{Kawaler_1988}
{Kawaler} S.~D.,  1988, \mn@doi [\apj] {10.1086/166740}, \href
  {https://ui.adsabs.harvard.edu/abs/1988ApJ...333..236K} {333, 236}

\bibitem[\protect\citeauthoryear{{Khodachenko}, {Shaikhislamov}, {Fossati},
  {Lammer}, {Rumenskikh}, {Berezutsky}, {Miroshnichenko}  \&
  {Efimof}}{{Khodachenko} et~al.}{2021a}]{Khodachenko_2021_107b_3d}
{Khodachenko} M.~L.,  {Shaikhislamov} I.~F.,  {Fossati} L.,  {Lammer} H.,
  {Rumenskikh} M.~S.,  {Berezutsky} A.~G.,  {Miroshnichenko} I.~B.,   {Efimof}
  M.~A.,  2021a, \mn@doi [\mnras] {10.1093/mnrasl/slab015}, \href
  {https://ui.adsabs.harvard.edu/abs/2021MNRAS.503L..23K} {503, L23}

\bibitem[\protect\citeauthoryear{{Khodachenko}, {Shaikhislamov}, {Lammer},
  {Miroshnichenko}, {Rumenskikh}, {Berezutsky}  \& {Fossati}}{{Khodachenko}
  et~al.}{2021b}]{Khodachenko_2021_B_field_hd209458b}
{Khodachenko} M.~L.,  {Shaikhislamov} I.~F.,  {Lammer} H.,  {Miroshnichenko}
  I.~B.,  {Rumenskikh} M.~S.,  {Berezutsky} A.~G.,   {Fossati} L.,  2021b,
  \mn@doi [\mnras] {10.1093/mnras/stab2366}, \href
  {https://ui.adsabs.harvard.edu/abs/2021MNRAS.507.3626K} {507, 3626}

\bibitem[\protect\citeauthoryear{{Kirk}, {Alam}, {L{\'o}pez-Morales}  \&
  {Zeng}}{{Kirk} et~al.}{2020}]{2020_Kirk_tail_107b}
{Kirk} J.,  {Alam} M.~K.,  {L{\'o}pez-Morales} M.,   {Zeng} L.,  2020, \mn@doi
  [\aj] {10.3847/1538-3881/ab6e66}, \href
  {https://ui.adsabs.harvard.edu/abs/2020AJ....159..115K} {159, 115}

\bibitem[\protect\citeauthoryear{{Koskinen}, {Harris}, {Yelle}  \&
  {Lavvas}}{{Koskinen} et~al.}{2013}]{Koskinen_2013}
{Koskinen} T.~T.,  {Harris} M.~J.,  {Yelle} R.~V.,   {Lavvas} P.,  2013,
  \mn@doi [\icarus] {10.1016/j.icarus.2012.09.027}, \href
  {https://ui.adsabs.harvard.edu/abs/2013Icar..226.1678K} {226, 1678}

\bibitem[\protect\citeauthoryear{Kramida, {Yu.~Ralchenko}, Reader  \& {and NIST
  ASD Team}}{Kramida et~al.}{2022}]{NIST_ASD}
Kramida A.,  {Yu.~Ralchenko} Reader J.,   {and NIST ASD Team} 2022, {NIST
  Atomic Spectra Database (ver. 5.10), [Online]. Available:
  {\tt{https://physics.nist.gov/asd}} [2022, December 7]. National Institute of
  Standards and Technology, Gaithersburg, MD.}

\bibitem[\protect\citeauthoryear{{Krishnamurthy} et~al.,}{{Krishnamurthy}
  et~al.}{2021}]{Krishnamurthy_2021_non_detections_trappist}
{Krishnamurthy} V.,  et~al., 2021, \mn@doi [\aj] {10.3847/1538-3881/ac0d57},
  \href {https://ui.adsabs.harvard.edu/abs/2021AJ....162...82K} {162, 82}

\bibitem[\protect\citeauthoryear{{Kubyshkina} \& {Fossati}}{{Kubyshkina} \&
  {Fossati}}{2022}]{Kubyshkina_Fossati2022}
{Kubyshkina} D.,  {Fossati} L.,  2022, \mn@doi [\aap]
  {10.1051/0004-6361/202244916}, \href
  {https://ui.adsabs.harvard.edu/abs/2022A&A...668A.178K} {668, A178}

\bibitem[\protect\citeauthoryear{{Kubyshkina} \& {Vidotto}}{{Kubyshkina} \&
  {Vidotto}}{2021}]{2021MNRAS.504.2034K}
{Kubyshkina} D.,  {Vidotto} A.~A.,  2021, \mn@doi [\mnras]
  {10.1093/mnras/stab897}, \href
  {https://ui.adsabs.harvard.edu/abs/2021MNRAS.504.2034K} {504, 2034}

\bibitem[\protect\citeauthoryear{{Kubyshkina} et~al.,}{{Kubyshkina}
  et~al.}{2018}]{Daria_2018_grid_1_40_earths}
{Kubyshkina} D.,  et~al., 2018, \mn@doi [\aap] {10.1051/0004-6361/201833737},
  \href {https://ui.adsabs.harvard.edu/abs/2018A&A...619A.151K} {619, A151}

\bibitem[\protect\citeauthoryear{{Kubyshkina}, {Vidotto}, {Fossati}  \&
  {Farrell}}{{Kubyshkina} et~al.}{2020}]{2020MNRAS.499...77K}
{Kubyshkina} D.,  {Vidotto} A.~A.,  {Fossati} L.,   {Farrell} E.,  2020,
  \mn@doi [\mnras] {10.1093/mnras/staa2815}, \href
  {https://ui.adsabs.harvard.edu/abs/2020MNRAS.499...77K} {499, 77}

\bibitem[\protect\citeauthoryear{{Lamp{\'o}n} et~al.,}{{Lamp{\'o}n}
  et~al.}{2020}]{Lampon_2020}
{Lamp{\'o}n} M.,  et~al., 2020, \mn@doi [\aap] {10.1051/0004-6361/201937175},
  \href {https://ui.adsabs.harvard.edu/abs/2020A&A...636A..13L} {636, A13}

\bibitem[\protect\citeauthoryear{{Lamp{\'o}n} et~al.,}{{Lamp{\'o}n}
  et~al.}{2021}]{Lampon_2021}
{Lamp{\'o}n} M.,  et~al., 2021, \mn@doi [\aap] {10.1051/0004-6361/202039417},
  \href {https://ui.adsabs.harvard.edu/abs/2021A&A...647A.129L} {647, A129}

\bibitem[\protect\citeauthoryear{{Linssen}, {Oklop{\v{c}}i{\'c}}  \&
  {MacLeod}}{{Linssen} et~al.}{2022}]{Linssen2022}
{Linssen} D.~C.,  {Oklop{\v{c}}i{\'c}} A.,   {MacLeod} M.,  2022, \mn@doi
  [\aap] {10.1051/0004-6361/202243830}, \href
  {https://ui.adsabs.harvard.edu/abs/2022A&A...667A..54L} {667, A54}

\bibitem[\protect\citeauthoryear{{Locci}, {Petralia}, {Micela}, {Maggio},
  {Ciaravella}  \& {Cecchi-Pestellini}}{{Locci}
  et~al.}{2022}]{Danielle_Locci_2022}
{Locci} D.,  {Petralia} A.,  {Micela} G.,  {Maggio} A.,  {Ciaravella} A.,
  {Cecchi-Pestellini} C.,  2022, \mn@doi [psj] {10.3847/PSJ/ac3f3c}, \href
  {https://ui.adsabs.harvard.edu/abs/2022PSJ.....3....1L} {3, 1}

\bibitem[\protect\citeauthoryear{{Loyd} et~al.,}{{Loyd}
  et~al.}{2016}]{2016ApJ...824..102L}
{Loyd} R.~O.~P.,  et~al., 2016, \mn@doi [\apj] {10.3847/0004-637X/824/2/102},
  \href {https://ui.adsabs.harvard.edu/abs/2016ApJ...824..102L} {824, 102}

\bibitem[\protect\citeauthoryear{{MacLeod} \& {Oklop{\v{c}}i{\'c}}}{{MacLeod}
  \& {Oklop{\v{c}}i{\'c}}}{2021}]{MacLeod_Oklopcic_2021}
{MacLeod} M.,  {Oklop{\v{c}}i{\'c}} A.,  2021, arXiv e-prints, \href
  {https://ui.adsabs.harvard.edu/abs/2021arXiv210707534M} {p. arXiv:2107.07534}

\bibitem[\protect\citeauthoryear{{Mazeh}, {Holczer}  \& {Faigler}}{{Mazeh}
  et~al.}{2016}]{Mazeh_2016}
{Mazeh} T.,  {Holczer} T.,   {Faigler} S.,  2016, \mn@doi [\aap]
  {10.1051/0004-6361/201528065}, \href
  {https://ui.adsabs.harvard.edu/abs/2016A&A...589A..75M} {589, A75}

\bibitem[\protect\citeauthoryear{{McCann}, {Murray-Clay}, {Kratter}  \&
  {Krumholz}}{{McCann} et~al.}{2019}]{2019ApJ...873...89M}
{McCann} J.,  {Murray-Clay} R.~A.,  {Kratter} K.,   {Krumholz} M.~R.,  2019,
  \mn@doi [\apj] {10.3847/1538-4357/ab05b8}, \href
  {https://ui.adsabs.harvard.edu/abs/2019ApJ...873...89M} {873, 89}

\bibitem[\protect\citeauthoryear{Murray-Clay, Chiang  \& Murray}{Murray-Clay
  et~al.}{2009}]{Murray-Clay2009}
Murray-Clay R.~A.,  Chiang E.~I.,   Murray N.,  2009, \mn@doi [Astrophysical
  Journal] {10.1088/0004-637X/693/1/23}, 693, 23

\bibitem[\protect\citeauthoryear{{Namekata}, {Toriumi}, {Airapetian}, {Shoda},
  {Watanabe}  \& {Notsu}}{{Namekata} et~al.}{2023}]{2023arXiv230210376N}
{Namekata} K.,  {Toriumi} S.,  {Airapetian} V.~S.,  {Shoda} M.,  {Watanabe} K.,
    {Notsu} Y.,  2023, \mn@doi [arXiv e-prints] {10.48550/arXiv.2302.10376},
  \href {https://ui.adsabs.harvard.edu/abs/2023arXiv230210376N} {p.
  arXiv:2302.10376}

\bibitem[\protect\citeauthoryear{{Norcross}}{{Norcross}}{1971}]{Norcross1971}
{Norcross} D.~W.,  1971, \mn@doi [Journal of Physics B Atomic Molecular
  Physics] {10.1088/0022-3700/4/5/006}, \href
  {https://ui.adsabs.harvard.edu/abs/1971JPhB....4..652N} {4, 652}

\bibitem[\protect\citeauthoryear{{Nortmann} et~al.,}{{Nortmann}
  et~al.}{2018}]{Nortmann_2018}
{Nortmann} L.,  et~al., 2018, \mn@doi [Science] {10.1126/science.aat5348},
  \href {https://ui.adsabs.harvard.edu/abs/2018Sci...362.1388N} {362, 1388}

\bibitem[\protect\citeauthoryear{{Oklop{\v{c}}i{\'c}}}{{Oklop{\v{c}}i{\'c}}}{2019}]{Oklopcic_2019_dep_st_rad}
{Oklop{\v{c}}i{\'c}} A.,  2019, \mn@doi [\apj] {10.3847/1538-4357/ab2f7f},
  \href {https://ui.adsabs.harvard.edu/abs/2019ApJ...881..133O} {881, 133}

\bibitem[\protect\citeauthoryear{{Oklop{\v{c}}i{\'c}} \&
  {Hirata}}{{Oklop{\v{c}}i{\'c}} \&
  {Hirata}}{2018}]{Oklopcic_2018_10839_window}
{Oklop{\v{c}}i{\'c}} A.,  {Hirata} C.~M.,  2018, \mn@doi [\apjl]
  {10.3847/2041-8213/aaada9}, \href
  {https://ui.adsabs.harvard.edu/abs/2018ApJ...855L..11O} {855, L11}

\bibitem[\protect\citeauthoryear{{Osterbrock} \& {Ferland}}{{Osterbrock} \&
  {Ferland}}{2006}]{2006agna.book.....O}
{Osterbrock} D.~E.,  {Ferland} G.~J.,  2006, {Astrophysics of gaseous nebulae
  and active galactic nuclei}

\bibitem[\protect\citeauthoryear{{Owen}}{{Owen}}{2019}]{Owen_EVOL_ATM_ESCAP_REVIEW_2019}
{Owen} J.~E.,  2019, \mn@doi [Annual Review of Earth and Planetary Sciences]
  {10.1146/annurev-earth-053018-060246}, \href
  {https://ui.adsabs.harvard.edu/abs/2019AREPS..47...67O} {47, 67}

\bibitem[\protect\citeauthoryear{{Owen} \& {Jackson}}{{Owen} \&
  {Jackson}}{2012}]{Owen_Jackson_2012}
{Owen} J.~E.,  {Jackson} A.~P.,  2012, \mn@doi [\mnras]
  {10.1111/j.1365-2966.2012.21481.x}, \href
  {https://ui.adsabs.harvard.edu/abs/2012MNRAS.425.2931O} {425, 2931}

\bibitem[\protect\citeauthoryear{{Owen} \& {Lai}}{{Owen} \&
  {Lai}}{2018}]{Owen_Dong_2018}
{Owen} J.~E.,  {Lai} D.,  2018, \mn@doi [\mnras] {10.1093/mnras/sty1760}, \href
  {https://ui.adsabs.harvard.edu/abs/2018MNRAS.479.5012O} {479, 5012}

\bibitem[\protect\citeauthoryear{{Parker}}{{Parker}}{1958}]{Parker}
{Parker} E.~N.,  1958, \mn@doi [apj] {10.1086/146579}, \href
  {http://adsabs.harvard.edu/abs/1958ApJ...128..664P} {128, 664}

\bibitem[\protect\citeauthoryear{{Pepe} et~al.,}{{Pepe}
  et~al.}{2011}]{2011A&A...534A..58P}
{Pepe} F.,  et~al., 2011, \mn@doi [\aap] {10.1051/0004-6361/201117055}, \href
  {https://ui.adsabs.harvard.edu/abs/2011A&A...534A..58P} {534, A58}

\bibitem[\protect\citeauthoryear{{Pezzotti}, {Eggenberger}, {Buldgen},
  {Meynet}, {Bourrier}  \& {Mordasini}}{{Pezzotti}
  et~al.}{2021a}]{2021A&A...650A.108P}
{Pezzotti} C.,  {Eggenberger} P.,  {Buldgen} G.,  {Meynet} G.,  {Bourrier} V.,
   {Mordasini} C.,  2021a, \mn@doi [\aap] {10.1051/0004-6361/202039652}, \href
  {https://ui.adsabs.harvard.edu/abs/2021A&A...650A.108P} {650, A108}

\bibitem[\protect\citeauthoryear{{Pezzotti}, {Attia}, {Eggenberger}, {Buldgen}
  \& {Bourrier}}{{Pezzotti} et~al.}{2021b}]{2021A&A...654L...5P}
{Pezzotti} C.,  {Attia} O.,  {Eggenberger} P.,  {Buldgen} G.,   {Bourrier} V.,
  2021b, \mn@doi [\aap] {10.1051/0004-6361/202141734}, \href
  {https://ui.adsabs.harvard.edu/abs/2021A&A...654L...5P} {654, L5}

\bibitem[\protect\citeauthoryear{{Poppenhaeger}}{{Poppenhaeger}}{2022}]{Poppenhaeger_2022}
{Poppenhaeger} K.,  2022, \mn@doi [\mnras] {10.1093/mnras/stac507}, \href
  {https://ui.adsabs.harvard.edu/abs/2022MNRAS.512.1751P} {512, 1751}

\bibitem[\protect\citeauthoryear{{Richey-Yowell} et~al.,}{{Richey-Yowell}
  et~al.}{2022}]{Richey_Yowell_2022}
{Richey-Yowell} T.,  et~al., 2022, \mn@doi [\apj] {10.3847/1538-4357/ac5f48},
  \href {https://ui.adsabs.harvard.edu/abs/2022ApJ...929..169R} {929, 169}

\bibitem[\protect\citeauthoryear{{Roberge} \& {Dalgarno}}{{Roberge} \&
  {Dalgarno}}{1982}]{Roberge_Dalgarno1982}
{Roberge} W.,  {Dalgarno} A.,  1982, \mn@doi [\apj] {10.1086/159849}, \href
  {https://ui.adsabs.harvard.edu/abs/1982ApJ...255..489R} {255, 489}

\bibitem[\protect\citeauthoryear{{Rumenskikh}, {Shaikhislamov}, {Khodachenko},
  {Lammer}, {Miroshnichenko}, {Berezutsky}  \& {Fossati}}{{Rumenskikh}
  et~al.}{2022}]{Rumenskikh_2022_3D_HD189733b}
{Rumenskikh} M.~S.,  {Shaikhislamov} I.~F.,  {Khodachenko} M.~L.,  {Lammer} H.,
   {Miroshnichenko} I.~B.,  {Berezutsky} A.~G.,   {Fossati} L.,  2022, \mn@doi
  [\apj] {10.3847/1538-4357/ac441d}, \href
  {https://ui.adsabs.harvard.edu/abs/2022ApJ...927..238R} {927, 238}

\bibitem[\protect\citeauthoryear{{Salz}, {Banerjee}, {Mignone}, {Schneider},
  {Czesla}  \& {Schmitt}}{{Salz} et~al.}{2015}]{Salz2015TPCI}
{Salz} M.,  {Banerjee} R.,  {Mignone} A.,  {Schneider} P.~C.,  {Czesla} S.,
  {Schmitt} J.~H.~M.~M.,  2015, \mn@doi [\aap] {10.1051/0004-6361/201424330},
  \href {https://ui.adsabs.harvard.edu/abs/2015A&A...576A..21S} {576, A21}

\bibitem[\protect\citeauthoryear{{Sanz-Forcada}}{{Sanz-Forcada}}{2022}]{2022AN....34320008S}
{Sanz-Forcada} J.,  2022, \mn@doi [Astronomische Nachrichten]
  {10.1002/asna.20220008}, \href
  {https://ui.adsabs.harvard.edu/abs/2022AN....34320008S} {343, e20008}

\bibitem[\protect\citeauthoryear{{Seager} \& {Sasselov}}{{Seager} \&
  {Sasselov}}{2000}]{Seager_Sasselov_2000}
{Seager} S.,  {Sasselov} D.~D.,  2000, \mn@doi [\apj] {10.1086/309088}, \href
  {https://ui.adsabs.harvard.edu/abs/2000ApJ...537..916S} {537, 916}

\bibitem[\protect\citeauthoryear{{Shaikhislamov}, {Khodachenko}, {Lammer},
  {Berezutsky}, {Miroshnichenko}  \& {Rumenskikh}}{{Shaikhislamov}
  et~al.}{2021}]{Shaikhislamov_2021_GJ3470b}
{Shaikhislamov} I.~F.,  {Khodachenko} M.~L.,  {Lammer} H.,  {Berezutsky} A.~G.,
   {Miroshnichenko} I.~B.,   {Rumenskikh} M.~S.,  2021, \mn@doi [\mnras]
  {10.1093/mnras/staa2367}, \href
  {https://ui.adsabs.harvard.edu/abs/2021MNRAS.500.1404S} {500, 1404}

\bibitem[\protect\citeauthoryear{{Shematovich}, {Ionov}  \&
  {Lammer}}{{Shematovich} et~al.}{2014}]{Shematovich2014}
{Shematovich} V.~I.,  {Ionov} D.~E.,   {Lammer} H.,  2014, \mn@doi [\aap]
  {10.1051/0004-6361/201423573}, \href
  {https://ui.adsabs.harvard.edu/abs/2014A&A...571A..94S} {571, A94}

\bibitem[\protect\citeauthoryear{{Skumanich}}{{Skumanich}}{1972}]{Skumanich_1972}
{Skumanich} A.,  1972, \mn@doi [\apj] {10.1086/151310}, \href
  {https://ui.adsabs.harvard.edu/abs/1972ApJ...171..565S} {171, 565}

\bibitem[\protect\citeauthoryear{{Spake} et~al.,}{{Spake}
  et~al.}{2018}]{Spake_2018_NATURE_He_in_atm}
{Spake} J.~J.,  et~al., 2018, \mn@doi [\nat] {10.1038/s41586-018-0067-5}, \href
  {https://ui.adsabs.harvard.edu/abs/2018Natur.557...68S} {557, 68}

\bibitem[\protect\citeauthoryear{{Spake}, {Oklop{\v{c}}i{\'c}}  \&
  {Hillenbrand}}{{Spake} et~al.}{2021}]{Spake_2021_tail_wasp107b}
{Spake} J.~J.,  {Oklop{\v{c}}i{\'c}} A.,   {Hillenbrand} L.~A.,  2021, \mn@doi
  [\aj] {10.3847/1538-3881/ac178a}, \href
  {https://ui.adsabs.harvard.edu/abs/2021AJ....162..284S} {162, 284}

\bibitem[\protect\citeauthoryear{{Spitzer}}{{Spitzer}}{1978}]{spitzer_78}
{Spitzer} L.,  1978, {Physical processes in the interstellar medium},
  \mn@doi{10.1002/9783527617722.
}

\bibitem[\protect\citeauthoryear{{Storey} \& {Hummer}}{{Storey} \&
  {Hummer}}{1995}]{Storey_Hummer1995}
{Storey} P.~J.,  {Hummer} D.~G.,  1995, \mn@doi [\mnras]
  {10.1093/mnras/272.1.41}, \href
  {https://ui.adsabs.harvard.edu/abs/1995MNRAS.272...41S} {272, 41}

\bibitem[\protect\citeauthoryear{Sun \& Hu}{Sun \&
  Hu}{2020}]{Sun_HU_2020_helium_spec}
Sun Y.~R.,  Hu S.-M.,  2020, \mn@doi [National Science Review]
  {10.1093/nsr/nwaa216}, 7, 1818

\bibitem[\protect\citeauthoryear{{Toriumi} \& {Airapetian}}{{Toriumi} \&
  {Airapetian}}{2022}]{2022ApJ...927..179T}
{Toriumi} S.,  {Airapetian} V.~S.,  2022, \mn@doi [\apj]
  {10.3847/1538-4357/ac5179}, \href
  {https://ui.adsabs.harvard.edu/abs/2022ApJ...927..179T} {927, 179}

\bibitem[\protect\citeauthoryear{{Verner}, {Ferland}, {Korista}  \&
  {Yakovlev}}{{Verner} et~al.}{1996}]{Verner1996}
{Verner} D.~A.,  {Ferland} G.~J.,  {Korista} K.~T.,   {Yakovlev} D.~G.,  1996,
  \mn@doi [\apj] {10.1086/177435}, \href
  {https://ui.adsabs.harvard.edu/abs/1996ApJ...465..487V} {465, 487}

\bibitem[\protect\citeauthoryear{{Vidotto} \& {Cleary}}{{Vidotto} \&
  {Cleary}}{2020}]{2020MNRAS.494.2417V}
{Vidotto} A.~A.,  {Cleary} A.,  2020, \mn@doi [\mnras] {10.1093/mnras/staa852},
  \href {https://ui.adsabs.harvard.edu/abs/2020MNRAS.494.2417V} {494, 2417}

\bibitem[\protect\citeauthoryear{{Vidotto} \& {Jatenco-Pereira}}{{Vidotto} \&
  {Jatenco-Pereira}}{2006}]{2006ApJ...639..416V}
{Vidotto} A.~A.,  {Jatenco-Pereira} V.,  2006, \mn@doi [\apj] {10.1086/499329},
  \href {https://ui.adsabs.harvard.edu/abs/2006ApJ...639..416V} {639, 416}

\bibitem[\protect\citeauthoryear{{Vidotto} et~al.,}{{Vidotto}
  et~al.}{2014}]{Vidotto_2014}
{Vidotto} A.~A.,  et~al., 2014, \mn@doi [\mnras] {10.1093/mnras/stu728}, \href
  {https://ui.adsabs.harvard.edu/abs/2014MNRAS.441.2361V} {441, 2361}

\bibitem[\protect\citeauthoryear{{Vidotto} et~al.,}{{Vidotto}
  et~al.}{2018}]{Vidotto_2018}
{Vidotto} A.~A.,  et~al., 2018, \mn@doi [\mnras] {10.1093/mnras/sty2130}, \href
  {https://ui.adsabs.harvard.edu/abs/2018MNRAS.481.5296V} {481, 5296}

\bibitem[\protect\citeauthoryear{{Villarreal D'Angelo}, {Schneiter}, {Costa},
  {Vel{\'a}zquez}, {Raga}  \& {Esquivel}}{{Villarreal D'Angelo}
  et~al.}{2014}]{2014MNRAS.438.1654V}
{Villarreal D'Angelo} C.,  {Schneiter} M.,  {Costa} A.,  {Vel{\'a}zquez} P.,
  {Raga} A.,   {Esquivel} A.,  2014, \mn@doi [\mnras] {10.1093/mnras/stt2303},
  \href {https://ui.adsabs.harvard.edu/abs/2014MNRAS.438.1654V} {438, 1654}

\bibitem[\protect\citeauthoryear{{Villarreal D'Angelo}, {Esquivel}, {Schneiter}
   \& {Sgr{\'o}}}{{Villarreal D'Angelo} et~al.}{2018}]{2018MNRAS.479.3115V}
{Villarreal D'Angelo} C.,  {Esquivel} A.,  {Schneiter} M.,   {Sgr{\'o}} M.~A.,
  2018, \mn@doi [\mnras] {10.1093/mnras/sty1544}, \href
  {https://ui.adsabs.harvard.edu/abs/2018MNRAS.479.3115V} {479, 3115}

\bibitem[\protect\citeauthoryear{{Vissapragada} et~al.,}{{Vissapragada}
  et~al.}{2021}]{2021_Vissapragada_search_He_in_V1298_tau_system}
{Vissapragada} S.,  et~al., 2021, \mn@doi [\aj] {10.3847/1538-3881/ac1bb0},
  \href {https://ui.adsabs.harvard.edu/abs/2021AJ....162..222V} {162, 222}

\bibitem[\protect\citeauthoryear{{Wang} \& {Dai}}{{Wang} \&
  {Dai}}{2021a}]{Wang_Dai2021_WASP-69b}
{Wang} L.,  {Dai} F.,  2021a, \mn@doi [\apj] {10.3847/1538-4357/abf1ee}, \href
  {https://ui.adsabs.harvard.edu/abs/2021ApJ...914...98W} {914, 98}

\bibitem[\protect\citeauthoryear{{Wang} \& {Dai}}{{Wang} \&
  {Dai}}{2021b}]{Wang_Dai2021_WASP107b}
{Wang} L.,  {Dai} F.,  2021b, \mn@doi [\apj] {10.3847/1538-4357/abf1ed}, \href
  {https://ui.adsabs.harvard.edu/abs/2021ApJ...914...99W} {914, 99}

\bibitem[\protect\citeauthoryear{{Wiese} \& {Fuhr}}{{Wiese} \&
  {Fuhr}}{2009}]{Wiese_Fuhr_2009_NIST_lines}
{Wiese} W.~L.,  {Fuhr} J.~R.,  2009, \mn@doi [Journal of Physical and Chemical
  Reference Data] {10.1063/1.3077727}, \href
  {https://ui.adsabs.harvard.edu/abs/2009JPCRD..38..565W} {38, 565}

\bibitem[\protect\citeauthoryear{{Yan}, {Seon}, {Guo}, {Chen}  \& {Li}}{{Yan}
  et~al.}{2022}]{Yan_2022_Wasp_52b}
{Yan} D.,  {Seon} K.-i.,  {Guo} J.,  {Chen} G.,   {Li} L.,  2022, \mn@doi
  [\apj] {10.3847/1538-4357/ac8793}, \href
  {https://ui.adsabs.harvard.edu/abs/2022ApJ...936..177Y} {936, 177}

\bibitem[\protect\citeauthoryear{{Yelle}}{{Yelle}}{2004}]{yelle_2004}
{Yelle} R.~V.,  2004, \mn@doi [icarus] {10.1016/j.icarus.2004.02.008}, \href
  {http://adsabs.harvard.edu/abs/2004Icar..170..167Y} {170, 167}

\bibitem[\protect\citeauthoryear{{Youngblood} et~al.,}{{Youngblood}
  et~al.}{2016}]{2016ApJ...824..101Y}
{Youngblood} A.,  et~al., 2016, \mn@doi [\apj] {10.3847/0004-637X/824/2/101},
  \href {https://ui.adsabs.harvard.edu/abs/2016ApJ...824..101Y} {824, 101}

\bibitem[\protect\citeauthoryear{{Zhang}, {Knutson}, {Wang}, {Dai}, {Oklopcic}
  \& {Hu}}{{Zhang} et~al.}{2021}]{Zhang_2021_55_cnc_e_non_detection}
{Zhang} M.,  {Knutson} H.~A.,  {Wang} L.,  {Dai} F.,  {Oklopcic} A.,   {Hu} R.,
   2021, \mn@doi [\aj] {10.3847/1538-3881/abe382}, \href
  {https://ui.adsabs.harvard.edu/abs/2021AJ....161..181Z} {161, 181}

\bibitem[\protect\citeauthoryear{{Zhang}, {Knutson}, {Dai}, {Wang}, {Ricker},
  {Schwarz}, {Mann}  \& {Collins}}{{Zhang}
  et~al.}{2022a}]{2022_Zhang_4_mini_Nep}
{Zhang} M.,  {Knutson} H.~A.,  {Dai} F.,  {Wang} L.,  {Ricker} G.~R.,
  {Schwarz} R.~P.,  {Mann} C.,   {Collins} K.,  2022a, arXiv e-prints, \href
  {https://ui.adsabs.harvard.edu/abs/2022arXiv220713099Z} {p. arXiv:2207.13099}

\bibitem[\protect\citeauthoryear{{Zhang}, {Knutson}, {Wang}, {Dai}  \&
  {Barrag{\'a}n}}{{Zhang} et~al.}{2022b}]{Zhang_2022_TOI_560}
{Zhang} M.,  {Knutson} H.~A.,  {Wang} L.,  {Dai} F.,   {Barrag{\'a}n} O.,
  2022b, \mn@doi [\aj] {10.3847/1538-3881/ac3fa7}, \href
  {https://ui.adsabs.harvard.edu/abs/2022AJ....163...67Z} {163, 67}

\bibitem[\protect\citeauthoryear{{Zhang} et~al.,}{{Zhang}
  et~al.}{2022c}]{2022_HD63433_upper_limit_non_detect}
{Zhang} M.,  et~al., 2022c, \mn@doi [\aj] {10.3847/1538-3881/ac3f3b}, \href
  {https://ui.adsabs.harvard.edu/abs/2022AJ....163...68Z} {163, 68}

\makeatother
\end{thebibliography}




\appendix

\section{Sensitivity to model free parameters}
\label{sec:append_sensitivity}

Two free parameters in our hydrodynamical models are the density and temperature at the base of the planetary atmosphere, assumed to be $\rho_0=4 \times 10^{-14}~$g cm$^{-3}$ and $T_0=1000~$K (see Section \ref{sec:fluid_dyn}). We now discuss the model's sensitivity to these values using two test models.

We test increasing the assumed base density by one order of magnitude, to $\rho_0=4 \times 10^{-13}~$g cm$^{-3}$. This larger base density causes the predicted mass-loss rate to increase by factors of 3.6 and 1.9 at ages of 16 and 5\,000 Myr, respectively. It only negligibly affects the terminal velocity (below 4\% in all models). Hence, the hydrodynamics do not change significantly by varying the base density within a factor of 10 (i.e., at most by a factor of 3.6 in mass-loss rate affecting the youngest model). 
This is in agreement with \citet{Murray-Clay2009} who find their model's hydrodynamic predictions to be insensitive to the assumed base density, so long as the chosen value results in an optical depth at the atmospheric base $\gg 1$. This is indeed true for the dominant H$^0$ sEUV photoionisation in our model for both of the base densities considered here.We also test raising  the assumed base temperature to $T_0= 2000$~K. This hotter base temperature causes the predicted mass-loss rate to increase by factors of 1.7 and 1.5 at young and old ages, respectively. As with the base density test, varying the base temperature only negligibly affects the terminal velocity (below 2\% in all models).

However, we find that the variations in base density can cause changes beyond the uncertainties of observations in transit depth and EW at younger ages: we found increases by factors of 2.7 (1.8) in the EW and 2.4 (1.7) in the phase-averaged peak transit depth at ages of 16 (5\,000) Myr. For example, for the youngest age model, the peak transit depth obtained by our fiducial F10\% model is 34\%, while  the increased-base density model results in a maximum depth of 81\%. This is a change that could be observed. The sensitivity is because most of the He($2^3S$) absorption takes place close to the planet. 
Because of such sensitivity, helium triplet transit observations can be used to constrain the atmospheric base density. However, it is interesting that this effect is not significant in the hydrodynamics (i.e., evaporation rates are not significantly altered by changes in base density). This is in stark contrast to isothermal models, in which the evaporation rate is directly proportional to base density. Raising the base temperature to 2000$~$K also affects the predicted observational results, albeit less so than the mentioned base density test. The increased base temperature leads to increases by factors of 1.6 (1.4) in the EW and 1.5 (1.4) in the peak transit depth at ages of 16 (5\,000) Myr.

\section{Additional data on a helium abundance of 10\%}
\label{sec:append_He_10}

\begin{figure*}
	\includegraphics[width=0.9\textwidth]{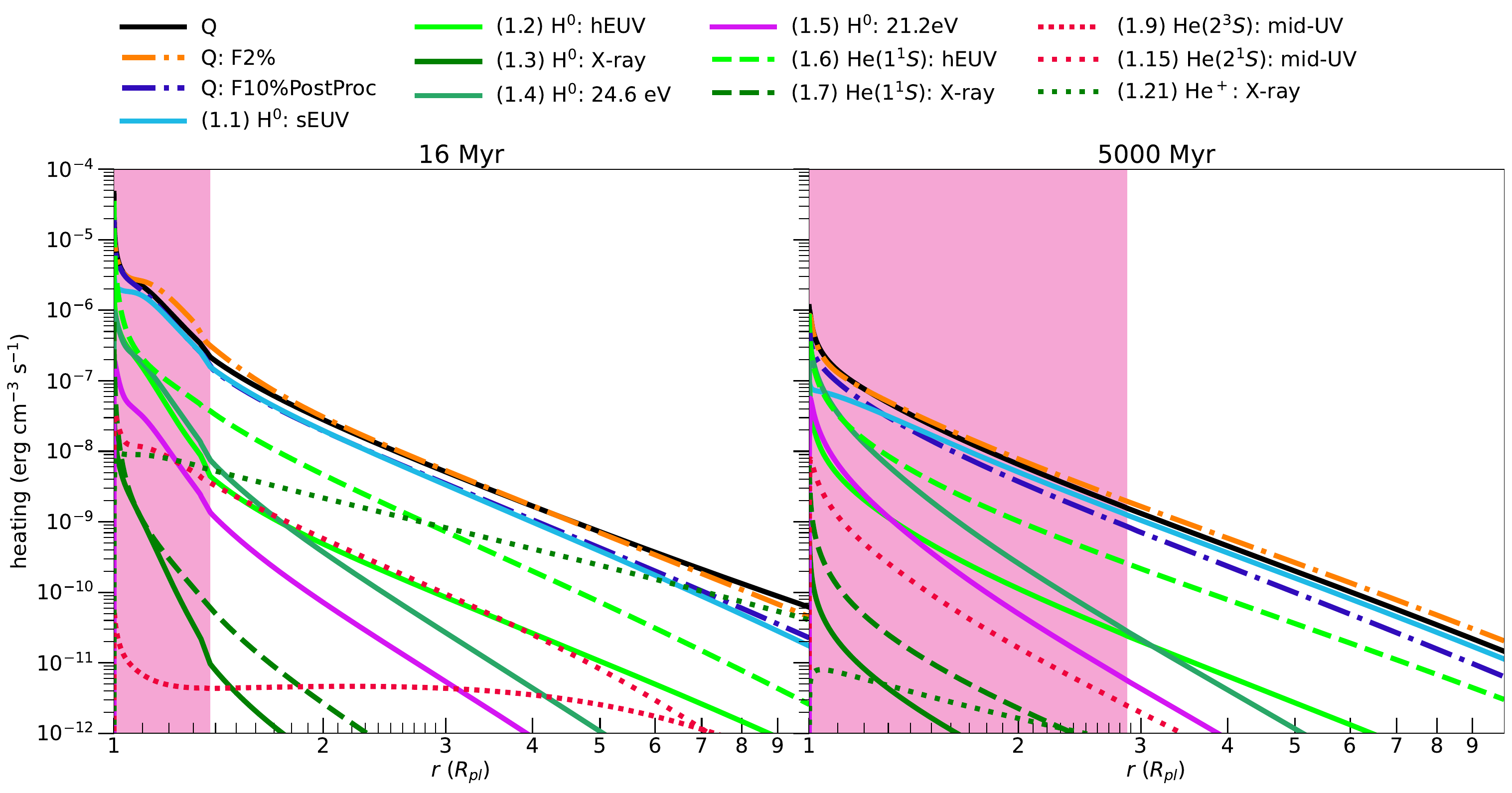}
	\includegraphics[width=0.9\textwidth]{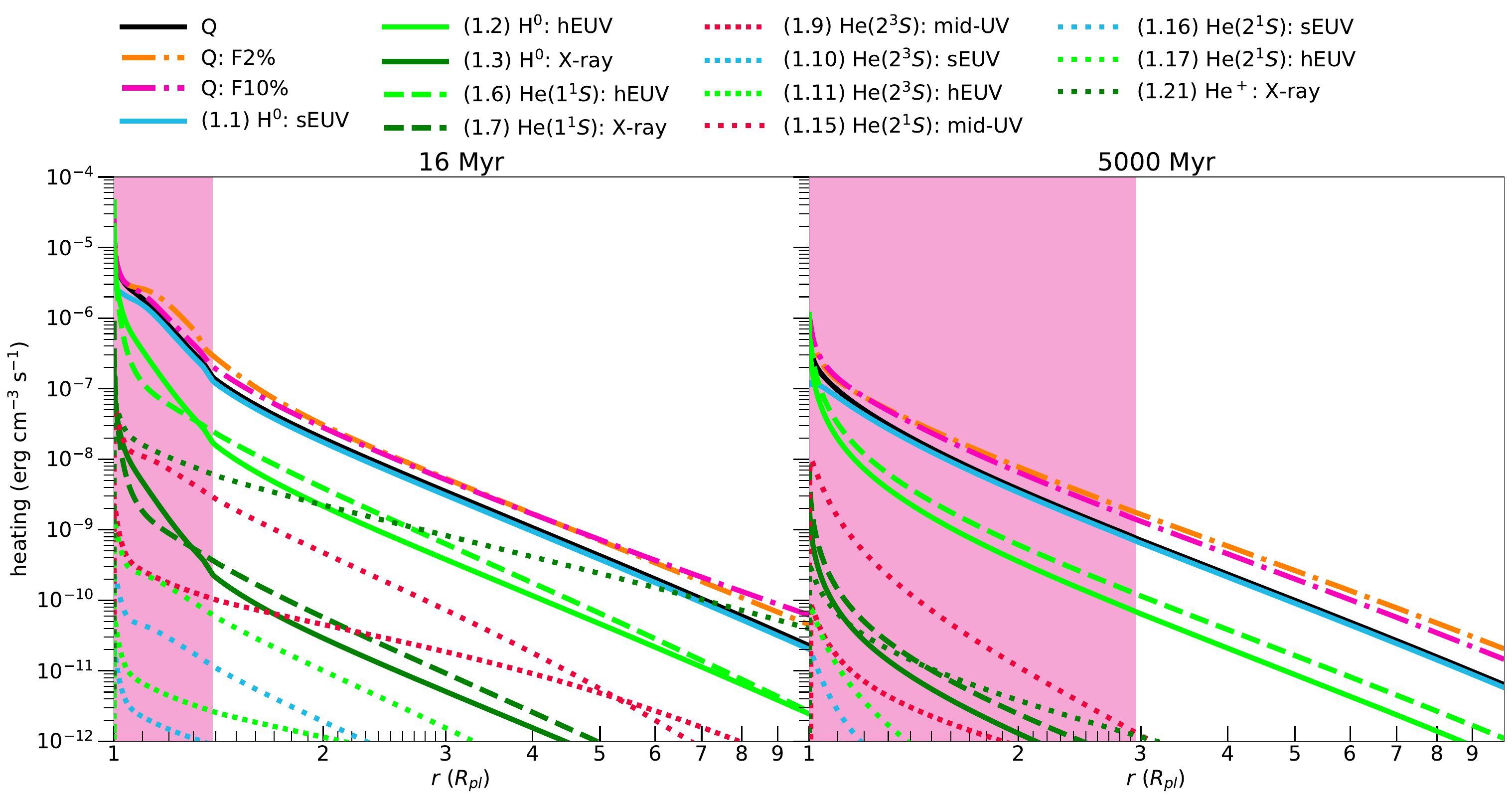}
    \caption{ The same as the upper-panel of Figure \ref{fig:heating_opt_depth}, now showing individual heating components for our F10\% (upper-panel) and F10\%PostProc (lower-panel) model.   } \label{fig:app1}
\end{figure*}

Figure \ref{fig:app1} shows the individual heating contributions for the F10\% model. Compared to the F2\% model (Figure \ref{fig:heating_opt_depth}), we see that heating due to helium unsurprisingly plays a more significant role for a greater helium abundance, however photoionisations of H$^0$ still dominates except in the very innermost atmosphere, where the photoionisation of He($1^1S$) takes over, at both young and old ages.

\begin{figure*}
    \includegraphics[width=0.45\textwidth]{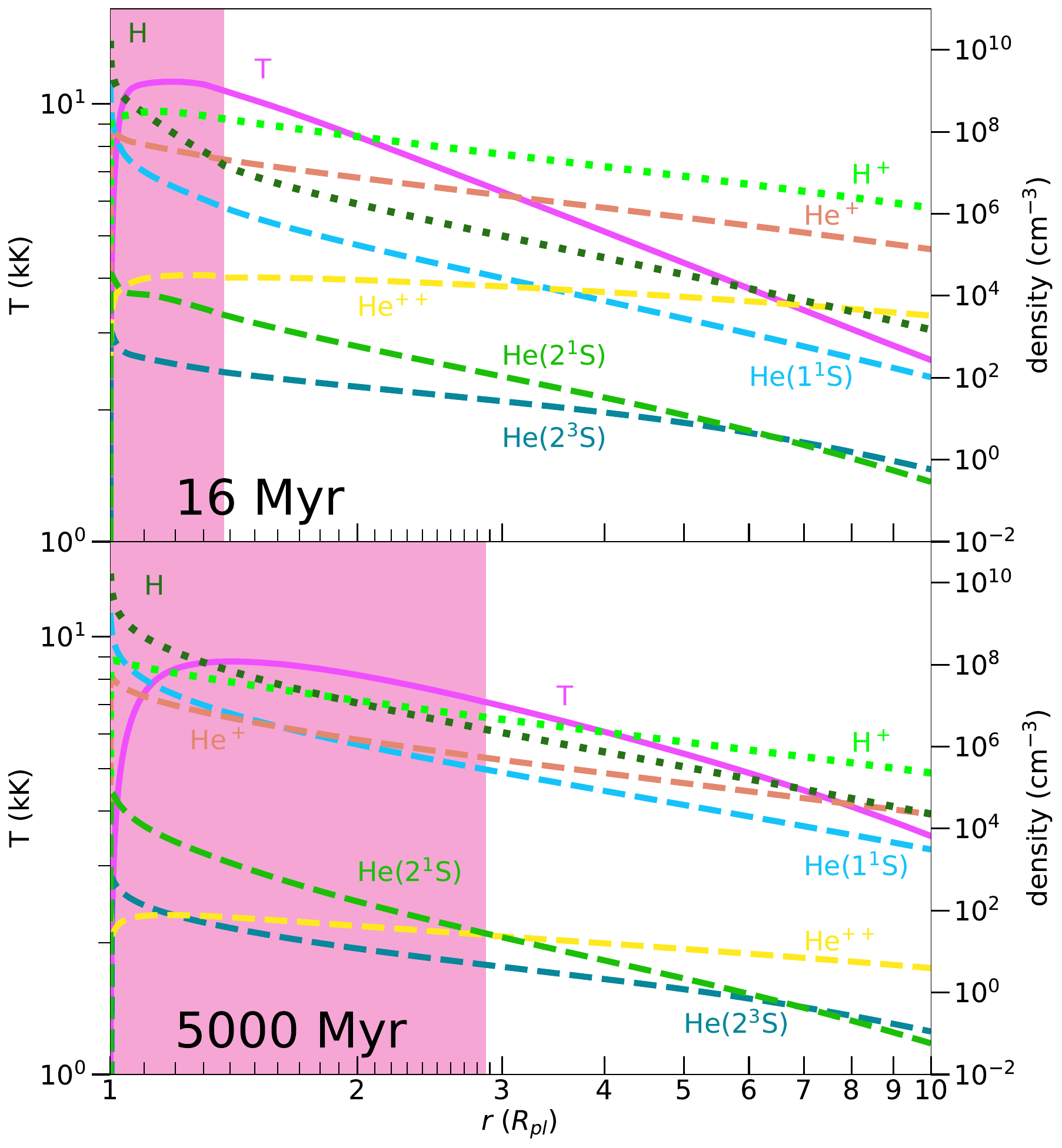}
	\includegraphics[width=0.45\textwidth]{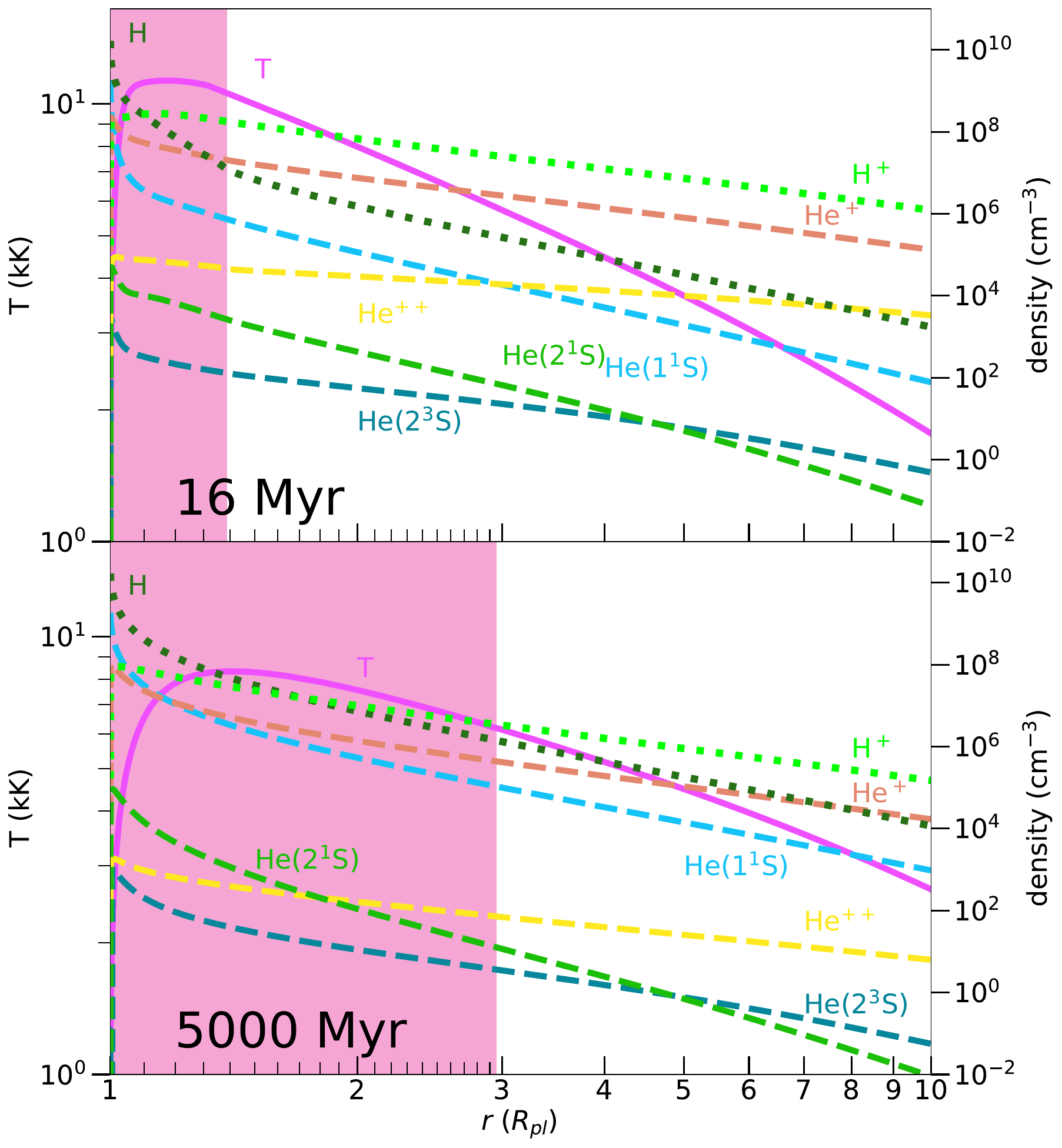}
    \caption{ The same as Figure \ref{fig:t_density_struct}, now for our F10\% (left-panel) and F10\%PostProc (right-panel) models.}  \label{fig:app2}
\end{figure*}

The left panel of Figure \ref{fig:app2} shows the resulting temperature profile for the F10\% model, where we see the temperature is very similar to the F2\% model (compare to Figure \ref{fig:t_density_struct}). The hydrodynamics of these models with two different abundances are very similar, reaching similar mass-loss rates, temperatures and velocities (with slightly lower speeds in the F10\% model, as discussed in Section \ref{sec:eff_He_frac} ). This is a result of hydrogen photoionisation being the dominant driver of the atmospheric escape in both cases as mentioned.

\section{Additional data on omitting helium energetics}
\label{sec:append_He_energetic}

The right panel of Figure \ref{fig:app2} shows the resulting temperature profile for the F10\%PostProc model, where now we see more substantial differences in the temperature profile. While the F10\% and F10\%PostProc reach similar maximum temperatures, the latter exhibits a faster temperature decay. The lowered availability of potential absorbers for photoionisation heating (neutral hydrogen only as opposed to neutral hydrogen and helium), and the omission of photons released in helium transitions, are responsible for the swifter temperature decay.

\section{Correction to v1}
\label{sec:correction}

This section describes the correction to the previous version of this work arXiv:2311.01313v1. This correction was accepted in MNRAS on March 28, 2025. As mentioned in its caption, the current Table \ref{tab:pop_depop} has the mentioned corrections applied. The figures are that of the original version.

 We correct the radiative decay rate from the helium 2$^1$S state. This was erroneously given previously as:  $$ A\left[\text{He}(2^1S \rightarrow 1^1S)\right]= 51.3 ~ 10^{-4} f_{2^1S}.$$
Instead, it now reads: $$ A\left[\text{He}(2^1S \rightarrow 1^1S)\right]= 51.3  f_{2^1S}.$$
This error does not affect the hydrodynamic predictions of atmospheric escape in the paper. Nor does it change the main conclusion of a weakening helium triplet signature as the atmospheric escape declines with evolution. However, it leads to slight reductions in the predicted helium triplet profiles for the theoretical modelled planets, with consistent decreases of $\sim$20 per cent in the He(2$^3$S) equivalent widths, at both the youngest and oldest ages, and at both of the assumed He/H fractions. While the population of the observationally important He(2$^3$S) state is hence only slightly affected by applying this correction, the population of the He(2$^1$S) state directly affected by the corrected decay rate is reduced by $\sim$2 orders of magnitude. The authors thank Dr.~Matth{\"a}us Schulik for bringing this error to our attention.

We also highlight typos present in the previous version of Table \ref{tab:pop_depop}: The exponent of 0.9 is now -0.9 in the recombination rate $\alpha_B\left[\text{H}^0\right]$. In the two charge exchange rows, the columns for `populates' and `depopulates' were correct, however the left-hand side of their rate equations erroneously had $\chi\left[\text{He}^+ \rightarrow \text{He}(1^1S)\right]$ in the place of $ \chi\left[\text{He}(1^1S) \rightarrow \text{He}^+\right]$, and vice versa. In the final row of collisional ionisations, $\Psi\left[{\text{H}^0}\right]$ is now corrected to $\Psi\left[{\text{He}^+}\right]$, and $f_{\text{He}^{++}}$ is now $f_{\text{He}^{+}}$. The corrections described in this paragraph are purely typographical and were not present in the model.


\bsp	
\label{lastpage}
\end{document}